# Transition to spatiotemporal chaos with multiple colliding pulse sequences of the nonlinear Schrödinger equation

Avner Peleg[1], Debananda Chakraborty[2]

[1]*Department of Mathematics, Azrieli College of Engineering, Jerusalem 9371207, Israel and*

[2]*Department of Mathematics and Technology,*

*Kean University, Union, New Jersey 07083, USA*

## Abstract

We present the first demonstration of transition to spatiotemporal chaos with multiple colliding pulse sequences in systems described by perturbed cubic nonlinear Schrödinger (NLS) equations. For this purpose, we consider propagation of multiple sequences of optical pulses in two distinct types of nonlinear waveguide arrays with cubic gain and loss. By employing a perturbation theory for NLS solitons, we show that the dynamics of pulse energies in the waveguide array systems is described by generalized Lotka-Volterra (LV) models, which exhibit dissipative chaos in a wide region in parameter space. We test the LV models' predictions for chaotic dynamics of pulse energies by extensive numerical simulations with perturbed systems of coupled-NLS equations. We find excellent agreement between the results of the LV and coupled-NLS models for energy dynamics in both types of waveguide array systems, despite the strong pulse pattern distortions and the strongly nonlinear nature of the dynamics.

Keywords: Soliton, nonlinear Schrödinger equation, Lotka-Volterra model, chaos, multisequence propagation, spatiotemporal chaos.

## I. INTRODUCTION

The rates of transmission of information in optical waveguide communication lines can be significantly enhanced by two main approaches. In the first approach, one employs very short pulses of light, while in the second approach, one employs multisequence transmission, where many pulse sequences are simultaneously transmitted through the same optical waveguide [1–4]. When the first approach is used, pulse duration is small, and as a result, the effects of nonlinear processes on the propagation become important [1–4]. The most important nonlinear process affecting the propagation of short pulses of light in optical waveguide communication lines is typically due to cubic (Kerr) nonlinearity. In this case, the propagation is described by the cubic nonlinear Schrödinger (NLS) equation [1–4]. It should be pointed out that the cubic NLS equation is one of the most widely used nonlinear wave models in science and engineering. Indeed, in addition to its extensive application in nonlinear waveguide optics it was successfully applied to describe the dynamics of Bose-Einstein condensates [5, 6], nonlinear waves in plasmas [7–9], and water wave dynamics [10–12]. The fundamental NLS solitons are the most important solutions of the cubic NLS equation due to their stability and shape preserving properties. Owing to these properties, fundamental NLS solitons are being considered for applications in numerous nonlinear optical waveguide systems, including optical waveguide communication lines, pulsed waveguide lasers, and optical switches [1–4, 13].

When the duration of the optical pulses is further decreased, the effects of additional physical processes, which can be regarded as perturbations to the cubic NLS equation, become important [1, 2, 7, 13, 14]. Principal examples for these perturbative processes are due to the effects of cubic loss [induced by two-photon absorption (2PA)] [15–17], cubic gain [induced by two-photon emission (2PE)] [18–20], delayed Raman response [1, 14], and third-order dispersion [1, 21, 22]. The perturbations can cause a variety of undesirable effects, including changes in the soliton's amplitude and energy, frequency, position, and phase, and distortion of the pulse shape due to emission of radiation (i.e., emission of unlocalized small-amplitude waves) [1, 2, 7, 23–26]. In fact, the cumulative impact of the perturbations at intermediate and long distances can be destructive even for single-soliton propagation [1, 25]. Furthermore, in the case of single-sequence soliton transmission, radiation emitted in the presence of perturbations induces long-range inter-soliton interaction, and this can cause the complete breakup of the soliton pattern [24, 27].

When multisequence transmission is employed, the pulses in each sequence propagate with the same central frequency and group velocity, but the central frequency and group velocity are different for pulses from different sequences [1, 3, 4]. As a result, intersequence pulse collisions are very frequent, and can lead to significant amplitude and position shifts, pulse pattern distortion due to radiation emission, transmission destabilization, and transmission errors. Consequently, major research efforts have been devoted to the study of intersequence pulse collisions in general [13, 28, 29], and to the investigation of intersequence collisions of NLS solitons in particular [1–4].

In several earlier works [30–39], we developed a collection of robust methods for stabilizing multisequence propagation of NLS solitons against the damaging effects of intersequence pulse collisions. The methods combine the following two ingredients: (1) stabilization against collision-induced amplitude (and energy) shifts, (2) stabilization against radiation emission. Stabilization against collision-induced amplitude (and energy) shifts was realized by showing that the dynamics of soliton amplitudes (and energies) in nonlinear $J$-sequence optical waveguide transmission systems can be approximately described by generalized $J$-dimensional Lotka-Volterra (LV) models. The specific form of the LV model is determined by the dissipative perturbation terms in the coupled-NLS model that describes the propagation. Stability and bifurcation analysis for the equilibrium points of the LV models was used to develop waveguide setups that lead to robust transmission stabilization in the presence of the main dissipative processes in nonlinear waveguides, including delayed Raman response and cubic gain and loss [33, 34, 36–39]. Stabilization against radiation emission was achieved by the following methods. In the first method, we used perturbation-induced shifting of the soliton's frequency together with frequency-dependent linear gain-loss [36]. In the second method, we applied nonlinear waveguides with a weak Ginzburg-Landau (GL) gain-loss profile, consisting of linear loss, cubic gain, and quintic loss [32–34, 39]. In the third method, we combined perturbation-induced shifting of the soliton's frequency with weak GL gain-loss [37]. The application of these stabilization methods enabled the observation of stable multisequence soliton transmission over distances of 1000 dispersion lengths or more in numerical simulations with perturbed coupled-NLS models [33–39]. Furthermore, the dynamics of soliton amplitudes (and energies) that was observed in the coupled-NLS simulations accurately coincided with the LV models' predictions.

The stabilization methods of multisequence transmission with solitons of the cubic NLS equation in Refs. [32–34, 36–39] are typically applicable in weakly nonlinear optical waveguide systems. Indeed, for intermediate or large perturbations, or for sufficiently large propagation distances, the soliton pattern distortions, observed in the coupled-NLS simulations, become significant [35–38]. Consequently, in this case, the results of the coupled-NLS simulations for amplitude dynamics deviate significantly from the LV models' predictions, and soliton transmission becomes unstable [35–38]. For these reasons, so far, all the studies of multisequence nonlinear optical waveguide transmission, which demonstrated stable propagation and LV amplitude (or energy) dynamics, were limited to weakly nonlinear amplitude (or energy) dynamics [30–37, 39], or to intermediate nonlinear amplitude (or energy) dynamics [38]. Furthermore, in all the previous studies, LV dynamics of pulse amplitudes was observed only with very weak pulse pattern distortions. It follows that all the previous studies in Refs. [30–39] failed to address the following major questions about the dynamics of pulse amplitudes (or energies) in multisequence nonlinear optical waveguide transmission systems. The first key question concerns the possibility of observing strongly nonlinear and chaotic dynamics of pulse amplitudes (or energies) in multisequence nonlinear optical waveguide transmission. The second key question concerns the feasibility of demonstrating LV dynamics of pulse amplitudes (or energies) in the presence of strong perturbations or in the presence of strong pulse pattern distortions. Due to the central roles of the cubic NLS and LV models in nonlinear science and engineering, these two questions are clearly of great importance in a more general context, namely, the context of chaotic dynamics in spatially extended systems and spatiotemporal chaos.

In the current paper, we address the two aforementioned key questions about the dynamics of pulse energies in systems described by perturbed coupled cubic NLS models. For this purpose, we consider propagation of multiple sequences of pulses of light in two distinct types of nonlinear waveguide arrays with cubic gain and loss. By employing a perturbation theory for the fundamental soliton of the cubic NLS equation, we show that the dynamics of pulse energies in these systems is described by two types of generalized LV models with conservation of the total energy, which exhibit dissipative chaos in a wide region in parameter space. We check the predictions of the LV models for chaotic dynamics of pulse energies by extensive numerical simulations with perturbed systems of coupled cubic NLS equations. We find excellent agreement between the results of the LV and perturbed coupled-NLS mod-els for energy dynamics in both types of waveguide array systems, despite the strong pulse pattern distortions and the strongly nonlinear nature of the dynamics. Thus, our study provides the first demonstration of transition to spatiotemporal chaos with multiple colliding pulse sequences in systems described by perturbed coupled-NLS models. Furthermore, due to the central roles of the cubic NLS and LV models in nonlinear science and engineering, our results are of broad interest in the context of studies of spatiotemporal chaos in general.

It is worth discussing some previous works in the general areas of nonlinear waves and pattern formation, which are closely but indirectly related to the current work. Many works considered chaotic scattering in slow collisions between solitons of conservative nonintegrable nonlinear wave models [40–43]. It was found that in many of these systems, the graph of the final relative velocity between the colliding solitons, $v_{out}$, vs the initial relative velocity, $v_{in}$, exhibits fractal resonant structure below some critical initial relative velocity [40–43]. Additionally, various asymptotic methods have been used to obtain iterated maps, which explained the chaotic nature of the soliton scattering, and the observed fractal structures in the graphs of $v_{out}$ vs $v_{in}$ [42, 43]. However, the nonlinear wave models and the reduced dynamical models in Refs. [40–43] were all conservative, while the perturbed coupled-NLS models and the generalized LV models in the current work are dissipative. Furthermore, the physical setups considered in Refs. [40–43] involved only single slow two-soliton collisions. In contrast, the physical setups considered in the current paper involve large numbers of fast two-pulse collisions occurring along ultra-long propagation distances, where the pulse shapes are often very different from the soliton shape due to strong pulse pattern distortions.

Chaotic dynamics in spatially extended setups was also widely investigated in systems described by the complex GL equation [44–48]. In particular, various types of spatiotemporal chaos were demonstrated with a single complex GL equation, including phase chaos, defect chaos, defect-hole chaos, and spatiotemporal intermittency [44, 45, 47]. Additionally, dissipative chaotic dynamics of a single breathing soliton-like state was demonstrated in Refs. [46, 48]. However, no explicit general low-dimensional dynamical model explaining the chaotic behavior was presented in these works. Furthermore, the chaotic dynamics in these systems was demonstrated with a single partial differential equation (PDE) model, and not with a system of coupled PDEs.

The results of the current paper are also indirectly related to the results of our previous works in Refs. [49–51]. More specifically, in Refs. [49–51], we showed that the stochastic dynamics of soliton parameters in certain soliton-based multisequence optical fiber transmission systems is intermittent in the sense that it is very sensitive to bit-pattern randomness. Furthermore, we established a surprising relation between the dynamics of soliton energy in a simple setup of the fiber optics transmission system, and the behavior of the local average of energy dissipation in random cascade models for fully developed turbulent flow [51]. However, no chaotic dynamics of the soliton energy was demonstrated in these works.

The other sections of the paper are organized in the following manner. In section II, we analyze the main effects of a single fast two-soliton collision in the presence of weak cubic loss or gain. In section III, we use the results of section II to derive unshifted and shifted LV models for energy dynamics in two distinct types of multisequence nonlinear waveguide array systems with cubic loss and gain. We also present the corresponding unshifted and shifted perturbed coupled-NLS models, which describe the propagation of multiple pulse sequences in these waveguide arrays. In section IV, we analyze the dynamics of pulse energies in the two unshifted LV models that were derived in section III, and demonstrate the dynamics by numerical solution of the two LV models. In section V, we present the results of the simulations with the perturbed coupled-NLS models for energy and pulse pattern dynamics. Furthermore, we compare the simulations results for energy dynamics with the predictions of the LV models. Our conclusions are summarized in section VI. In appendix A, we present the theoretical prediction for the pulse patterns, which is used in the analysis of the coupled-NLS simulations results for pulse pattern dynamics.

## II. THE EFFECTS OF A SINGLE FAST TWO-SOLITON COLLISION IN THE PRESENCE OF WEAK CUBIC LOSS OR GAIN

Let us analyze the main effects of a single fast collision between two fundamental solitons of the unperturbed cubic NLS equation in the presence of weak cubic loss or cubic gain. The cubic loss or cubic gain arises due to 2PA or 2PE, respectively. The derivation is somewhat similar to the one presented in Ref. [31] with some important generalizations, which help motivate the coupled-NLS models that we use, and provide insight into the dynamics observed in the coupled-NLS simulations.

Each of the colliding solitons satisfies the unperturbed NLS equation

$$i\partial_z\psi_j + \partial_t^2\psi_j + 2|\psi_j|^2\psi_j = 0\,, \tag{1}$$

where $j = 1$ for pulse 1 and $j = 2$ for pulse 2. In Eq. (1), $\psi_j$ is proportional to the envelope of the electric field of the $j$th pulse, $z$ is propagation distance, and $t$ is time [52]. The fundamental soliton solution of Eq. (1) is given by

$$\psi_{sj}(t, z) = \Psi_{sj}(t, z) \exp(i\chi_j) = \eta_j \exp(i\chi_j)/\cosh(x_j), \tag{2}$$

where $x_j = \eta_j (t - y_j + 2\beta_j z)$ and $\chi_j = \alpha_j - \beta_j(t - y_j) + \left(\eta_j^2 - \beta_j^2\right) z$. The parameters $\eta_j$, $\beta_j$, $y_j$, and $\alpha_j$ are the $j$th soliton's amplitude, frequency, position, and phase. The parameters $\eta_j$ and $\beta_j$ are also related to the $j$th soliton's energy and group velocity, respectively.

We consider a fast collision between two NLS solitons in the presence of weak cubic loss or cubic gain with slow variation of the cubic loss or gain coefficient. Since we are interested in a fast collision, the frequency spacing between the solitons $\Delta\beta = \beta_2 - \beta_1$ satisfies $|\Delta\beta| \gg 1$. We assume that the effects of degenerate (intra-pulse) 2PA or 2PE are much weaker than the effects of nondegenerate (inter-pulse) 2PA or 2PE, a situation that can be realized, for example, in certain direct-gap semiconductors [53–55]. We also assume that the nondegenerate (inter-pulse) cubic nonlinearity is much weaker than the nondegenerate 2PA or 2PE, which can also be realized in the same semiconductors [54, 56, 57]. We therefore neglect the effects of intra-pulse 2PA and 2PE and the effects of inter-pulse cubic nonlinearity in the coupled-NLS model. We point out that the analysis in Refs. [54–57] was based on the Kramers-Krönig relations, and therefore, all our assumptions are fully consistent with these relations. Thus, under these assumptions, the dynamics of the collision is described by the following system of weakly perturbed coupled-NLS equations:

$$\begin{aligned} i\partial_z\psi_1 + \partial_t^2\psi_1 + 2|\psi_1|^2\psi_1 &= i\epsilon_3 g_{12}(z)|\psi_2|^2\psi_1 \\ i\partial_z\psi_2 + \partial_t^2\psi_2 + 2|\psi_2|^2\psi_2 &= i\epsilon_3 g_{21}(z)\epsilon|\psi_1|^2\psi_2. \end{aligned} \tag{3}$$

In Eq. (3), $\epsilon_3$ is the cubic loss or gain coefficient, which satisfies $0 < \epsilon_3 \ll 1$, and $g_{12}(z)$ and $g_{21}(z)$ are slowly varying with $z$ [58]. Note that $g_{12}(z) < 0$ and $g_{21}(z) < 0$ for cubic loss, and $g_{12}(z) > 0$ and $g_{21}(z) > 0$ for cubic gain. The second and third terms on the left hand sides of Eq. (3) describe second-order dispersion effects and intra-pulse effects due to cubic nonlinearity, respectively. The terms on the right hand sides of Eq. (3) describe the effects of inter-pulse interaction due to weak cubic loss or gain with slow variation of the cubic loss or gain coefficient.

In addition to the assumptions $0 < \epsilon_3 \ll 1$ and $|\Delta\beta| \gg 1$, we assume a complete collision, such that the two solitons are well separated at $z = 0$ and at the final distance $z = z_f$. Under

these assumptions, we can employ a variant of the perturbation method, developed in Refs. [59, 60], and successfully applied for studying soliton collisions in the presence of third-order dispersion [59], quintic nonlinearity [60], delayed Raman response [61], cubic loss [31], and quintic loss [32]. In accordance with the perturbation method, we look for a solution of the perturbed coupled-NLS model (3) in the form

$$\psi_j(t, z) = \psi_{sj}(t, z) + \phi_j(t, z), \tag{4}$$

where $j = 1, 2$, $\psi_{sj}$ are the soliton solutions of Eq. (1), and $\phi_j$ describe corrections to $\psi_{sj}$ due to weak collision-induced effects [62]. We substitute relation (4) into Eq. (3) and obtain equations for the $\phi_j$. We focus attention on the calculation of $\phi_1$, as the calculation of $\phi_2$ is similar. Taking into account only leading-order effects of the collision, we can neglect terms containing $\phi_j$ on the right hand side of the resulting equation. We therefore obtain:

$$i\partial_z\phi_1 + \partial_t^2\phi_1 + 4|\psi_{s1}|^2\phi_1 + 2\psi_{s1}^2\phi_1^* = i\epsilon_3 g_{12}(z)|\psi_{s2}|^2\psi_{s1}. \tag{5}$$

We now substitute $\psi_{sj}(t, z) = \Psi_{sj}(t, z)\exp[i\chi_j(t, z)]$ and $\phi_1(t, z) = \Phi_1(t, z)\exp[i\chi_1(t, z)]$, where $\Psi_{sj}$ and $\chi_j$ are real-valued, into Eq. (5). We obtain the following equation for $\Phi_1$:

$$i\partial_z\Phi_1 + (\partial_t^2 - \eta_1^2)\Phi_1 + 4\Psi_{s1}^2\Phi_1 + 2\Psi_{s1}^2\Phi_1^* = i\epsilon_3 g_{12}(z)\Psi_{s2}^2\Psi_{s1}. \tag{6}$$

The term on the right hand side of Eq. (6) is of order $\epsilon_3$. Additionally, since the collision length $\Delta z_c$ is of order $1/|\Delta\beta|$, the term $i\partial_z\Phi_1$ is of order $|\Delta\beta| \times O(\Phi_1)$. Equating the orders of $i\partial_z\Phi_1$ and $i\epsilon_3 g_{12}(z)\Psi_{s2}^2\Psi_{s1}$, we find that $\Phi_1$ is of order $\epsilon_3/|\Delta\beta|$. We also notice that all other terms on the left hand side of Eq. (6) are of order $\epsilon_3/|\Delta\beta|$ or higher, and can therefore be neglected. As a result, the equation for $\Phi_1$ in the leading order of the perturbative calculation is [31]:

$$\partial_z\Phi_1 = \epsilon_3 g_{12}(z)\Psi_{s2}^2\Psi_{s1}. \tag{7}$$

Integration with respect to $z$ yields [31]

$$\Delta\Phi_1^{(c)}(t, z_c) = \frac{\epsilon_3 g_{12}(z_c)\eta_1(0)\eta_2(0)}{|\Delta\beta|\cosh(x_1)}, \tag{8}$$

where $z_c$ is the collision distance, i.e., the distance at which the maxima of $|\psi_j(t, z)|$ coincide. From Eq. (8) it follows that the only effect of the collision on the four soliton parameters in the leading order of the perturbative calculation is an amplitude shift $\Delta\eta_1^{(c)}$, which is given by [31]:

$$\Delta\eta_1^{(c)} = 2\epsilon_3 g_{12}(z_c)\eta_1(0)\eta_2(0)/|\Delta\beta|. \tag{9}$$

Since $\Delta\eta_1^{(c)}$ is proportional to $\epsilon_3/|\Delta\beta|$, in $J$-sequence transmission systems, the cumulative amplitude shift due to the collisional cubic loss or gain effects is proportional to $J\epsilon_3/T$, where $T$ is the temporal separation between adjacent solitons within each sequence [31]. Consequently, in multisequence transmission systems with large $J$ values, the collisional cubic loss or gain effects are dominant compared with the effects of cubic loss or gain on single-pulse propagation. This behavior was confirmed by extensive numerical simulations with perturbed coupled-NLS models in Refs. [37, 38].

It is useful to write Eq. (9) in terms of the soliton energies, $\int_{-\infty}^{\infty} dt\, |\psi_{sj}(t,z)|^2$. We denote $U_j(z) = \int_{-\infty}^{\infty} dt\, |\psi_{sj}(t,z)|^2/2 = \eta_j(z)$. Using this relation, we obtain

$$\Delta U_1^{(c)} = 2\epsilon_3 g_{12}(z_c)U_1(0)U_2(0)/|\Delta\beta| \tag{10}$$

for the collisional energy shift of soliton 1. As we will see in section V, in the numerical simulations of multisequence propagation, the soliton sequences undergo very strong distortions. As a result, one cannot identify the amplitudes of the highly distorted pulses, and Eq. (9) is less useful in the analysis of the simulations results. In contrast, one can accurately measure the total energies of the pulse sequences in the simulations. Moreover, it turns out that Eq. (10) is highly useful in the analysis and in the interpretation of the numerical simulations results despite the strong pulse pattern distortions.

## III. LV AND COUPLED-NLS MODELS FOR DYNAMICS IN MULTISEQUENCE NONLINEAR WAVEGUIDE ARRAY SYSTEMS

In this section, we use the results of section II to derive unshifted and shifted generalized LV models for energy dynamics in two different types of multisequence nonlinear waveguide array transmission systems with cubic loss and gain. We then present the corresponding unshifted and shifted perturbed coupled-NLS models, which describe the propagation of multiple pulse sequences in these waveguide array systems. The perturbed coupled-NLS models that we present are expected to yield the same energy dynamics as the generalized LV models.

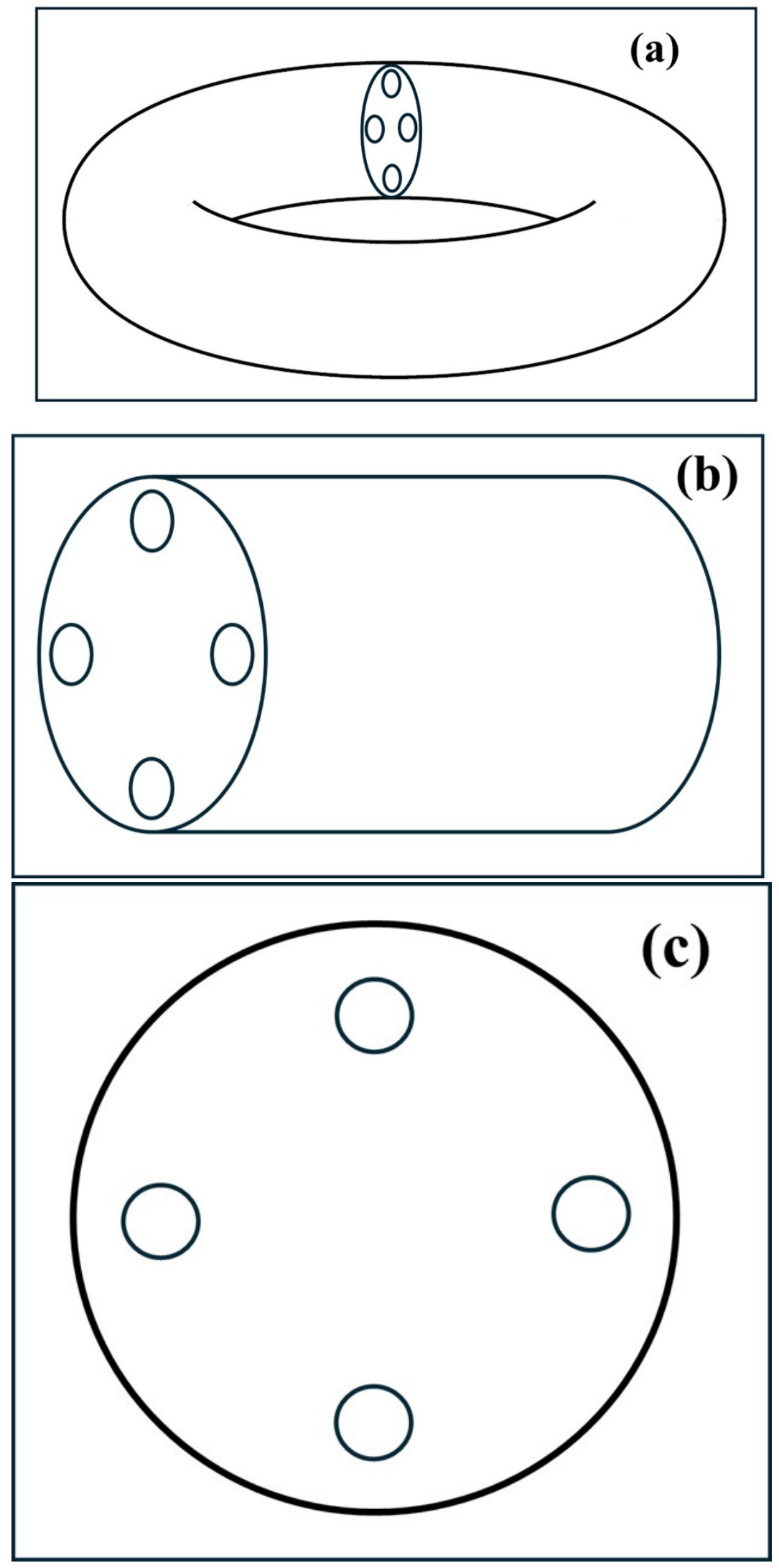

FIG. 1: A sketch of the waveguide array loops considered in the paper. (a) A side view. (b) A closer side view. (c) The cross section.

### A. The two unshifted Di Cera models for energy dynamics

We consider transmission of $J$ periodic soliton sequences in a nonlinear optical waveguide array loop, consisting of $J$ close waveguides with a circular arrangement. A sketch of the waveguide array loop is shown in Fig. 1 for the case $J = 4$. Each soliton sequence propagates inside its own waveguide in the presence of broadband cubic loss or gain. Due to the broadband nature of the cubic gain-loss, the solitons in each sequence interact with solitons from other sequences during intersequence collisions. Similar to Ref. [39], we assume that only the interaction between solitons from nearest-neighbor (NN) waveguides is significant (see also section III B). We also assume that the temporal separation between adjacent solitons within each sequence $T$ (the time-slot width) is a large constant, $T \gg 1$, and that the constant frequency difference between adjacent sequences $|\Delta\beta|$ satisfies $|\Delta\beta| \gg 1$. Both assumptions represent the typical situation in broadband multisequence soliton-based transmission [63–65]. To derive the LV models for dynamics of soliton energies, we employ the following supplementary assumptions, which were also used in Refs. [30–34, 36–39]. (1) The energies are equal for all solitons from the same sequence, but are not necessarily equal for solitons from different sequences. This setup corresponds, for example, to phase-shift-keyed soliton transmission [66]. (2) Since the soliton sequences propagate in a waveguide array loop, they are subject to periodic boundary conditions [4, 63, 64]. (3) Since $T \gg 1$, intrasequence interaction is exponentially weak and is neglected. (4) High-order effects due to radiation emission are also neglected.

Under the assumptions listed in the preceding paragraph, the solitons sequences remain periodic throughout the propagation. As a result, the energies of all solitons in a given sequence follow the same dynamics. We can therefore describe the spatiotemporal dynamics of the $J$ pulse sequences by a reduced system of $J$ ordinary differential equations for the soliton energies, which takes the form of a $J$-dimensional generalized LV model. We derive the generalized LV model by employing Eq. (10) for the energy shift in a single fast two-soliton collision in the presence of weak cubic loss or gain, collision-rate calculations similar to the ones in Refs. [30–32], and the NN interaction property. The derivation yields the following generalized LV model for the dynamics of soliton energies:

$$\dot{U}_j = \frac{4\epsilon_3 U_j}{T}\left[g_{j\,j-1}(z)U_{j-1} + g_{j\,j+1}(z)U_{j+1}\right] \quad (11)$$

for sequences $2 \le j \le J-1$,

$$\dot{U}_1 = \frac{4\epsilon_3 U_1}{T}\left[g_{1J}(z)U_J + g_{12}(z)U_2\right] \quad (12)$$

for sequence $j = 1$, and

$$\dot{U}_J = \frac{4\epsilon_3 U_J}{T}\left[g_{J\,J-1}(z)U_{J-1} + g_{J1}(z)U_1\right] \quad (13)$$

for sequence $j = J$, where $\dot{U}_j \equiv dU_j/dz$.

We now consider two four-dimensional (4D) LV models for energy dynamics in four-sequence waveguide array transmission. The models are equivalent to the two 4D LV models that were studied by Di Cera et al. in Ref. [67] and by Mori and Di Cera in Ref. [68] in the context of cyclic networks of autocatalytic chemical reactions. We will refer to these LV models as the unshifted 1989 Di Cera model and the unshifted 1990 Di Cera model, respectively. As shown in Refs. [67, 68] and as will be demonstrated in section IV of the current paper, the two models exhibit dissipative chaotic dynamics in a wide region in parameter space. As a result, these models appear as good candidates for the demonstration of robust transition to spatiotemporal chaos in the spatially extended waveguide array systems that are studied in the current paper.

#### 1. *The unshifted 1989 Di Cera model*

The 4D unshifted 1989 Di Cera model is obtained by assigning the following values to the $g_{jk}(z)$ coefficients:

$$\begin{aligned} &g_{14}(z) = \tilde{b}_{14}, \;\; g_{12}(z) = -\tilde{b}_{12}\left[m - \eta_2(z)\right], \;\; g_{21}(z) = \tilde{b}_{12}\left[m - \eta_2(z)\right], \\ &g_{23}(z) = -\tilde{b}_{23}, \; g_{32}(z) = \tilde{b}_{23}, \; g_{34}(z) = -\tilde{b}_{34}, \; g_{43}(z) = \tilde{b}_{34}, \; g_{41}(z) = -\tilde{b}_{14}, \end{aligned} \tag{14}$$

where $m$, $\tilde{b}_{14}$, $\tilde{b}_{12}$, $\tilde{b}_{23}$, and $\tilde{b}_{34}$ are constants. Thus, in this system, sequence 1 propagates in the presence of cubic gain interaction with sequence 4 and cubic loss or gain interaction with sequence 2. Sequence 2 propagates in the presence of cubic gain or loss interaction with sequence 1 and cubic loss interaction with sequence 3. Sequence 3 propagates in the presence of cubic gain interaction with sequence 2 and cubic loss interaction with sequence 4. Sequence 4 propagates in the presence of cubic gain interaction with sequence 3 and cubic loss interaction with sequence 1. Substituting the coefficients from Eq. (14) into Eqs. (11)-(13), we arrive at the unshifted 1989 Di Cera model:

$$\begin{aligned} \dot{U}_1 &= \frac{4\epsilon_3 U_1}{T}\left(\tilde{b}_{14}U_4 - m\tilde{b}_{12}U_2 + \tilde{b}_{12}U_2^2\right) \\ \dot{U}_2 &= \frac{4\epsilon_3 U_2}{T}\left(m\tilde{b}_{12}U_1 - \tilde{b}_{12}U_1U_2 - \tilde{b}_{23}U_3\right) \\ \dot{U}_3 &= \frac{4\epsilon_3 U_3}{T}\left(\tilde{b}_{23}U_2 - \tilde{b}_{34}U_4\right) \\ \dot{U}_4 &= \frac{4\epsilon_3 U_4}{T}\left(\tilde{b}_{34}U_3 - \tilde{b}_{14}U_1\right). \end{aligned} \tag{15}$$

It is straightforward to show that the total energy, $M = \sum_{j=1}^{4} U_j(z)$, is a conserved quantity of Eq. (15) [67].

*2. The unshifted 1990 Di Cera model*

The 4D unshifted 1990 Di Cera model is obtained by assigning the following values to the $g_{jk}(z)$ coefficients:

$$
\begin{aligned}
& g_{14}(z) = \tilde{b}_{14}, \;\; g_{12}(z) = -\tilde{b}_{12}\left[m - \eta_2(z)\right], \;\; g_{21}(z) = \tilde{b}_{12}\left[m - \eta_2(z)\right], \\
& g_{23}(z) = -\tilde{b}_{23}, \;\; g_{32}(z) = \tilde{b}_{23}, \;\; g_{34}(z) = -\tilde{b}_{34}\left[m - \eta_4(z)\right], \\
& g_{43}(z) = \tilde{b}_{34}\left[m - \eta_4(z)\right], \;\; g_{41}(z) = -\tilde{b}_{14}. \qquad (16)
\end{aligned}
$$

We see that in the current waveguide array system, the propagation setup for sequences 1 and 2 is similar to the one in subsection III A 1. Additionally, sequence 3 propagates in the presence of cubic gain interaction with sequence 2 and cubic loss or gain interaction with sequence 4. Furthermore, sequence 4 propagates in the presence of cubic gain or loss interaction with sequence 3 and cubic loss interaction with sequence 1. Substitution of the coefficients from Eq. (16) into Eqs. (11)-(13) yields the unshifted 1990 Di Cera model:

$$
\begin{aligned}
\dot{U}_1 &= \frac{4\epsilon_3 U_1}{T}\left(\tilde{b}_{14}U_4 - m\tilde{b}_{12}U_2 + \tilde{b}_{12}U_2^2\right) \\
\dot{U}_2 &= \frac{4\epsilon_3 U_2}{T}\left(m\tilde{b}_{12}U_1 - \tilde{b}_{12}U_1U_2 - \tilde{b}_{23}U_3\right) \\
\dot{U}_3 &= \frac{4\epsilon_3 U_3}{T}\left(\tilde{b}_{23}U_2 - m\tilde{b}_{34}U_4 + \tilde{b}_{34}U_4^2\right) \\
\dot{U}_4 &= \frac{4\epsilon_3 U_4}{T}\left(m\tilde{b}_{34}U_3 - \tilde{b}_{34}U_3U_4 - \tilde{b}_{14}U_1\right). \qquad (17)
\end{aligned}
$$

Similar to the situation in Eq. (15), the total energy $M$ is a conserved quantity of Eq. (17) [68].

## B. The corresponding unshifted perturbed coupled-NLS models

We consider propagation of $J$ sequences of pulses of light in a nonlinear optical waveguide array loop. The array consists of $J$ waveguides and has the circular arrangement, which is shown in Fig. 1 for the case $J = 4$. Each pulse sequence propagates within its own waveguide in the presence of second-order dispersion, narrowband (degenerate) cubic nonlinearity, and broadband (nondegenerate) cubic loss or gain. Due to the broadband character of the cubic gain-loss, the pulses in each sequence interact with pulses from other sequences during intersequence collisions. However, similar to Ref. [39], we assume that the magnitude of the electric field of the pulses from a given sequence decays sufficiently fast with increasing distance from the pulse sequence's waveguide, such that only the interaction between pulses from the two NN waveguides is significant, while all other intersequence interactions are negligible. Under these assumptions, the propagation of the $J$ pulse sequences is described by the following perturbed coupled-NLS model:

$$i\partial_z\psi_j + \partial_t^2\psi_j + 2|\psi_j|^2\psi_j = i\epsilon_3 g_{jj-1}(z)|\psi_{j-1}|^2\psi_j + i\epsilon_3 g_{jj+1}(z)|\psi_{j+1}|^2\psi_j, \tag{18}$$

for $2 \le j \le J-1$,

$$i\partial_z\psi_1 + \partial_t^2\psi_1 + 2|\psi_1|^2\psi_1 = i\epsilon_3 g_{1J}(z)|\psi_J|^2\psi_1 + i\epsilon_3 g_{12}(z)|\psi_2|^2\psi_1, \tag{19}$$

for $j = 1$, and

$$i\partial_z\psi_J + \partial_t^2\psi_J + 2|\psi_J|^2\psi_J = i\epsilon_3 g_{JJ-1}(z)|\psi_{J-1}|^2\psi_J + i\epsilon_3 g_{J1}(z)|\psi_1|^2\psi_J, \tag{20}$$

for $j = J$ [69]. Under the assumptions stated in subsection III A, we expect that the generalized LV model (11)-(13) would provide a good approximation to the dynamics of pulse energies in the full (spatially extended) coupled-NLS model (18)-(20).

Let us consider specific perturbed coupled-NLS models for four-sequence waveguide array transmission, which can potentially yield the energy dynamics that is described by the unshifted 1989 and 1990 Di Cera models (15) and (17). When we use the $g_{jk}(z)$ values of Eq. (14), we obtain the perturbed coupled-NLS model corresponding to the unshifted 1989 Di Cera model:

$$\begin{aligned}
&i\partial_z\psi_1 + \partial_t^2\psi_1 + 2\,|\psi_1|^2\,\psi_1 = i\epsilon_3\tilde{b}_{14}\,|\psi_4|^2\,\psi_1 + i\epsilon_3\tilde{b}_{12}\,[\eta_2(z) - m]\,|\psi_2|^2\,\psi_1\\
&i\partial_z\psi_2 + \partial_t^2\psi_2 + 2\,|\psi_2|^2\,\psi_2 = i\epsilon_3\tilde{b}_{12}\,[m - \eta_2(z)]\,|\psi_1|^2\,\psi_2 - i\epsilon_3\tilde{b}_{23}\,|\psi_3|^2\,\psi_2\\
&i\partial_z\psi_3 + \partial_t^2\psi_3 + 2\,|\psi_3|^2\,\psi_3 = i\epsilon_3\tilde{b}_{23}\,|\psi_2|^2\,\psi_3 - i\epsilon_3\tilde{b}_{34}\,|\psi_4|^2\,\psi_3\\
&i\partial_z\psi_4 + \partial_t^2\psi_4 + 2\,|\psi_4|^2\,\psi_4 = i\epsilon_3\tilde{b}_{34}\,|\psi_3|^2\,\psi_4 - i\epsilon_3\tilde{b}_{14}\,|\psi_1|^2\,\psi_4\,.
\end{aligned} \tag{21}$$

Additionally, when we use the $g_{jk}(z)$ values of Eq. (16), we arrive at the perturbed coupled-NLS model corresponding to the unshifted 1990 Di Cera model:

$$\begin{aligned}
&i\partial_z\psi_1 + \partial_t^2\psi_1 + 2\,|\psi_1|^2\,\psi_1 = i\epsilon_3\tilde{b}_{14}\,|\psi_4|^2\,\psi_1 + i\epsilon_3\tilde{b}_{12}\,[\eta_2(z) - m]\,|\psi_2|^2\,\psi_1\\
&i\partial_z\psi_2 + \partial_t^2\psi_2 + 2\,|\psi_2|^2\,\psi_2 = i\epsilon_3\tilde{b}_{12}\,[m - \eta_2(z)]\,|\psi_1|^2\,\psi_2 - i\epsilon_3\tilde{b}_{23}\,|\psi_3|^2\,\psi_2\\
&i\partial_z\psi_3 + \partial_t^2\psi_3 + 2\,|\psi_3|^2\,\psi_3 = i\epsilon_3\tilde{b}_{23}\,|\psi_2|^2\,\psi_3 + i\epsilon_3\tilde{b}_{34}\,[\eta_4(z) - m]\,|\psi_4|^2\,\psi_3\\
&i\partial_z\psi_4 + \partial_t^2\psi_4 + 2\,|\psi_4|^2\,\psi_4 = i\epsilon_3\tilde{b}_{34}\,[m - \eta_4(z)]\,|\psi_3|^2\,\psi_4 - i\epsilon_3\tilde{b}_{14}\,|\psi_1|^2\,\psi_4\,.
\end{aligned} \tag{22}$$

In what follows, we will also refer to Eqs. (21) and (22) as the unshifted 1989 coupled-NLS model and the unshifted 1990 coupled-NLS model, respectively. We emphasize that the two studies in Refs. [67] and [68] considered only LV models of the form (15) and (17). No spatially extended PDE models were considered in these works.

### C. The two shifted Di Cera models for energy dynamics

The two unshifted Di Cera models (15) and (17) exhibit chaotic dynamics in a wide region in parameter space and therefore, the corresponding coupled-NLS models (21) and (22) are good candidates for the demonstration of transition to spatiotemporal chaos. However, in the context of the multisequence waveguide array systems that are considered in the current paper, the two unshifted Di Cera models suffer from the following deficiency. In the chaotic dynamics region, the total energies $U_j(z)$ attain very small values at some distances. It follows that at these distances, the optical pulses should be rather flat and possess small amplitudes. In this case, the optical pulses might be subject to strong modulation instability, and the energy dynamics that is obtained in numerical simulations with Eqs. (21) and (22) might deviate from the dynamics in the two unshifted Di Cera models. As a result, the observation of transition to spatiotemporal chaos with the unshifted coupled-NLS models (21) and (22) might become unattainable. In section V, we show by numerical simulations with Eqs. (21) and (22) that this is indeed the case. That is, it is not possible to realize transition to spatiotemporal chaos with the models (21) and (22) due to strong deviations of the numerically obtained $U_j(z)$ curves from the curves expected by the reduced unshifted Di Cera models (15) and (17).

A possible way for overcoming the disadvantages of the two unshifted Di Cera models is by means of a translation transformation of the form $U'_j = U_j + a_j$, where $0 < a_j < 1$ for $1 \le j \le 4$. This transformation can also be viewed as a shift of the origin, such that the origin of the old phase space (the $U_j$ phase space) is located in the first orthant upward and to the right relative to the origin of the new phase space (the $U'_j$ phase space). For this reason, we refer to the new dynamical models for the $U'_j$ as the shifted 1989 and 1990 Di Cera models. Clearly, for any initial condition $U_j(0) > 0$ for $1 \le j \le 4$, the $U'_j(z)$ satisfy $U'_j(z) \ge a_j > 0$ for all $z$, and thus, the problem of small $U_j(z)$ values is easily circumvented. Additionally, since the transformation is a simple translation, it does not change the dynamics. Therefore, the dynamics of the $U'_j$ is just a shifted version (in phase space) of the dynamics of the $U_j$. In particular, the two shifted Di Cera models for the $U'_j$ should exhibit the same chaotic dynamics as the two unshifted Di Cera models, and this behavior should be observed in the same region in parameter space. Additionally, the total energy, $M' = \sum_{j=1}^{4} U'_j(z) = M + \sum_{j=1}^{4} a_j$, is a conserved quantity of the two shifted Di Cera models.

We now substitute the relations $U_j = U'_j - a_j$ into Eq. (15). Performing this substitution and dropping the prime notation, we arrive at the shifted 1989 Di Cera model:

$$\begin{aligned}
\dot{U}_1 &= \frac{4\epsilon_3}{T}(U_1 - a_1)\left[\tilde{b}_{14}(U_4 - a_4) - m\tilde{b}_{12}(U_2 - a_2) + \tilde{b}_{12}(U_2 - a_2)^2\right] \\
\dot{U}_2 &= \frac{4\epsilon_3}{T}(U_2 - a_2)\left[m\tilde{b}_{12}(U_1 - a_1) - \tilde{b}_{12}(U_1 - a_1)(U_2 - a_2) - \tilde{b}_{23}(U_3 - a_3)\right] \\
\dot{U}_3 &= \frac{4\epsilon_3}{T}(U_3 - a_3)\left[\tilde{b}_{23}(U_2 - a_2) - \tilde{b}_{34}(U_4 - a_4)\right] \\
\dot{U}_4 &= \frac{4\epsilon_3}{T}(U_4 - a_4)\left[\tilde{b}_{34}(U_3 - a_3) - \tilde{b}_{14}(U_1 - a_1)\right].
\end{aligned} \tag{23}$$

It is possible to write this model in the form

$$\begin{aligned}
\dot{U}_1 &= \frac{4\epsilon_3}{T}C_1(U_1)U_1\left[\tilde{b}_{14}C_4(U_4)U_4 - m\tilde{b}_{12}C_2(U_2)U_2 + \tilde{b}_{12}C_2^2(U_2)U_2^2\right] \\
\dot{U}_2 &= \frac{4\epsilon_3}{T}C_2(U_2)U_2\left[m\tilde{b}_{12}C_1(U_1)U_1 - \tilde{b}_{12}C_1(U_1)C_2(U_2)U_1U_2 - \tilde{b}_{23}C_3(U_3)U_3\right] \\
\dot{U}_3 &= \frac{4\epsilon_3}{T}C_3(U_3)U_3\left[\tilde{b}_{23}C_2(U_2)U_2 - \tilde{b}_{34}C_4(U_4)U_4\right] \\
\dot{U}_4 &= \frac{4\epsilon_3}{T}C_4(U_4)U_4\left[\tilde{b}_{34}C_3(U_3)U_3 - \tilde{b}_{14}C_1(U_1)U_1\right],
\end{aligned} \tag{24}$$

where we defined: $C_j(U_j) = 1 - a_j/U_j$.

The shifted 1990 Di Cera model is obtained in a similar manner. That is, we substitute $U_j = U'_j - a_j$ into Eq. (17), drop the prime notation, and rearrange terms in a similar way to Eq. (24). These calculations yield the following form of the shifted 1990 Di Cera model:

$$\begin{aligned}
\dot{U}_1 &= \frac{4\epsilon_3}{T}C_1(U_1)U_1\left[\tilde{b}_{14}C_4(U_4)U_4 - m\tilde{b}_{12}C_2(U_2)U_2 + \tilde{b}_{12}C_2^2(U_2)U_2^2\right] \\
\dot{U}_2 &= \frac{4\epsilon_3}{T}C_2(U_2)U_2\left[m\tilde{b}_{12}C_1(U_1)U_1 - \tilde{b}_{12}C_1(U_1)C_2(U_2)U_1U_2 - \tilde{b}_{23}C_3(U_3)U_3\right] \\
\dot{U}_3 &= \frac{4\epsilon_3}{T}C_3(U_3)U_3\left[\tilde{b}_{23}C_2(U_2)U_2 - m\tilde{b}_{34}C_4(U_4)U_4 + \tilde{b}_{34}C_4^2(U_4)U_4^2\right] \\
\dot{U}_4 &= \frac{4\epsilon_3}{T}C_4(U_4)U_4\left[m\tilde{b}_{34}C_3(U_3)U_3 - \tilde{b}_{34}C_3(U_3)C_4(U_4)U_3U_4 - \tilde{b}_{14}C_1(U_1)U_1\right].
\end{aligned} \tag{25}$$

### D. The corresponding shifted perturbed coupled-NLS models

Let us describe the perturbed coupled-NLS models that can yield the energy dynamics of the two *shifted* Di Cera models (24) and (25), *including chaotic dynamics*. The optical waveguide array setup is similar to the one considered in subsection III B. That is, we consider propagation of four sequences of optical pulses in a four-waveguide array, where each sequence propagates in its own waveguide in the presence of second-order dispersion, narrowband cubic nonlinearity, and broadband cubic loss or gain. Additionally, due to the broadband cubic gain-loss, the pulses in each sequence experience dissipative cubic interaction with pulses from the sequences in the two NN waveguides. The only difference from the waveguides in subsection III B is in the more complicated form of the $z$-dependent coefficients of the dissipative cubic interaction terms on the right hand side of the coupled-NLS models. The form of the $z$-dependent coefficients in the new coupled-NLS models is deduced based on the form of the shifted Di Cera models (24) and (25), the expression for the collisional energy shift in a single collision (10), and simple collision rate calculations similar to the ones in Refs. [30–32]. These simple arguments and calculations yield the new coupled-NLS models, which are expected to exhibit the same energy dynamics as the two *shifted* Di Cera models.

The perturbed coupled-NLS model, which can yield the energy dynamics of the shifted 1989 Di Cera model is

$$
\begin{aligned}
& i\partial_z\psi_1 + \partial_t^2\psi_1 + 2\left|\psi_1\right|^2\psi_1 = i\epsilon_3\tilde{b}_{14}C_1\left(\eta_1\right)C_4\left(\eta_4\right)\left|\psi_4\right|^2\psi_1 \\
& +i\epsilon_3\tilde{b}_{12}C_1\left(\eta_1\right)C_2\left(\eta_2\right)\left[C_2\left(\eta_2\right)\eta_2 - m\right]\left|\psi_2\right|^2\psi_1 \\
& i\partial_z\psi_2 + \partial_t^2\psi_2 + 2\left|\psi_2\right|^2\psi_2 = i\epsilon_3\tilde{b}_{12}C_1\left(\eta_1\right)C_2\left(\eta_2\right)\left[m - C_2\left(\eta_2\right)\eta_2\right]\left|\psi_1\right|^2\psi_2 \\
& -i\epsilon_3\tilde{b}_{23}C_2\left(\eta_2\right)C_3\left(\eta_3\right)\left|\psi_3\right|^2\psi_2 \\
& i\partial_z\psi_3 + \partial_t^2\psi_3 + 2\left|\psi_3\right|^2\psi_3 = i\epsilon_3\tilde{b}_{23}C_2\left(\eta_2\right)C_3\left(\eta_3\right)\left|\psi_2\right|^2\psi_3 \\
& -i\epsilon_3\tilde{b}_{34}C_3\left(\eta_3\right)C_4\left(\eta_4\right)\left|\psi_4\right|^2\psi_3 \\
& i\partial_z\psi_4 + \partial_t^2\psi_4 + 2\left|\psi_4\right|^2\psi_4 = i\epsilon_3\tilde{b}_{34}C_3\left(\eta_3\right)C_4\left(\eta_4\right)\left|\psi_3\right|^2\psi_4 \\
& -i\epsilon_3\tilde{b}_{14}C_1\left(\eta_1\right)C_4\left(\eta_4\right)\left|\psi_1\right|^2\psi_4 \,.
\end{aligned}
\tag{26}
$$

Additionally, the perturbed coupled-NLS model, which is expected to exhibit the energy dynamics of the shifted 1990 Di Cera model is

$$
\begin{aligned}
&i\partial_z\psi_1 + \partial_t^2\psi_1 + 2\left|\psi_1\right|^2\psi_1 = i\epsilon_3\tilde{b}_{14}C_1\left(\eta_1\right)C_4\left(\eta_4\right)\left|\psi_4\right|^2\psi_1 \\
&+i\epsilon_3\tilde{b}_{12}C_1\left(\eta_1\right)C_2\left(\eta_2\right)\left[C_2\left(\eta_2\right)\eta_2 - m\right]\left|\psi_2\right|^2\psi_1 \\
&i\partial_z\psi_2 + \partial_t^2\psi_2 + 2\left|\psi_2\right|^2\psi_2 = i\epsilon_3\tilde{b}_{12}C_1\left(\eta_1\right)C_2\left(\eta_2\right)\left[m - C_2\left(\eta_2\right)\eta_2\right]\left|\psi_1\right|^2\psi_2 \\
&-i\epsilon_3\tilde{b}_{23}C_2\left(\eta_2\right)C_3\left(\eta_3\right)\left|\psi_3\right|^2\psi_2 \\
&i\partial_z\psi_3 + \partial_t^2\psi_3 + 2\left|\psi_3\right|^2\psi_3 = i\epsilon_3\tilde{b}_{23}C_2\left(\eta_2\right)C_3\left(\eta_3\right)\left|\psi_2\right|^2\psi_3 \\
&+i\epsilon_3\tilde{b}_{34}C_3\left(\eta_3\right)C_4\left(\eta_4\right)\left[C_4\left(\eta_4\right)\eta_4 - m\right]\left|\psi_4\right|^2\psi_3 \\
&i\partial_z\psi_4 + \partial_t^2\psi_4 + 2\left|\psi_4\right|^2\psi_4 = i\epsilon_3\tilde{b}_{34}C_3\left(\eta_3\right)C_4\left(\eta_4\right)\left[m - C_4\left(\eta_4\right)\eta_4\right]\left|\psi_3\right|^2\psi_4 \\
&-i\epsilon_3\tilde{b}_{14}C_1\left(\eta_1\right)C_4\left(\eta_4\right)\left|\psi_1\right|^2\psi_4\,.
\end{aligned}
\tag{27}
$$

We refer to Eqs. (26) and (27) as the shifted 1989 coupled-NLS model and the shifted 1990 coupled-NLS model, respectively. We emphasize again that in the two studies in Refs. [67] and [68], no spatially extended PDE models were considered.

## IV. ANALYSIS OF ENERGY DYNAMICS IN THE TWO UNSHIFTED DI CERA MODELS (15) AND (17)

We now turn to analyze the dynamics of pulse energies in the two unshifted Di Cera models (15) and (17). Our treatment follows the presentation in Refs. [67] and [68] with some important additions, corrections, and generalizations. To simplify the comparison with the analysis of Refs. [67] and [68], in the current section, we adopt the parameter notation that was used in Refs. [67] and [68]. That is, we replace the $4\tilde{b}_{jk}/T$ factors by the corresponding chemical reaction rate coefficients $k_j$ of Refs. [67] and [68]. We also demonstrate the different types of dynamics observed in the two models, including chaotic dynamics, by numerical solution of the models.

### A. Energy dynamics in the unshifted 1989 Di Cera model

#### *1. Stability and bifurcation analysis for the equilibrium points of the model*

We first write the unshifted 1989 Di Cera model using the parameter notation of Ref. [67]:

$$\begin{aligned}
\dot{U}_1 &= \epsilon_3 U_1 \left(k_4 U_4 - m k_1 U_2 + k_1 U_2^2\right) \\
\dot{U}_2 &= \epsilon_3 U_2 \left(m k_1 U_1 - k_1 U_1 U_2 - k_2 U_3\right) \\
\dot{U}_3 &= \epsilon_3 U_3 \left(k_2 U_2 - k_3 U_4\right) \\
\dot{U}_4 &= \epsilon_3 U_4 \left(k_3 U_3 - k_4 U_1\right) .
\end{aligned} \tag{28}$$

In the context of the cyclic networks of autocatalytic chemical reactions that were considered in Ref. [67], the $U_j$ represent chemical concentrations, the $k_j$ are the reaction rate coefficients, $\dot{U}_j$ are the time-derivatives of $U_j$, and $M = \sum_{j=1}^{4} U_j(z)$ is the total mass. The $k_j$ coefficients are related to the $\tilde{b}_{jk}$ coefficients in Eq. (15) via

$$k_1 = 4\tilde{b}_{12}/T, \quad k_2 = 4\tilde{b}_{23}/T, \quad k_3 = 4\tilde{b}_{34}/T, \quad k_4 = 4\tilde{b}_{14}/T . \tag{29}$$

Eq. (28) possesses one equilibrium point with nonzero energies for all four sequences. We denote this equilibrium point by $\mathbf{U_e^{(1)}}$. The four coordinates of $\mathbf{U_e^{(1)}}$ are given by [67]:

$$\begin{aligned}
&U_{e1}^{(1)} = \frac{M - (1 + k_2/k_3)\left[m - k_2 k_4/(k_1 k_3)\right]}{(1 + k_4/k_3)}, \quad U_{e2}^{(1)} = m - \frac{k_2 k_4}{k_1 k_3}, \\
&U_{e3}^{(1)} = k_4 U_{e1}^{(1)}/k_3, \quad U_{e4}^{(1)} = k_2 U_{e2}^{(1)}/k_3 .
\end{aligned} \tag{30}$$

The equilibrium point $\mathbf{U_e^{(1)}}$ exists in the following region in parameter space:

$$\frac{k_2 k_4}{k_1 k_3} + \frac{M k_3}{k_2 + k_3} > m > \frac{k_2 k_4}{k_1 k_3}. \tag{31}$$

Linear stability analysis for $\mathbf{U_e^{(1)}}$ shows that this equilibrium point undergoes a supercritical Hopf bifurcation at $m_H = k_2(k_3 + 2k_4)/(k_1 k_3)$ [67]. For $m < m_H$, $\mathbf{U_e^{(1)}}$ is a stable focus, while for

$$m > m_H = \frac{2 k_2 k_4}{k_1 k_3} + \frac{k_2}{k_1} \tag{32}$$

it is unstable, and a stable limit cycle about $\mathbf{U_e^{(1)}}$ appears. Combining inequalities (31) and (32), we find the region in parameter space, where stable limit-cycle behavior can exist:

$$\frac{k_2 k_4}{k_1 k_3} + \frac{M k_3}{k_2 + k_3} > m > \frac{2 k_2 k_4}{k_1 k_3} + \frac{k_2}{k_1}. \tag{33}$$

Additionally, from inequality (33) if follows that the necessary condition for the occurrence of the supercritical Hopf bifurcation is [67]:

$$Mk_1k_3^2 > k_2^2k_3 + k_2^2k_4 + k_2k_3k_4 + k_2k_3^2. \tag{34}$$

In addition to the equilibrium point $\mathbf{U}_{\mathbf{e}}^{(\mathbf{1})}$, Eq. (28) possesses two line segments of fuzzy equilibrium points of the form $(s_1, 0, M-s_1, 0)$ and $(0, s_2, 0, M-s_2)$, where $0 \leq s_1 \leq M$ and $0 \leq s_2 \leq M$ [67]. The fuzzy equilibrium points $(s_1, 0, M - s_1, 0)$ are all unstable, but the situation is quite different for the $(0, s_2, 0, M-s_2)$ equilibrium points. More specifically, when the value of $m$ exceeds a certain threshold value that we denote by $m_P$, a line segment of stable equilibrium points, which is contained in the entire line segment $(0, s_2, 0, M-s_2),\ 0 \leq s_2 \leq M$, appears. The appearance of this line segment of stable fuzzy equilibrium points is associated with a period-doubling (PD) bifurcation, and with the emergence of chaotic dynamics via a PD cascade [67]. In view of the importance of the fuzzy equilibrium points $(0, s_2, 0, M - s_2)$ in the emergence of chaotic dynamics, it is worth providing some details about the stability analysis of these equilibrium points [70].

The eigenvalues of the Jacobian matrix for the linearization of Eq. (28) around the equilibrium points $(0, s_2, 0, M - s_2)$ are

$$\lambda_1 = \lambda_2 = 0,\ \lambda_3 = k_1s_2^2 - (mk_1 + k_4)s_2 + Mk_4,\ \lambda_4 = (k_2 + k_4)s_2 - Mk_3. \tag{35}$$

For stability, we require $\lambda_3 < 0$ and $\lambda_4 < 0$. The requirement $\lambda_4 < 0$ is satisfied on the following interval of $s_2$ values:

$$0 \leq s_2 < Mk_3/(k_2 + k_3)\,. \tag{36}$$

Additionally, the requirement $\lambda_3 < 0$ is satisfied on the interval [67]

$$\frac{mk_1 + k_4 - \Delta^{1/2}}{2k_1} < s_2 < \frac{mk_1 + k_4 + \Delta^{1/2}}{2k_1}\,, \tag{37}$$

where

$$\Delta \equiv (mk_1 + k_4)^2 - 4Mk_1k_4 > 0\,. \tag{38}$$

Inequality (38) provides the link between the appearance of the line segment of stable fuzzy equilibrium points and the PD bifurcation. Indeed, from this inequality it follows that the line segment (37) exists only when the value of $m$ exceeds the PD bifurcation value $m_P$ [67].

Setting $\Delta = 0$ we find: $m_P = 2(Mk_4/k_1)^{1/2} - k_4/k_1$. Thus, the line segment (37) exists provided that [67]

$$m > m_P = 2(Mk_4/k_1)^{1/2} - k_4/k_1\,. \tag{39}$$

The conditions $\lambda_3 < 0$ and $\lambda_4 < 0$ are simultaneously satisfied only when there is some overlap between the two intervals in Eqs. (36) and (37), i.e., only when the intersection of these two intervals is not the empty set. Therefore, we must require:

$$\frac{Mk_3}{k_2 + k_3} > \frac{mk_1 + k_4 - \Delta^{1/2}}{2k_1}\,. \tag{40}$$

This requirement yields

$$m < \frac{2Mk_3}{k_2 + k_3} - \frac{k_4}{k_1}\,. \tag{41}$$

Using Eqs. (39) and (41), we find that the region in parameter space, where the stable equilibrium points $(0, s_2, 0, M - s_2)$ and the PD cascade can exist is

$$\frac{2Mk_3}{k_2 + k_3} - \frac{k_4}{k_1} > m > 2\left(\frac{Mk_4}{k_1}\right)^{1/2} - \frac{k_4}{k_1}\,. \tag{42}$$

In addition, inequality (42) yields the following condition on the parameters $k_j$ and $M$ for the existence of stable fuzzy equilibrium points of the form $(0, s_2, 0, M - s_2)$:

$$Mk_1k_3^2 > k_2^2k_4 + k_3^2k_4 + 2k_2k_3k_4. \tag{43}$$

Combining the results of the current paragraph with the results of the second paragraph in the current subsection, we conclude that the region in parameter space where chaotic dynamics can be observed is given by inequalities (33) and (42). Furthermore, inequalities (34) and (43) provide the conditions on the parameters $k_j$ and $M$ for the appearance of chaotic dynamics via Hopf and PD bifurcations.

### *2. Numerical solution of the model*

We now demonstrate the dynamics in the unshifted 1989 Di Cera model by numerical solution of the model for a representative physical parameter setup. We consider the case where $\epsilon_3 = 1.0$, $M = 1.0$, $k_1 = 10.0$, and $k_2 = k_3 = k_4 = 1.0$. In this case, conditions (34) and (43) for the appearance of chaotic dynamics via Hopf and PD bifurcations are satisfied. We take the initial condition as $U_1(0) = 0.55$, $U_2(0) = U_3(0) = U_4(0) = 0.15$, and remark that similar dynamical behavior is obtained for other initial conditions. We illustrate the dynamics by considering the following four $m$ values: $m = 0.25$, $m = 0.35$, $m = 0.55$, and $m = 0.575$. According to inequalities (33) and (42), the expected dynamical behaviors are decaying oscillations for $m = 0.25$, limit-cycle oscillations with period 1 for $m = 0.35$, limit-cycle oscillations with period 2 for $m = 0.55$, and chaos for $m = 0.575$.

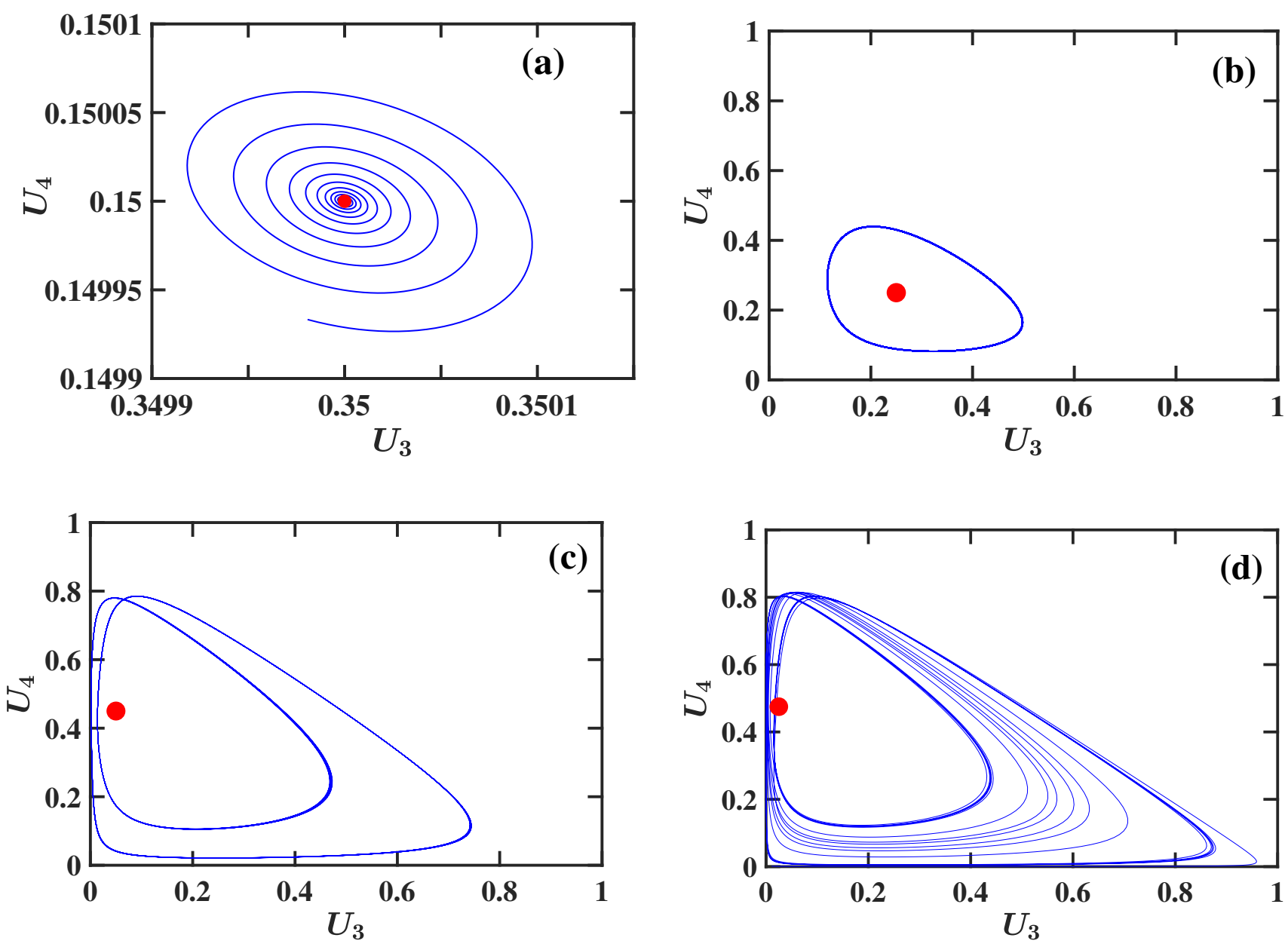

FIG. 2: Projections of the numerically obtained trajectories of the unshifted 1989 Di Cera model (28) on the $U_3 - U_4$ plane for $m = 0.25$ (a), $m = 0.35$ (b), $m = 0.55$ (c), and $m = 0.575$ (d). The values of the other physical parameters are $\epsilon_3 = 1.0$, $M = 1.0$, $k_1 = 10.0$, and $k_2 = k_3 = k_4 = 1.0$.

Fig. 2 shows the projections of the numerical solution's trajectories in phase space on the $U_3 - U_4$ plane for the four aforementioned $m$ values. The numerically obtained trajectories are in excellent agreement with the predictions of subsection IV A 1. More specifically, for $m = 0.25$, we observe a spiral towards the stable focus of Eq. (30), while for $m = 0.35$, we find a stable period-1 periodic orbit about the unstable equilibrium point of Eq. (30). Furthermore, for $m = 0.55$, we find a stable period-2 periodic orbit about the same equilibrium point, while for $m = 0.575$, the trajectory acquires an intertwined structure that is typical for chaotic dynamics in low-dimensional dissipative dynamical systems [71–73].

The $z$-dependences of the pulse energies $U_j$ that are obtained by numerical solution of Eq. (28) are shown in Figs. 3 and 4. For $m = 0.25$, we find that the $U_j$ exhibit decaying oscillations and approach their equilibrium values $U_{ej}^{(1)}$, which are given by Eq. (30). For $m = 0.35$, the energies exhibit oscillations that tend to limit-cycle oscillations with period 1, in accordance with the stability and bifurcation analysis in subsection IV A 1. For $m = 0.55$, the energies exhibit stable limit-cycle oscillations with period 2, in agreement with the predictions of subsection IV A 1. Moreover, for $m = 0.575$, the $z$-dependence of the $U_j$ consists of irregular nonperiodic oscillations, which are typical to chaotic dynamics in low-dimensional dissipative nonlinear dynamical systems [71–74]. Thus, the $z$-dependences of the $U_j$ in Figs. 3 and 4 clearly verify the theoretical predictions of subsection IV A 1 and of Ref. [67]. Furthermore, the results in Figs. 3 and 4 demonstrate that the 1989 Di Cera model exhibits chaotic dynamics, where chaos emerges via Hopf and PD bifurcations.

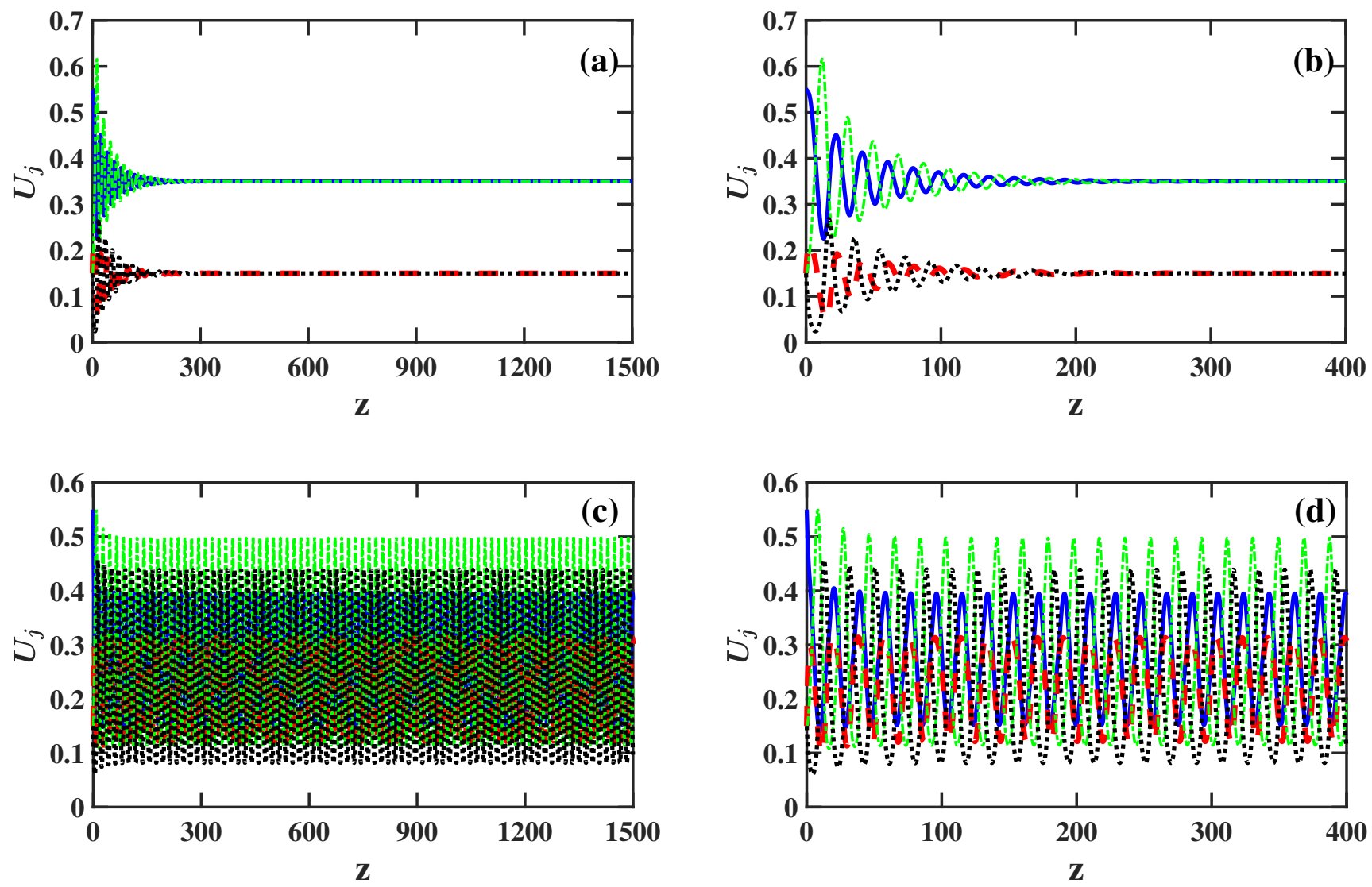

FIG. 3: The $z$-dependences of the pulse energies $U_j$ in the unshifted 1989 Di Cera model (28) for $m = 0.25$ [(a) and (b)], and for $m = 0.35$ [(c) and (d)]. The initial condition is $U_1(0) = 0.55$, $U_2(0) = U_3(0) = U_4(0) = 0.15$, and the other parameter values are the same as in Fig. 2. The solid blue, dashed red, dashed-dotted green, and dotted black curves represent $U_j(z)$, obtained by numerical solution of Eq. (28).

Further insight into the dynamics at higher $m$ values is obtained by analyzing the distance $S(z)$ between two trajectories $\mathbf{U}(z)$ and $\mathbf{U}'(z)$ with close initial conditions [71, 73, 75]. This quantity is defined as [71]: $S(z) = \left\{\sum_{j=1}^{4}\left[U_j(z) - U'_j(z)\right]^2\right\}^{1/2}$. It is a measure of the sensitive dependence of the trajectory in phase space on the initial conditions [71]. Fig. 5 shows the $z$-dependence of $S$ that is obtained by numerical solution of Eq. (28) for $m = 0.575$. The two initial conditions are $U_1(0) = 0.55$, $U_2(0) = U_3(0) = U_4(0) = 0.15$, and $U'_1(0) = 0.552$, $U'_2(0) = U'_3(0) = 0.15$, $U'_4(0) = 0.148$. We observe that $S(z)$ varies in an irregular nonperiodic manner, and attains relatively large values (i.e., values of order 1) at many distances. This observation provides additional key evidence for the chaotic nature of the dynamics in the 1989 Di Cera model at $m = 0.575$. Moreover, the claim for chaotic dynamics of the 1989 Di Cera model at $m = 0.575$ is supported by calculation of its Lyapunov exponents. Indeed, using the standard numerical methods for Lyapunov exponent calculation, which are described in Refs. [71, 76], we find that the largest Lyapunov exponent of Eq. (28) is $\lambda_1 = 0.0087$, in full accordance with the chaotic nature of the dynamics that was deduced from Figs. 2, 4, and 5.

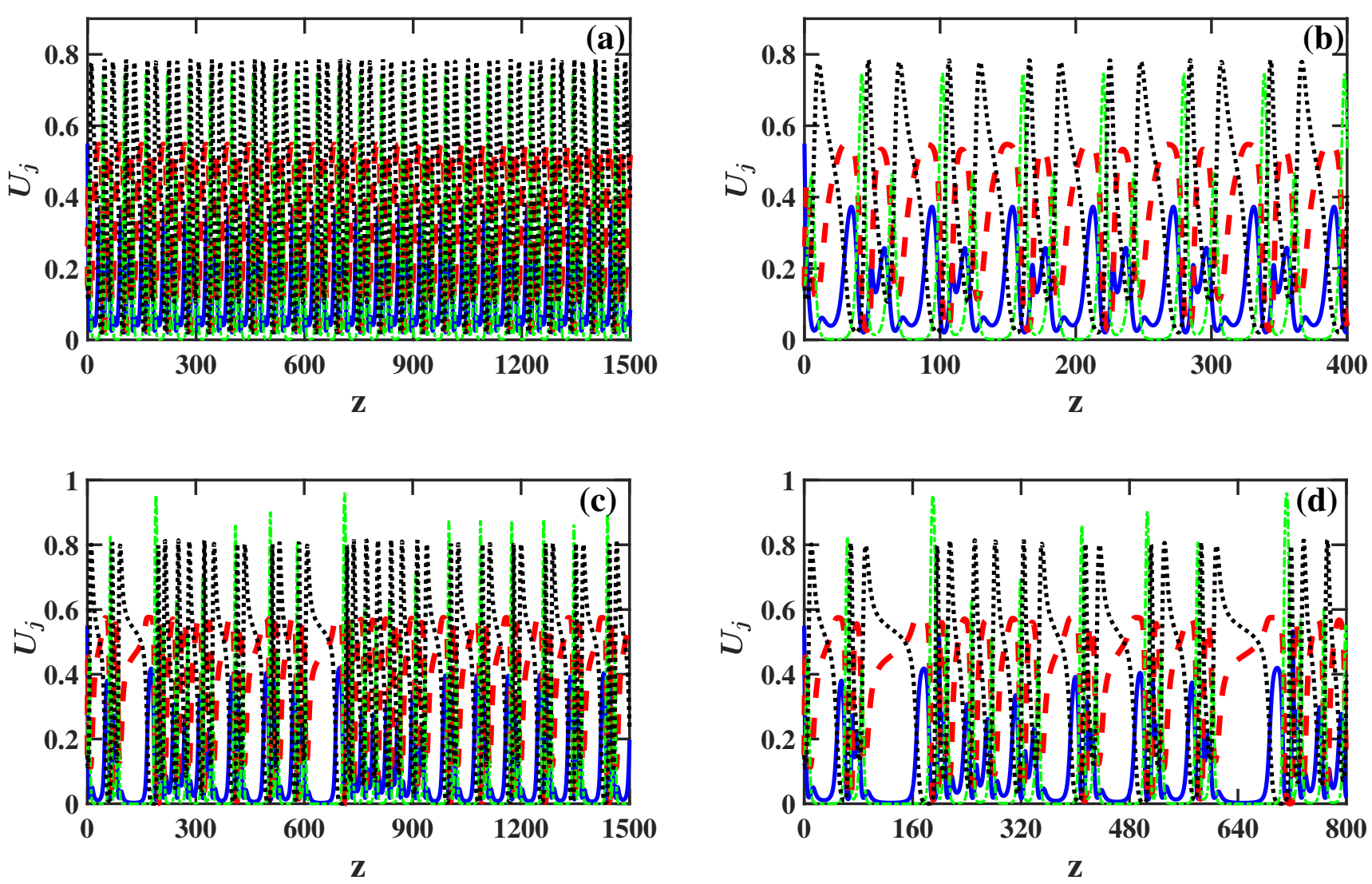

FIG. 4: The $z$-dependences of the pulse energies $U_j$ in the unshifted 1989 Di Cera model (28) for $m = 0.55$ [(a) and (b)], and for $m = 0.575$ [(c) and (d)]. The initial condition is $U_1(0) = 0.55$, $U_2(0) = U_3(0) = U_4(0) = 0.15$, and the other parameter values are the same as in Figs. 2 and 3. The solid blue, dashed red, dashed-dotted green, and dotted black curves represent $U_j(z)$, obtained by numerical solution of Eq. (28).

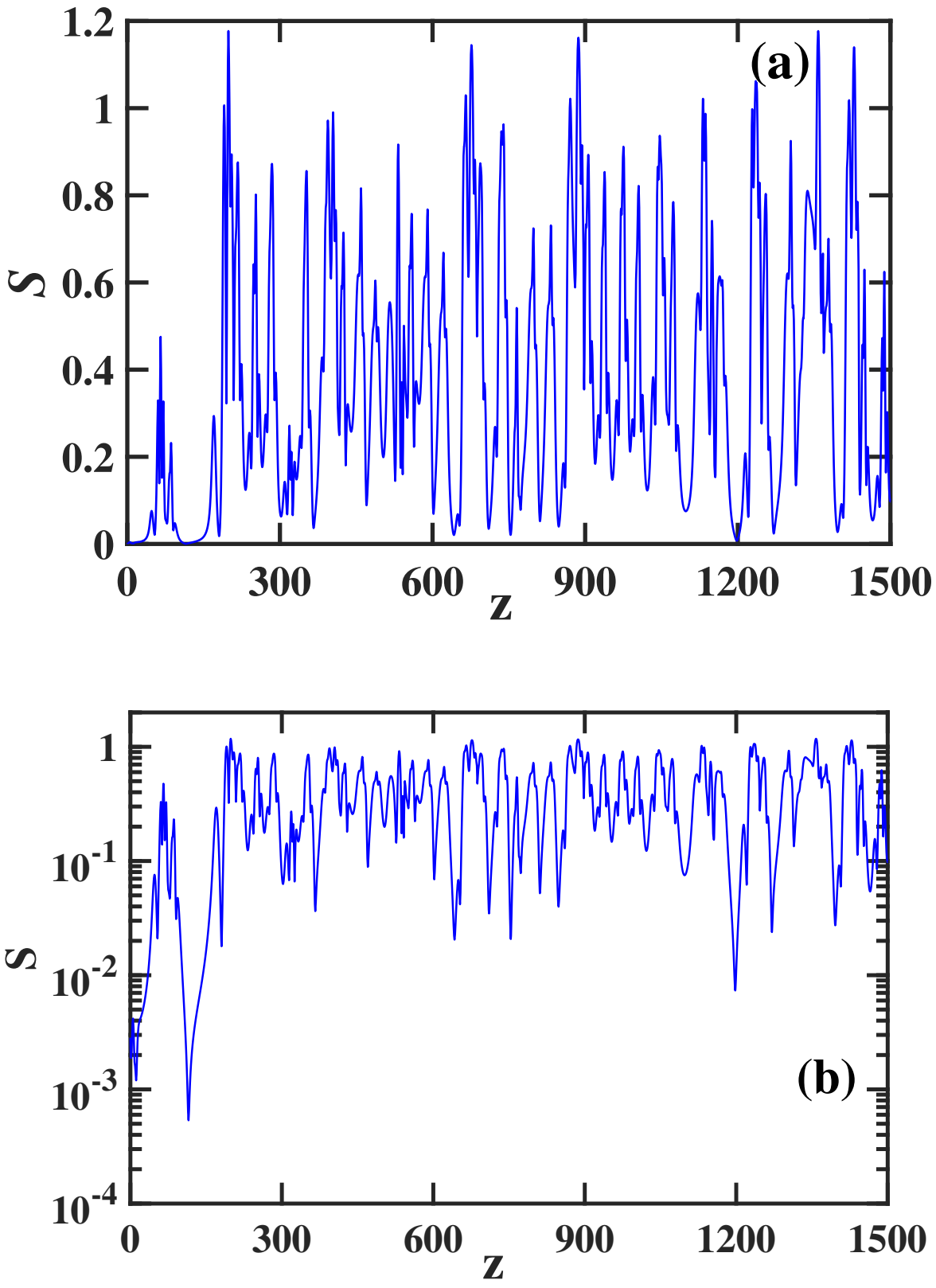

FIG. 5: The distance between two trajectories with close initial conditions $S$ vs propagation distance $z$ for the unshifted 1989 Di Cera model (28) at $m = 0.575$. Graph (a) shows $S(z)$ on a linear scale, and graph (b) shows $S(z)$ on a logarithmic scale. The initial conditions are $U_1(0) = 0.55$, $U_2(0) = U_3(0) = U_4(0) = 0.15$, and $U_1'(0) = 0.552$, $U_2'(0) = U_3'(0) = 0.15$, $U_4'(0) = 0.148$. The other parameter values are the same as in Fig. 2.

### B. Energy dynamics in the unshifted 1990 Di Cera model

#### 1. *Stability and bifurcation analysis for the equilibrium points of the model*

Using the parameter notation of Ref. [68], we can write the unshifted 1990 Di Cera model as:

$$\begin{aligned}
\dot{U}_1 &= \epsilon_3 U_1 \left(k_4 U_4 - m k_1 U_2 + k_1 U_2^2\right) \\
\dot{U}_2 &= \epsilon_3 U_2 \left(m k_1 U_1 - k_1 U_1 U_2 - k_2 U_3\right) \\
\dot{U}_3 &= \epsilon_3 U_3 \left(k_2 U_2 - m k_3 U_4 + k_3 U_4^2\right) \\
\dot{U}_4 &= \epsilon_3 U_4 \left(m k_3 U_3 - k_3 U_3 U_4 - k_4 U_1\right).
\end{aligned} \tag{44}$$

The interpretation of the symbols $U_j$, $k_j$, $\dot{U}_j$, and $M$ in the context of the autocatalytic chemical reactions that were studied in Ref. [68] is identical to the one specified in subsection IV A 1. Additionally, the relations between the $k_j$ coefficients and the $\tilde{b}_{jk}$ coefficients of Eq. (17) are still given by Eq. (29).

We consider a symmetric setup of Eq. (44), in which $k_1 = k_3$ and $k_2 = k_4$. This setup is far more general than the one that was analyzed in Ref. [68]. In this case, Eq. (44) possesses three equilibrium points with nonzero energies for all four sequences, which we denote by $\mathbf{U_e^{(1)}}$, $\mathbf{U_e^{(2)}}$, and $\mathbf{U_e^{(3)}}$. The coordinates of these equilibrium points are given by [68]:

$$U_{e1}^{(1)} = U_{e3}^{(1)} = (M + 2k_2/k_1 - 2m)/2, \quad U_{e2}^{(1)} = U_{e4}^{(1)} = m - k_2/k_1, \tag{45}$$

$$\begin{aligned} U_{e1}^{(2)} &= \frac{k_2 (M - m - k_2/k_1)}{k_1 U_{e4}^{(2)}}, \quad U_{e2}^{(2)} = \left(m + k_2/k_1 + \Delta_1^{1/2}\right)/2, \\ U_{e3}^{(2)} &= \frac{k_2 (M - m - k_2/k_1)}{k_1 U_{e2}^{(2)}}, \quad U_{e4}^{(2)} = \left(m + k_2/k_1 - \Delta_1^{1/2}\right)/2, \end{aligned} \tag{46}$$

and

$$\begin{aligned} U_{e1}^{(3)} &= \frac{k_2 (M - m - k_2/k_1)}{k_1 U_{e4}^{(3)}}, \quad U_{e2}^{(3)} = \left(m + k_2/k_1 - \Delta_1^{1/2}\right)/2, \\ U_{e3}^{(3)} &= \frac{k_2 (M - m - k_2/k_1)}{k_1 U_{e2}^{(3)}}, \quad U_{e4}^{(3)} = \left(m + k_2/k_1 + \Delta_1^{1/2}\right)/2, \end{aligned} \tag{47}$$

where $\Delta_1 \equiv (m + k_2/k_1)(m - 3k_2/k_1)$.

The equilibrium point $\mathbf{U_e^{(1)}}$ exists with positive energy values for all four coordinates provided that

$$\frac{M}{2} + \frac{k_2}{k_1} > m > \frac{k_2}{k_1}. \tag{48}$$

However, as we show below, when the values of $U_{e1}^{(1)}$ and $U_{e3}^{(1)}$ become negative as $m$ becomes larger than $M/2 + k_2/k_1$, a line segment of stable fuzzy equilibrium points in the $U_2 - U_4$ plane appears. As a result, chaotic dynamics can still be observed for $m > M/2 + k_2/k_1$ in the physically relevant regime. Thus, only the right hand side of inequality (48) should be taken into account when considering the conditions for the appearance of chaotic dynamics. Additionally, linear stability analysis shows that $\mathbf{U_e^{(1)}}$ is a saddle (i.e., that it is unstable) for $M/2 + k_2/k_1 > m > 3k_2/k_1$ [68], and that it is an unstable node for $M - k_2/k_1 > m > M/2 + k_2/k_1$. For $m = M/2 + k_2/k_1$, we cannot use linear stability analysis to determine its stability.

For the symmetric setup of Eq. (44) with $k_1 = k_3$ and $k_2 = k_4$, the existence, stability, and bifurcation analysis for the equilibrium points $\mathbf{U_e^{(2)}}$ and $\mathbf{U_e^{(3)}}$ can be done within the same calculation [68]. We find that the two equilibrium points exist with positive energy values for all four coordinates provided that

$$M - \frac{k_2}{k_1} > m \geq \frac{3k_2}{k_1}. \tag{49}$$

It follows that the condition for the existence of $\mathbf{U_e^{(2)}}$ and $\mathbf{U_e^{(3)}}$ on the parameters $M$, $k_1$, and $k_2$ is:

$$M > \frac{4k_2}{k_1}. \tag{50}$$

Linear stability analysis shows that the equilibrium points $\mathbf{U_e^{(2)}}$ and $\mathbf{U_e^{(3)}}$ undergo a supercritical Hopf bifurcation at [68]

$$m_H = \frac{1}{4}\left\{M + \frac{2k_2}{k_1} + \left[\left(M - \frac{2k_2}{k_1}\right)^2 + \frac{32k_2^2}{k_1^2}\right]^{1/2}\right\}. \tag{51}$$

For $m < m_H$, $\mathbf{U_e^{(2)}}$ and $\mathbf{U_e^{(3)}}$ are stable foci, while for $m > m_H$, they are unstable, and stable limit cycles about $\mathbf{U_e^{(2)}}$ and $\mathbf{U_e^{(3)}}$ appear [68]. Combining inequalities (49), (50), and $m > m_H$, we find that the region in parameter space, where $\mathbf{U_e^{(2)}}$ and $\mathbf{U_e^{(3)}}$ exist and are unstable due to the supercritical Hopf bifurcation is

$$M - \frac{k_2}{k_1} > m > m_H. \tag{52}$$

As we will see in subsection IV B 2, this is also the region in parameter space, where chaotic dynamics can exist.

The unshifted 1990 Di Cera model (44) also possesses two line segments of fuzzy equilibrium points of the form $(s_1, 0, M - s_1, 0)$ and $(0, s_2, 0, M - s_2)$, where $0 \leq s_1 \leq M$ and $0 \leq s_2 \leq M$. The fuzzy equilibrium points $(s_1, 0, M - s_1, 0)$ are all unstable in the region in parameter space where $\mathbf{U_e^{(2)}}$ and $\mathbf{U_e^{(3)}}$ exist. In contrast, when

$$m > \frac{M}{2} + \frac{k_2}{k_1} \tag{53}$$

a line segment of stable equilibrium points $(0, s_2, 0, M - s_2)$, which is contained in the entire line segment $(0, s_2, 0, M - s_2)$, $0 \leq s_2 \leq M$, appears. Similar to the situation in the unshifted 1989 Di Cera model (28), the appearance of the stable fuzzy equilibrium points for $m > M/2 + k_2/k_1$ enables the existence of chaotic dynamics in this region in parameter space as well. The condition (53) is obtained by linear stability analysis for the equilibrium points $(0, s_2, 0, M - s_2)$. Indeed, the eigenvalues of the Jacobian matrix for the linearization of Eq. (44) about the equilibrium points $(0, s_2, 0, M - s_2)$ are

$$\lambda_1 = \lambda_2 = 0, \ \lambda_3 = k_1 s_2^2 - (mk_1 + k_2)s_2 + Mk_2,$$
$$\lambda_4 = k_1(M - s_2)^2 - (mk_1 + k_2)(M - s_2) + Mk_2. \tag{54}$$

Stability is realized when $\lambda_3 < 0$ and $\lambda_4 < 0$. The condition $\lambda_3 < 0$ is satisfied on the line segment

$$(0, s_2, 0, M - s_2), \quad \frac{mk_1 + k_2 - \Delta_3^{1/2}}{2k_1} < s_2 < \frac{mk_1 + k_2 + \Delta_3^{1/2}}{2k_1}, \tag{55}$$

where $\Delta_3 \equiv (mk_1 + k_2)^2 - 4Mk_1k_2 > 0$. Additionally, the condition $\lambda_4 < 0$ is satisfied on the line segment

$$(0, s_2, 0, M - s_2), \quad \frac{mk_1 + k_2 - \Delta_3^{1/2}}{2k_1} < M - s_2 < \frac{mk_1 + k_2 + \Delta_3^{1/2}}{2k_1}. \tag{56}$$

It is straightforward to show that when Eqs. (49) and (50) are satisfied, the line segments (55) and (56) have some overlap, provided that condition (53) holds. Thus, when Eqs. (49) and (50) are satisfied, stable fuzzy equilibrium points of the form $(0, s_2, 0, M - s_2)$ exist in the region in parameter space that is defined by Eq. (53), and as a result, chaotic dynamics can exist in this region.

We note that from Eqs. (45) and (48) it follows that when $m > M/2 + k_2/k_1$, the first and third components of $\mathbf{U_e^{(1)}}$ become negative. However, as shown in the preceding paragraph, this does not lead to the exclusion of chaotic dynamic behavior for $m > M/2 + k_2/k_1$ due to the emergence of stable fuzzy equilibrium points in this region. Moreover, from the analysis in the preceding paragraphs it follows that the region in parameter space where chaotic dynamics can exist, $M - k_2/k_1 > m > m_H$, can be divided into two main subregions based on the presence or absence of the saddle $\mathbf{U_e^{(1)}}$ and of the stable fuzzy equilibrium points. Indeed, Eqs. (52), (48), and (53) imply that for $M/2 + k_2/k_1 > m > m_H$, chaotic dynamics can occur in the presence of the saddle $\mathbf{U_e^{(1)}}$, and in the absence of stable fuzzy equilibrium points. The same equations also imply that for $M - k_2/k_1 > m > M/2 + k_2/k_1$, chaotic dynamics can occur in the absence of the saddle $\mathbf{U_e^{(1)}}$, and in the presence of stable fuzzy equilibrium points. For $m = M/2 + k_2/k_1$, $\mathbf{U_e^{(1)}}$ becomes a fuzzy equilibrium point of the form $(0, s_2, 0, M - s_2)$, whose stability cannot be determined by linear stability analysis. The numerical solution of Eq. (48) in subsection IV B 2 shows that chaotic dynamics can occur in this case as well.

#### *2. Numerical solution of the model*

Let us demonstrate the dynamics in the unshifted 1990 Di Cera model by numerical solution of Eq. (48) for a characteristic physical parameter setup. We choose, for example, the case where $\epsilon_3 = 1.0$, $M = 1.0$, $k_1 = 10.0$, and $k_2 = 1.0$. In this case, the condition (52) for the possible appearance of chaotic dynamics via a supercritical Hopf bifurcation is satisfied for $0.9 > m > m_H$, where $m_H = 0.54495...$. Based on the analysis in subsection IV B 1, for $0.6 > m > m_H$, chaotic dynamics can occur in the presence of the saddle $\mathbf{U_e^{(1)}}$, while for $0.9 > m > 0.6$, it can occur in the presence of stable fuzzy equilibrium points. In the latter case, the trajectories in phase space can alternatively approach one of the stable fuzzy equilibrium points $(0, s_2, 0, M - s_2)$. We consider the initial condition $U_1(0) = 0.12$, $U_2(0) = 0.16$, $U_3(0) = 0.2$, and $U_4(0) = 0.52$, and note that similar dynamical behavior is obtained for other initial conditions. We demonstrate the dynamics with the following $m$ values: $m = 0.52$, $m = 0.58$, $m = 0.6$, $m = 0.61$, $m = 0.7$, and $m = 0.8$. According to the analysis in the preceding subsection, we expect to observe decaying oscillations for $m = 0.52$, while for the other $m$ values, chaotic dynamics is possible. Additionally, for $m = 0.61$, $m = 0.7$, and $m = 0.8$, the trajectory in phase space can alternatively approach one of the stable fuzzy equilibrium points $(0, s_2, 0, M - s_2)$.

The projections of the numerically obtained trajectories of the 1990 Di Cera model (44) on the $U_3 - U_4$ plane for the six aforementioned $m$ values are shown in Fig. 6. We observe that the trajectories are in excellent agreement with the stability and bifurcation analysis of subsection IV B 1. In particular, for $m = 0.52$, the trajectory is a spiral towards the stable focus $\mathbf{U_e^{(3)}}$ of Eq. (47) with $U_{e3}^{(3)} = 0.30318...$ and $U_{e4}^{(3)} = 0.43832...$. Furthermore, for $m = 0.58$, $m = 0.6$, and $m = 0.61$, the trajectory has a convoluted structure, which is typical to chaotic dynamics in low-dimensional dissipative dynamical systems [71–73]. This finding also validates the prediction from subsection IV B 1 that chaotic dynamics can be observed both in the presence of the saddle $\mathbf{U_e^{(1)}}$ and in the presence of stable fuzzy equilibrium points of the form $(0, s_2, 0, M - s_2)$. For $m = 0.7$, we see that the trajectory starts by "encircling" the unstable focus $\mathbf{U_e^{(3)}}$, but ends up at a stable fuzzy equilibrium point $(0, s_2, 0, M - s_2)$, where $M - s_2 = 0.43832....$ Additionally, for $m = 0.8$, the trajectory tends directly to another stable fuzzy equilibrium point $(0, s_2, 0, M - s_2)$, where $M - s_2 = 0.67045....$

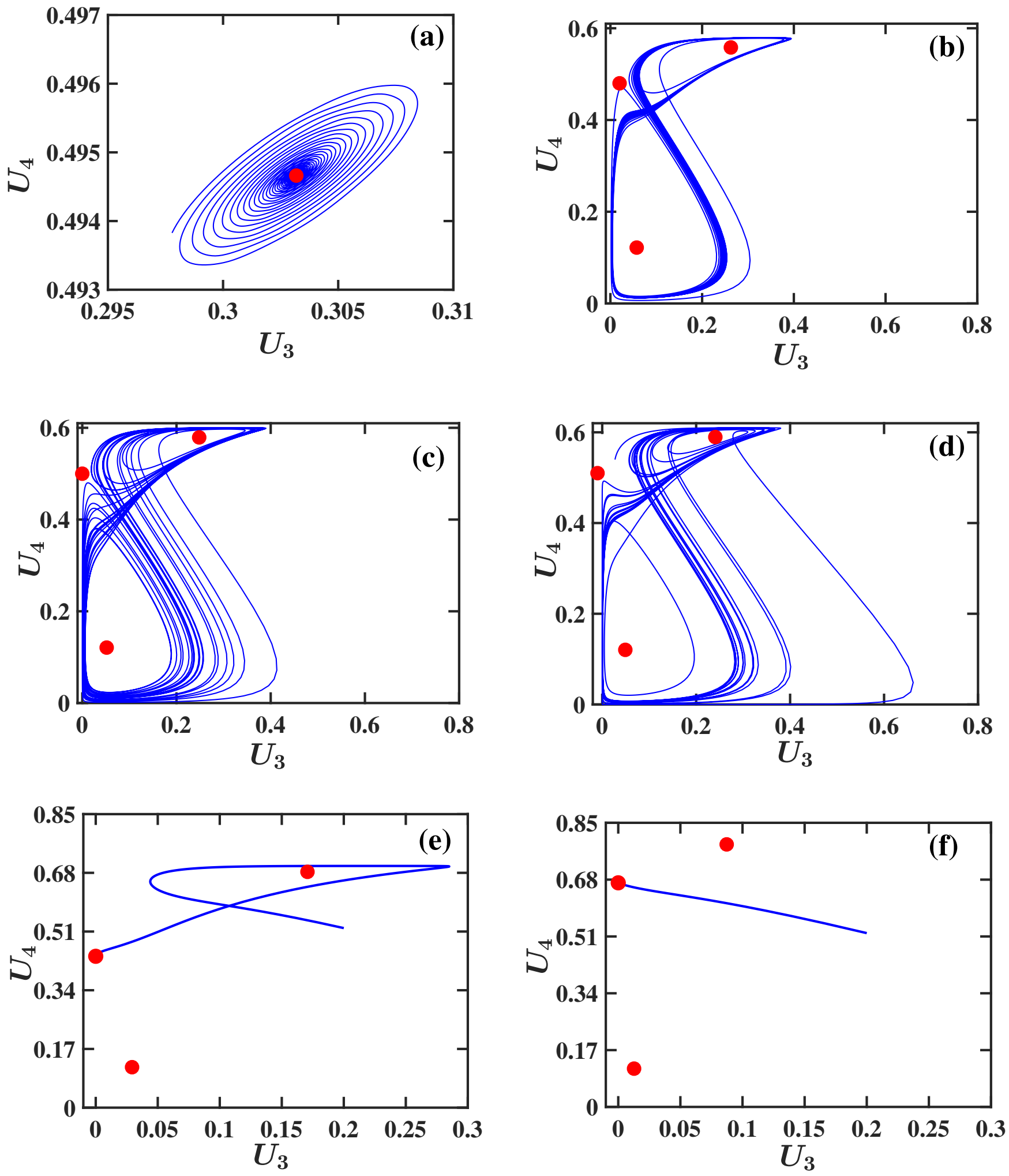

FIG. 6: Projections of the numerically obtained trajectories of the unshifted 1990 Di Cera model (44) on the $U_3 - U_4$ plane for $m = 0.52$ (a), $m = 0.58$ (b), $m = 0.6$ (c), $m = 0.61$ (d), $m = 0.7$ (e), and $m = 0.8$ (f). The values of the other physical parameters in the model are $\epsilon_3 = 1.0$, $M = 1.0$, $k_1 = 10.0$, and $k_2 = 1.0$.

The numerically obtained $z$-dependences of the pulse energies $U_j$ are shown in Figs. 7 and 8. For $m = 0.52$, the $U_j$ exhibit decaying oscillations and approach their equilibrium values $U_{ej}^{(3)}$, which are given by Eq. (47), in accordance with the fact that $\mathbf{U_e^{(3)}}$ is a stable focus. Furthermore, for $m = 0.58$, $m = 0.6$, and $m = 0.61$, the $U_j$ exhibit irregular nonperiodic oscillations, which are characteristic to dissipative chaos in low-dimensional dissipative dynamical systems [71–74]. For $m = 0.7$, the $U_j(z)$ curves start with a single oscillation, but then tend to the equilibrium values of an equilibrium point $(0, s_2, 0, M - s_2)$ with $M - s_2 = 0.43832...$. Finally, for $m = 0.8$, the $U_j(z)$ curves tend to the equilibrium values of an equilibrium point $(0, s_2, 0, M - s_2)$ with $M - s_2 = 0.67045...$ without any oscillations. These results and similar results obtained with other initial conditions support the analysis in subsection IV B 1 and show that chaotic dynamics can indeed exist for $m$ values satisfying inequality (52).

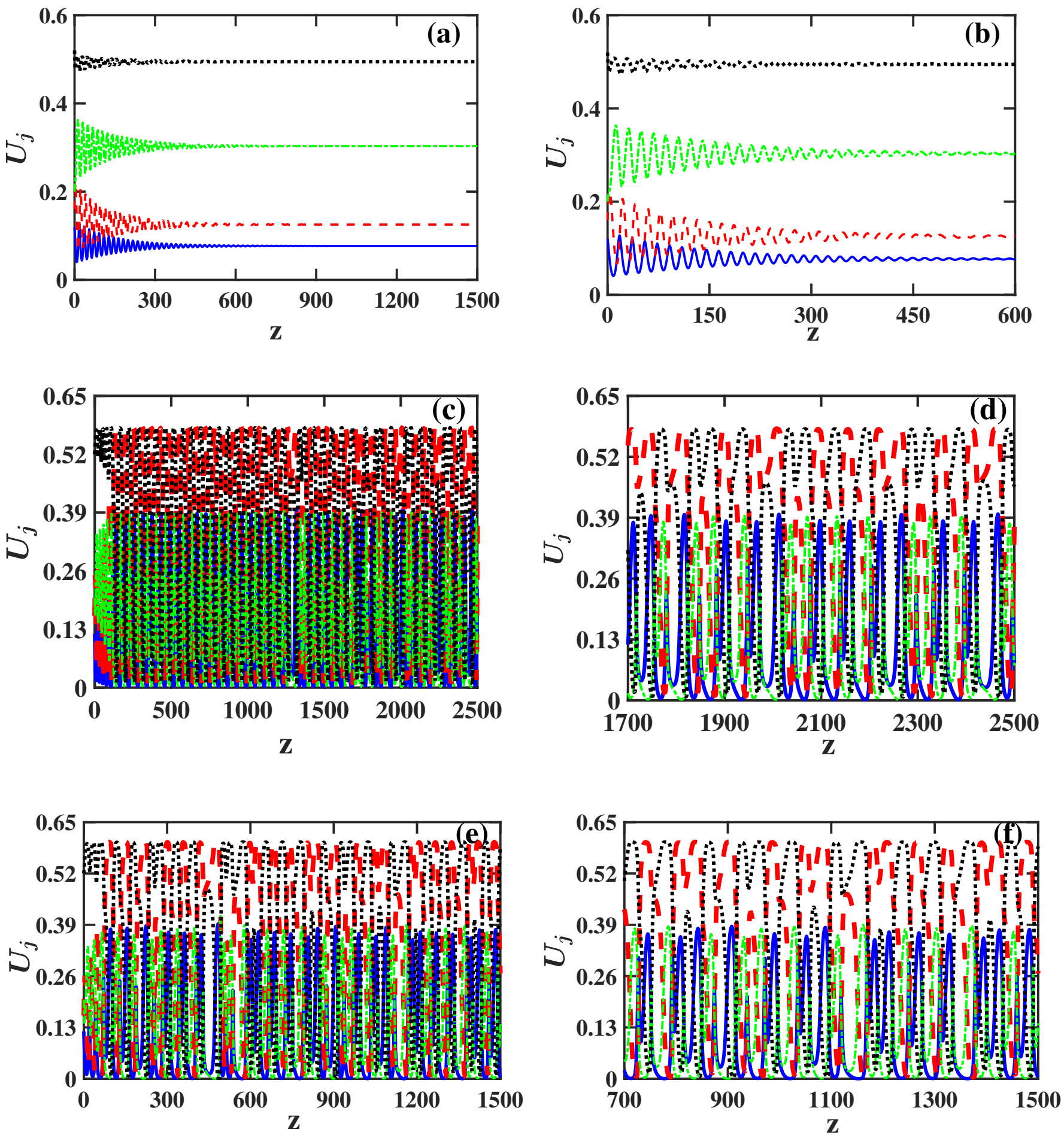

FIG. 7: The $z$-dependences of the pulse energies $U_j$ in the unshifted 1990 Di Cera model (44) for $m = 0.52$ [(a) and (b)], $m = 0.58$ [(c) and (d)], and $m = 0.6$ [(e) and (f)]. The initial condition is $U_1(0) = 0.12$, $U_2(0) = 0.16$, $U_3(0) = 0.2$, and $U_4(0) = 0.52$, and the other parameter values are the same as in Fig. 6. The solid blue, dashed red, dashed-dotted green, and dotted black curves represent $U_j(z)$, obtained by numerical solution of Eq. (44).

In order to gain further insight into the chaotic dynamics in the 1990 Di Cera model (44) it is useful to analyze the dynamics of the distance $S(z)$ between two trajectories with adjacent initial conditions [71, 73, 75]. Fig. 9 shows the numerically obtained $z$-dependence of $S$ for $m = 0.58$, $m = 0.6$, and $m = 0.61$. The two initial conditions used in the calculation of $S(z)$ are $U_1(0) = 0.12$, $U_2(0) = 0.16$, $U_3(0) = 0.2$, $U_4(0) = 0.52$, and $U'_1(0) = 0.122$, $U'_2(0) = 0.162$, $U'_3(0) = 0.198$, $U'_4(0) = 0.518$. It is seen that $S(z)$ varies in an irregular and nonperiodic manner for all three $m$ values. This finding demonstrates the sensitive dependence of the trajectories in phase space on the initial conditions for $m = 0.58$, $m = 0.6$, and $m = 0.61$, and in this manner, provides additional essential evidence for chaotic dynamics in these cases. Furthermore, the claim that the dynamics in the 1990 Di Cera model at these $m$ values is chaotic is fully supported by Lyapunov exponent calculations. Indeed, we find that the largest Lyapunov exponents of Eq. (44) at $m = 0.58$, $m = 0.6$, and $m = 0.61$ are $\lambda_1 = 0.0100$, $\lambda_1 = 0.0152$, and $\lambda_1 = 0.0181$, respectively, in full alignment with the conclusions about chaotic dynamics that were drawn from Figs. 6−9.

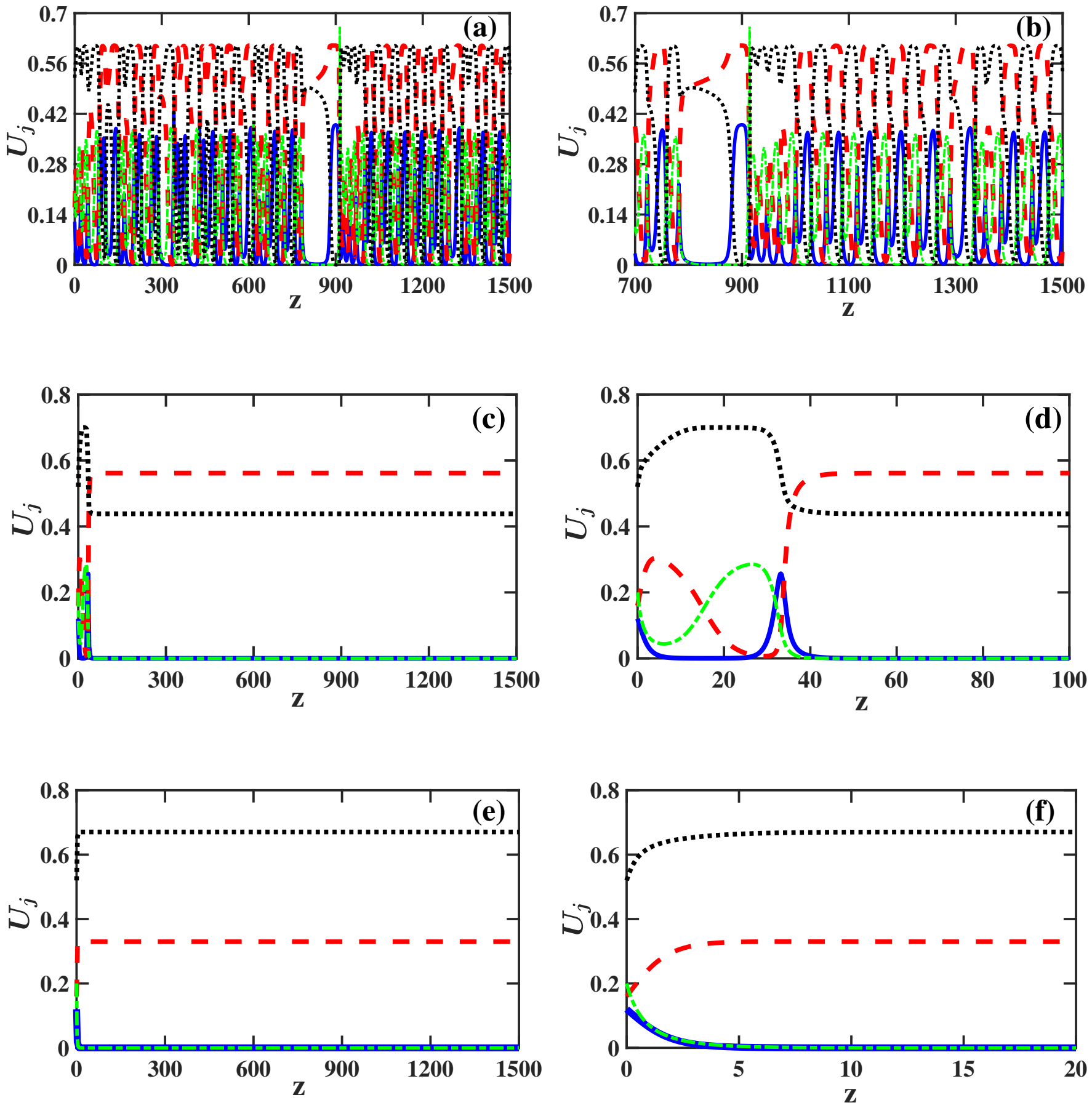

FIG. 8: The $z$-dependences of the pulse energies $U_j$ in the unshifted 1990 Di Cera model (44) for $m = 0.61$ [(a) and (b)], $m = 0.7$ [(c) and (d)], and $m = 0.8$ [(e) and (f)]. The initial condition is $U_1(0) = 0.12$, $U_2(0) = 0.16$, $U_3(0) = 0.2$, and $U_4(0) = 0.52$, and the other parameter values are the same as in Figs. 6 and 7. The solid blue, dashed red, dashed-dotted green, and dotted black curves represent $U_j(z)$, obtained by numerical solution of Eq. (44).

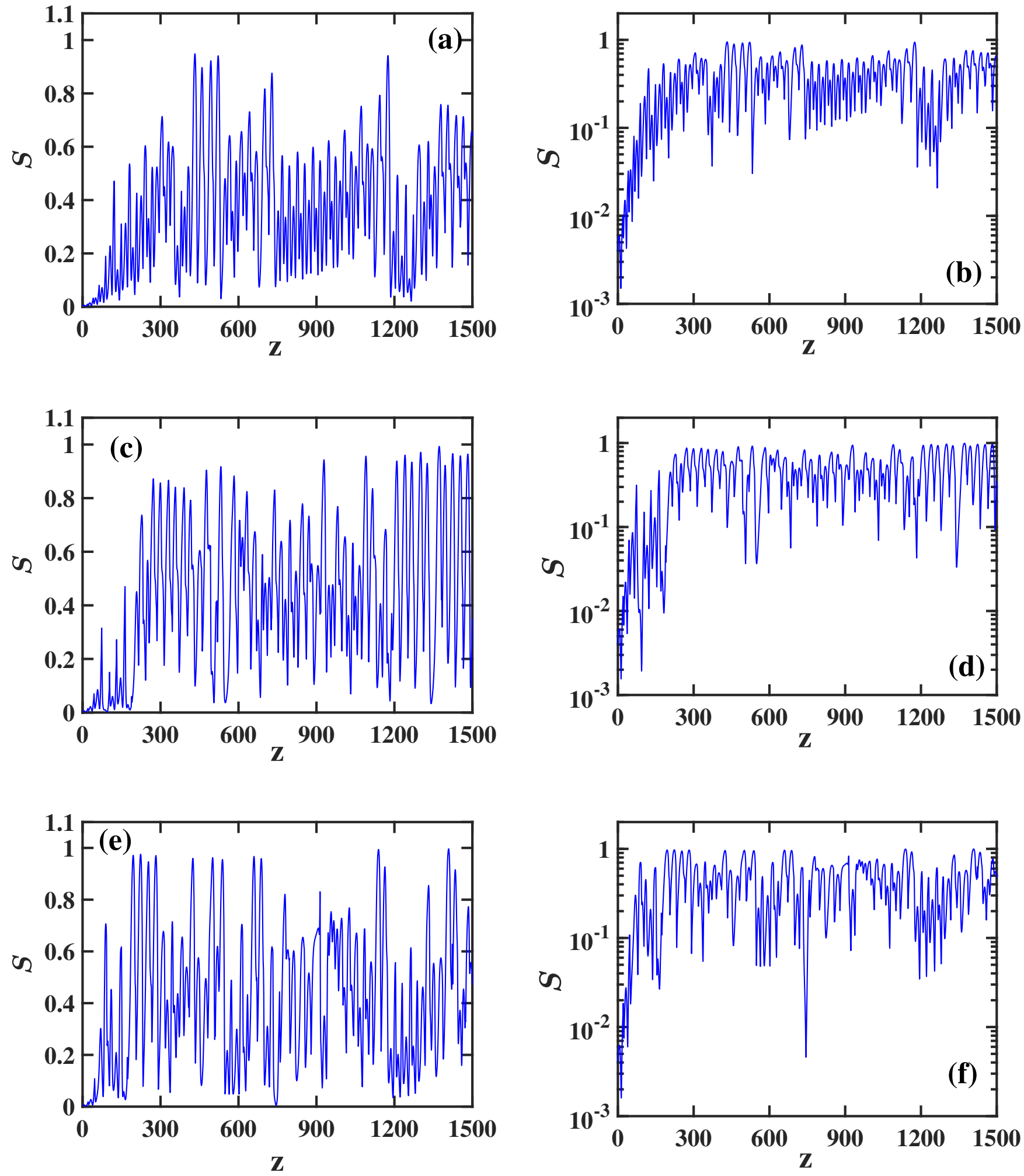

FIG. 9: The distance between two trajectories with adjacent initial conditions $S$ vs propagation distance $z$ for the unshifted 1990 Di Cera model (44) at $m = 0.58$ [(a) and (b)], $m = 0.6$ [(c) and (d)], and $m = 0.61$ [(e) and (f)]. Graphs (a), (c), and (e) show $S(z)$ on a linear scale, while graphs (b), (d), and (f) show $S(z)$ on a logarithmic scale. The initial conditions are $U_1(0) = 0.12$, $U_2(0) = 0.16$, $U_3(0) = 0.2$, $U_4(0) = 0.52$, and $U'_1(0) = 0.122$, $U'_2(0) = 0.162$, $U'_3(0) = 0.198$, $U'_4(0) = 0.518$. The other parameter values are the same as in Fig. 6.

# V. NUMERICAL SIMULATIONS WITH THE PERTURBED COUPLED-NLS MODELS

## A. Introduction

The main goal of the current paper is to demonstrate transition to spatiotemporal chaos in systems described by perturbed coupled-NLS models, where the chaotic dynamics of pulse energies is described by the reduced LV models (15), (17), (24), and (25). We therefore turn to describe the setups and the results of the numerical simulations with the coupled-NLS models (21), (22), (26), and (27). We point out that the reduced LV models (15), (17), (24), and (25) are based on several simplifying assumptions, which were described in sections II and III A. One of these assumptions is actually not satisfied in the coupled-NLS simulation setups of our paper, and the validity of two other assumptions might break down at intermediate propagation distances. More specifically, we first note that in the current coupled-NLS simulation setups, some of the physical parameter values characterizing the strength of the dissipative cubic interaction are of order 1, and therefore, the assumption of weak dissipative perturbation is not satisfied. Second, the LV models (15), (17), (24), and (25) neglect the effects of pulse-pattern distortion due to radiation emission. However, cumulative radiation emission effects can induce severe pulse-pattern distortion at intermediate distances, and this can lead to large deviations of the energy dynamics in the coupled-NLS simulations from the chaotic dynamics expected by the LV models. Third, the interplay between radiation emission and intrasequence interaction effects, which are also neglected by the LV models, can break the periodic structure of the pulse sequences. This phenomenon might also lead to large differences between the results of the LV and coupled-NLS models, and might prevent the observation of chaotic dynamics in the coupled-NLS simulations. For these reasons, it is important to check the predictions of the reduced LV models (15), (17), (24), and (25) for chaotic dynamics of pulse energies by numerical simulations with the full perturbed coupled-NLS models (21), (22), (26), and (27). In the current section, we take on this essential task.

The perturbed coupled-NLS models (21), (22), (26), and (27) are solved numerically on a computational domain $[t_{min}, t_{max}]$ using the split-step method with periodic boundary conditions [1, 77]. Since we use periodic boundary conditions, the simulations represent propagation of the pulse sequences in a closed waveguide-array loop, similar to the one shown in Fig. 1(a). We emphasize that many long-distance multisequence transmission experiments are performed in closed waveguide loops [4, 28, 63, 64], and therefore, the simulation setups that we use are very pertinent from both the physical point of view and the experimental point of view. The initial condition for the simulations is in the form of four periodic sequences of $2N + 1$ NLS solitons with amplitudes $\eta_j(0)$, frequencies $\beta_j(0)$, and zero phases. To enforce the periodic boundary conditions at $z = 0$ with good accuracy, we add $N_s = 10$ solitons on both sides of the computational domain, as was also done in our previous works on multisequence soliton-based propagation [33–39]. Thus, the initial condition is:

$$\psi_j(t,0) = \sum_{n=-N-N_s}^{N+N_s} \frac{\eta_j(0)\exp[-i\beta_j(0)(t-nT)]}{\cosh[\eta_j(0)(t-nT)]} \quad \text{for } t_{min} \le t \le t_{max}, \tag{57}$$

where $1 \le j \le 4$, and $\Delta\beta = \beta_{j+1}(0) - \beta_j(0) \gg 1$. For concreteness, we present here the results of numerical simulations with $\epsilon_3 = 0.02$, $T = 20$, $\Delta\beta = 40$, $N = 1$, and $[t_{min}, t_{max}] = [-30, 30]$. Additionally, the total energy is taken as $M = 1.0$ in the simulations with the two unshifted coupled-NLS models, and as $M = 2.6$ in the simulations with the two shifted coupled-NLS models. We emphasize, however, that similar results are obtained with other values of these physical parameters.

In some of the coupled-NLS simulations reported in the current paper, the soliton sequences experience very strong distortions at intermediate and large distances. As a result, in this case one cannot even identify the amplitudes of the highly distorted pulses, and it becomes necessary to measure the energies of the pulse sequences instead. We define the numerically obtained energy of the $j$th pulse sequence at distance $z$ as:

$$U_j^{(num)}(z) = (4N+2)^{-1} \int_{t_{min}}^{t_{max}} dt\, |\psi_j^{(num)}(t,z)|^2, \tag{58}$$

where $\psi^{(num)}(t,z)$ is the numerically obtained electric field of the $j$th sequence. Thus, $U_j^{(num)}(z)$ is actually half of the energy of the $j$th sequence at $z$ per initial pulse. The division by an extra factor of 2 in Eq. (58) ensures that in the case of weakly perturbed well-separated soliton sequences with intermediate or large amplitude values $\eta_j(z)$, the equation approximately reduces to the simple relation $U_j(z) = \eta_j(z)$. Indeed, exactly the same equation was successfully used in Refs. [33–39] for measuring the $\eta_j(z)$ values from the results of coupled-NLS simulations in the case of weakly perturbed soliton sequences.

Another aspect of the simulation setups that deserves attention is the determination of the values of the $U_j(0)$ and the $\eta_j(0)$ in the initial conditions for the coupled-NLS simulations. Indeed, in some of the initial conditions used in the simulations, the energy and amplitude values are small. As a result, there is strong overlap between adjacent solitons within the same sequence at $z = 0$, and the relation $U_j(0) = \eta_j(0)$, which is valid for a single fundamental NLS soliton on an infinite domain, is highly inaccurate. This is especially important in the case of chaotic dynamics, due to the sensitive dependence of the dynamics on the initial conditions. We therefore develop the following self-consistent procedure for determining the values of $U_j(0)$ and $\eta_j(0)$ in the initial conditions for the simulations.

1. We first prescribe preliminary initial energy values, $\tilde{U}_j(0)$, based on preliminary numerical solution of one of the LV models (15), (17), (24), or (25).

2. We then find the corresponding $\eta_j(0)$ values by numerical solution of the equation
$$(4N+2)^{-1}\int_{t_{min}}^{t_{max}} dt\,|\psi_j(t,0)|^2 = \tilde{U}_j(0), \qquad (59)$$
where $\psi_j(t,0)$ is given by Eq. (57). These $\eta_j(0)$ values are used in Eq. (57) for the initial condition of the coupled-NLS simulation.

3. The values of the $U_j(0)$ that are used in the actual numerical solution of the LV model are recalculated as:
$$U_j(0) = (4N+2)^{-1}\int_{t_{min}}^{t_{max}} dt\,|\psi_j(t,0)|^2. \qquad (60)$$
This step is performed for (formal) consistency, since the $\eta_j(0)$ values that are found in step 2 are only numerical approximations to the exact solutions of Eq. (59). However, as we show below (in Tables 1-4), the values of the $U_j(0)$ and $\tilde{U}_j(0)$ are very close. Therefore, for the parameter setups considered in our paper, the results of the LV simulations obtained with the $U_j(0)$ are essentially the same as the results obtained with the $\tilde{U}_j(0)$.

### B. Simulations with the two 1989 coupled-NLS models

#### 1. *Simulations results for the unshifted 1989 coupled-NLS model*

We first discuss the results of the numerical simulations with the *unshifted* 1989 coupled-NLS model (21). To enable a clear comparison between the results of the LV and coupled-NLS models (15) and (21), we use parameter values that are similar to the ones used in section IV A. More specifically, the values of the $\tilde{b}_{jk}$ coefficients are $\tilde{b}_{12} = 50$, and $\tilde{b}_{14} = \tilde{b}_{23} = \tilde{b}_{34} = 5$. By Eq. (29), this set corresponds to the set $k_1 = 10.0$, and $k_2 = k_3 = k_4 = 1.0$ that was used in section IV A 2. The preliminary initial energies $\tilde{U}_j(0)$ are identical to the initial energies used in section IV A 2, i.e., $\tilde{U}_1(0) = 0.55$, and $\tilde{U}_2(0) = \tilde{U}_3(0) = \tilde{U}_4(0) = 0.15$. The corresponding $\eta_j(0)$ values and the actual $U_j(0)$ values that are used in the simulations of the current subsection are listed in Table 1. We run the simulations with $m = 0.35$ and $m = 0.575$ up to the final distances $z_f = 10^4$ and $z_f = 3 \times 10^4$, respectively. We verify the accuracy of the numerical solution by monitoring the numerically obtained value of the total energy $M$. We find that the deviation of $M$ from the expected value of 1.0 is always smaller than 0.0072. By the analysis and the simulations in section IV A, we expect to observe stable limit-cycle oscillations with period 1 for $m = 0.35$, and transition to chaos for $m = 0.575$. We also note that the value of the product $\epsilon_3\tilde{b}_{12}$ in the current simulation setup is 1.0. As a result, the dissipative cubic interaction terms $i\epsilon_3\tilde{b}_{12}\left[\eta_2(z) - m\right]|\psi_2|^2\psi_1$ and $i\epsilon_3\tilde{b}_{12}\left[m - \eta_2(z)\right]|\psi_1|^2\psi_2$ in Eq. (21) cannot be regarded as weak perturbation terms, and the weak perturbation assumption of the derivation of the LV model (15) is not satisfied.

TABLE I: The $\eta_j(0)$ values and the actual $U_j(0)$ values used in the simulations with Eqs. (21) and (15).

| $j$ | $\tilde{U}_j(0)$ | $\eta_j(0)$ | $U_j(0)$ |
|---|---|---|---|
| 1 | 0.55 | 0.5495950786121190 | 0.5499999999695991 |
| 2 | 0.15 | 0.1864453719228507 | 0.1499999999472344 |
| 3 | 0.15 | 0.1864453719228507 | 0.1499999999472344 |
| 4 | 0.15 | 0.0147973723089574 | 0.1499999999958060 |

The $z$-dependences of the pulse energies obtained by the numerical simulations with Eq.

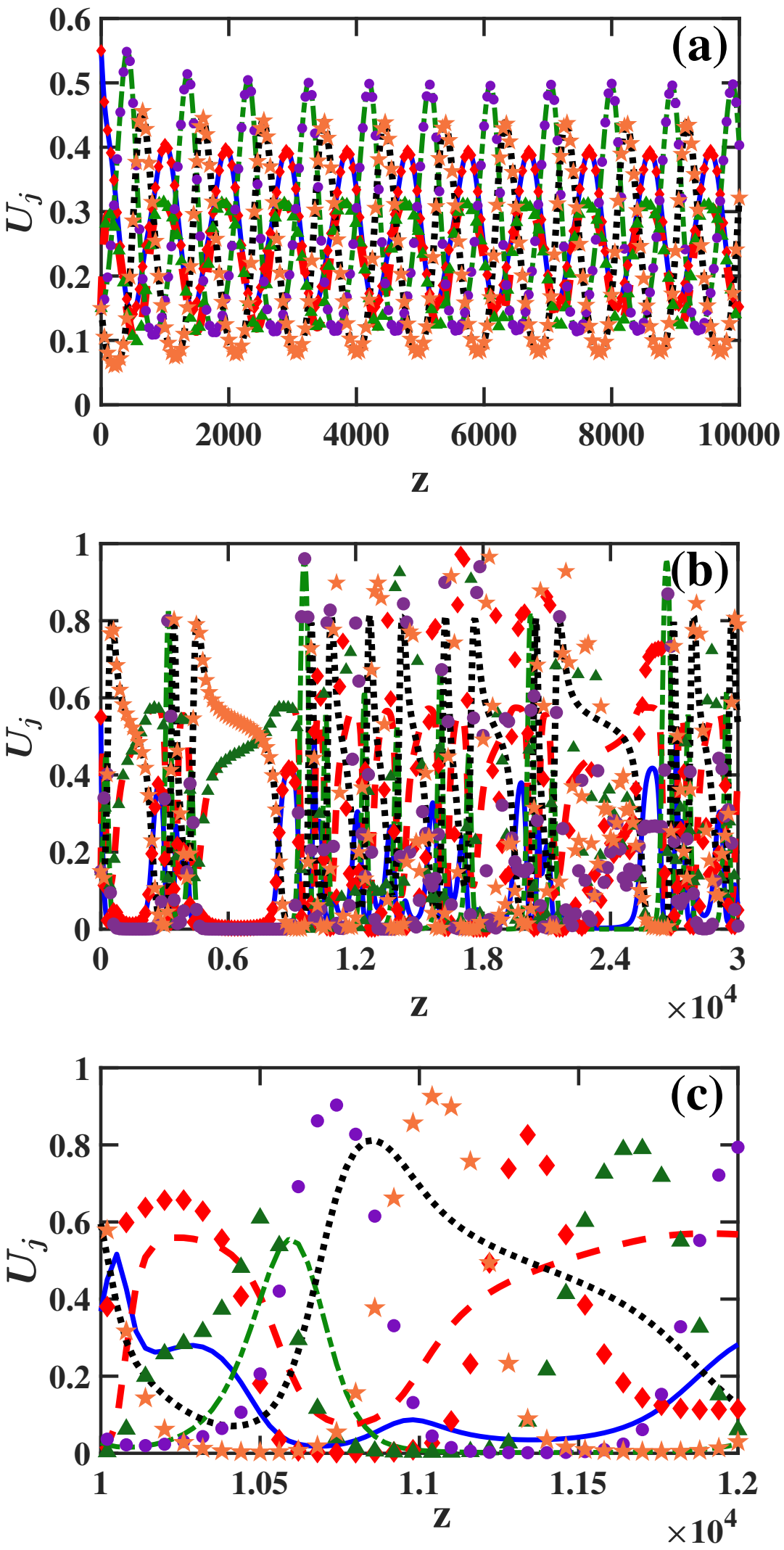

FIG. 10: The $z$-dependences of the pulse energies $U_j$ obtained with the unshifted 1989 coupled-NLS and Di Cera models (21) and (15) for $m = 0.35$ (a), and $m = 0.575$ [(b) and (c)]. The solid blue, dashed red, dashed-dotted green, and dotted black curves represent $U_j(z)$ with $j = 1, 2, 3, 4$, obtained by numerical solution of Eq. (21). The red diamonds, green triangles, purple circles, and orange stars represent $U_j(z)$ with $j = 1, 2, 3, 4$, obtained with Eq. (15).

(21) are shown in Fig. 10 together with the results obtained with the unshifted Di Cera model (15). As seen in Fig. 10(a), for $m = 0.35$, the numerically obtained $U_j(z)$ curves exhibit stable limit-cycle oscillations in the entire propagation interval $0 \le z \le 10^4$, in very good agreement with the predictions of Eq. (15). However, the situation is quite different for $m = 0.575$. Indeed, as seen in Fig. 10(b), in this case we observe good agreement between the results of Eqs. (21) and (15) only up to around $z \sim 8500$. At this distance, the coupled-NLS model's result starts to deviate significantly from the LV model's result, and the deviation keeps growing with increasing $z$. As a result, at larger distances, there is a complete mismatch between the results of the coupled-NLS and LV models, as illustrated in Fig. 10(c) for the interval $10^4 \le z \le 1.2 \times 10^4$. This result together with similar results that are obtained with other sets of physical parameter values strongly indicates that chaotic dynamics of pulse energies cannot be realized in the unshifted 1989 coupled-NLS model (21).

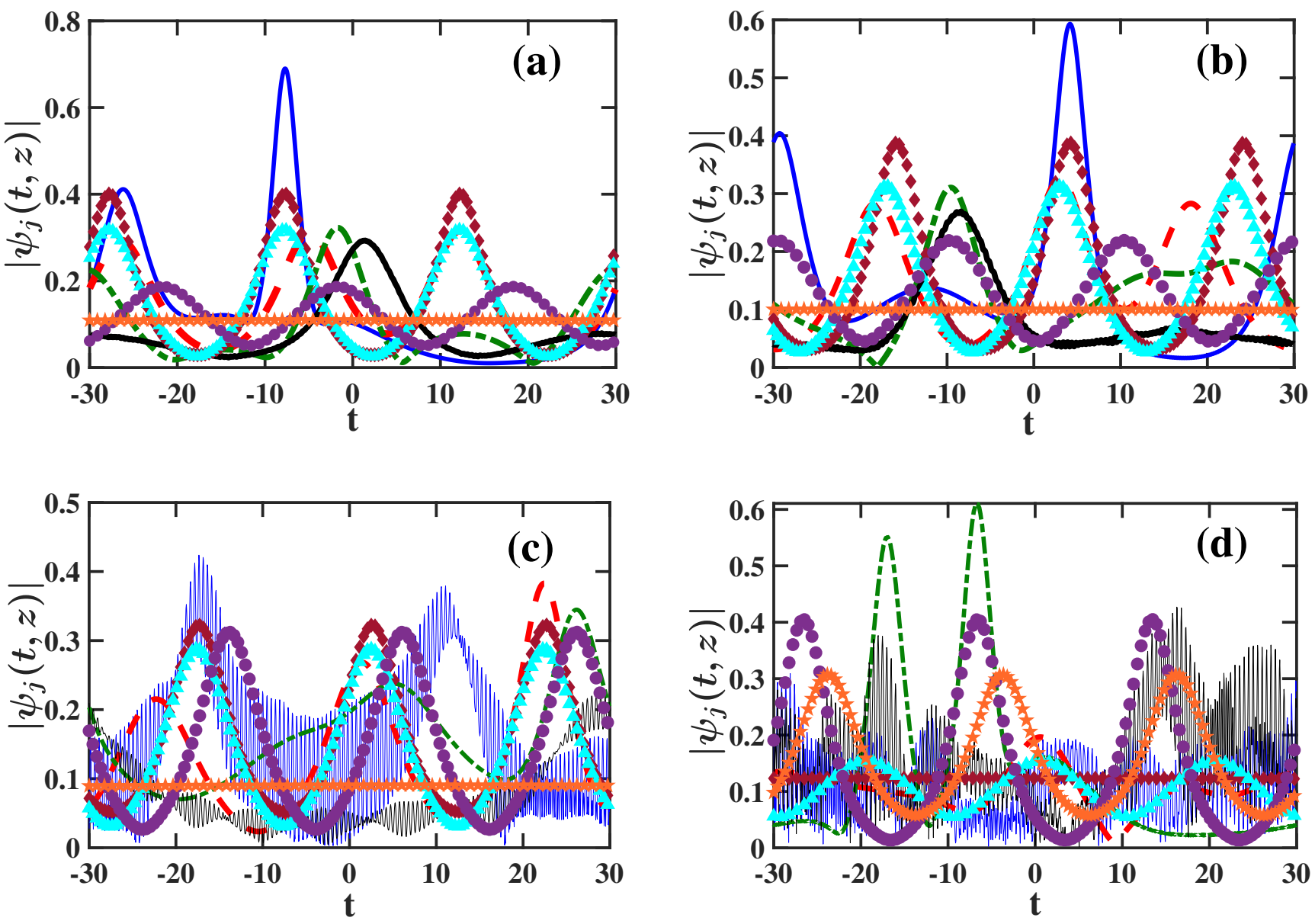

FIG. 11: The pulse patterns $|\psi_j(t,z)|$ obtained with the unshifted 1989 coupled-NLS model (21) for $m = 0.35$ at $z = 1000$ (a), $z = 2000$ (b), $z = 4000$ (c), and $z = 10^4$ (d). The solid blue, dashed red, dashed-dotted green, and solid black curves represent $|\psi_j(t,z)|$ with $j = 1, 2, 3, 4$, obtained by numerical solution of Eq. (21). The brown diamonds, cyan triangles, purple circles, and orange stars represent the theoretical predictions for $|\psi_j(t,z)|$ with $j = 1, 2, 3, 4$, obtained with Eq. (A1).

Further insight into the spatiotemporal dynamics is obtained by analyzing the $t$-dependences of the pulse patterns $|\psi_j(t,z)|$ at different distances. Fig. 11 shows the pulse patterns that are obtained in the numerical simulation with Eq. (21) for $m = 0.35$ at four representative distances. The figure also shows the theoretical prediction of Eq. (A1), which is based on the assumption of weakly distorted periodic soliton sequences. As seen in the figure, there is strong disagreement between the theoretical prediction and the result of the coupled-NLS simulation already at $z = 1000$. In particular, the numerically obtained pulse

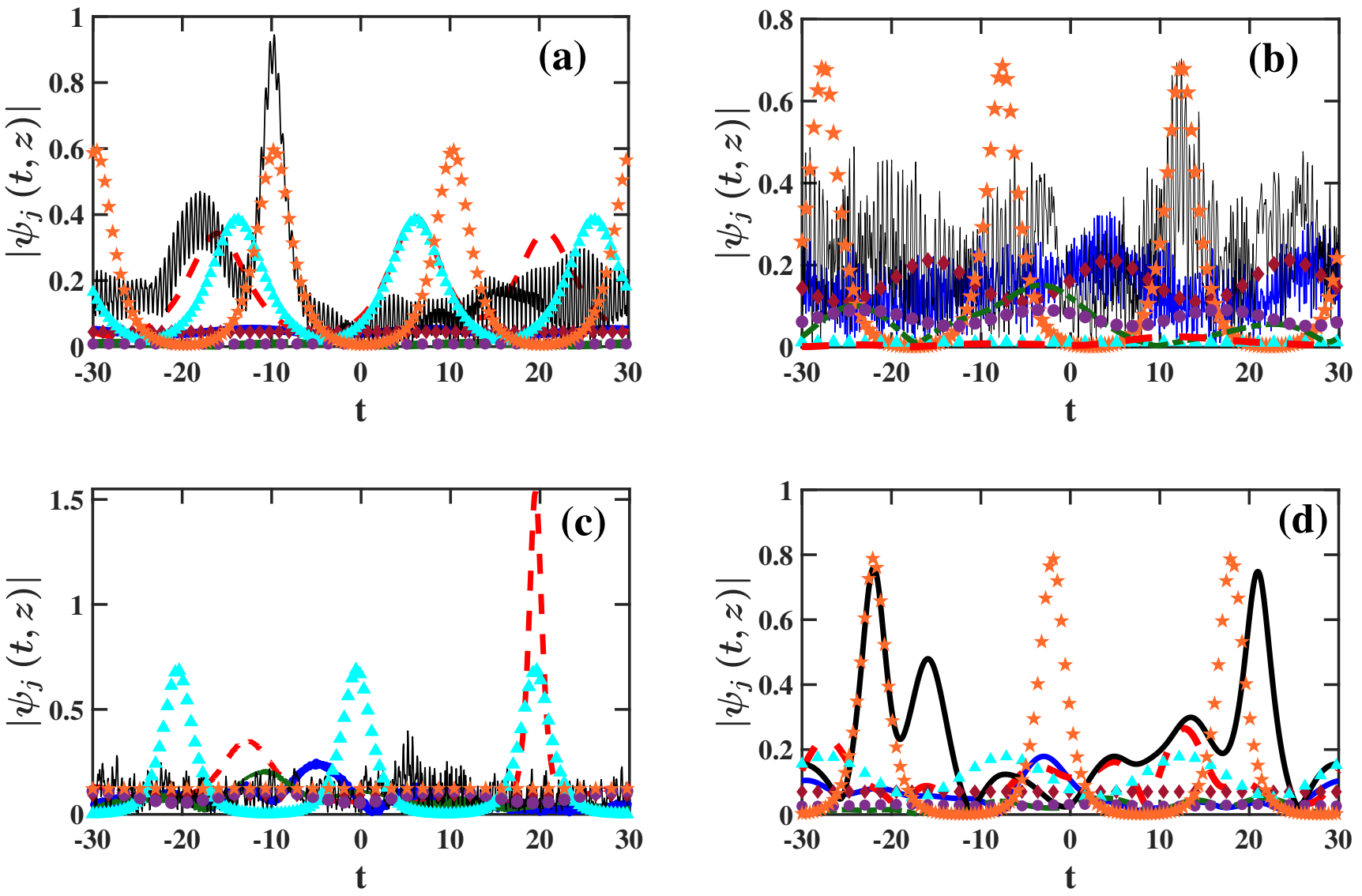

FIG. 12: The pulse patterns $|\psi_j(t, z)|$ obtained with the unshifted 1989 coupled-NLS model (21) for $m = 0.575$ at $z = 1000$ (a), $z = 10^4$ (b), $z = 1.5 \times 10^4$ (c), and $z = 3 \times 10^4$ (d). The solid blue, dashed red, dashed-dotted green, and solid black curves represent $|\psi_j(t, z)|$ with $j = 1, 2, 3, 4$, obtained by numerical solution of Eq. (21). The brown diamonds, cyan triangles, purple circles, and orange stars represent the theoretical predictions for $|\psi_j(t, z)|$ with $j = 1, 2, 3, 4$, obtained with Eq. (A1).

patterns of sequences 1 and 3 consist of of two pulses instead of three, and the numerical pulse pattern of sequence 4 contains only a single pulse. Additionally, significant radiative tails exist for the same pulse sequences, but these radiative tails are not oscillatory. At larger distances, $4000 \le z \le 10^4$, the disagreement between the theoretical and the numerical results is even stronger due to the appearance of new distortion characteristics in the form of fast modulations of the $|\psi_j(t, z)|$. Indeed, as seen in Figs. 11(c) and 11(d), at these distances, the numerically obtained pulse patterns of sequences 1 and 4 are highly modulated. In contrast, the pulse patterns of sequences 2 and 3 are strongly distorted by radiative tails, but do not exhibit any fast modulations or other highly oscillatory features.

Similar interesting spatiotemporal dynamics of the pulse patterns $|\psi_j(t, z)|$ is observed in Fig. 12 for $m = 0.575$. More specifically, the disagreement between the theoretical and the numerical pulse patterns is very strong already at $z = 1000$. Furthermore, the strong pulse pattern modulation of sequence 4, which appeared at $z \ge 4000$ for $m = 0.35$, exists here at $z = 1000$. The strong pulse pattern modulations of sequences 1 and 4 persist at intermediate distances [see Fig. 12(b)]. However, at larger distances, the pulse pattern modulations weaken, and the temporary generation of a single high-peak soliton in sequence 2 is observed [see Fig. 12(c)]. Eventually, the high-frequency pulse pattern modulations disappear [see Fig. 12(d)]. As a result, at $z = 3 \times 10^4$, the pulse patterns are strongly distorted, but no significant highly oscillatory features are observed.

### *2. Simulations results for the shifted 1989 coupled-NLS model*

We saw in the preceding subsection that chaotic dynamics is not realized in the unshifted coupled-NLS model (21). Therefore, we must study the energy and pulse pattern dynamics in the *shifted* coupled-NLS model (26). The values of the $\tilde{b}_{jk}$ coefficients in the numerical simulations with Eq. (26) are identical to the ones used in the simulations with Eq. (21). Additionally, the values of the shift parameters are $a_j = 0.4$ for $1 \le j \le 4$. Thus, the preliminary initial energy values are increased by 0.4 relative to the values used in the simulations with Eq. (21), i.e., we now use $\tilde{U}_1(0) = 0.95$, and $\tilde{U}_2(0) = \tilde{U}_3(0) = \tilde{U}_4(0) = 0.55$. The corresponding $\eta_j(0)$ values and the actual $U_j(0)$ values are listed in Table 2. We present here the results of the simulations with $m = 0.35$, $m = 0.55$, and $m = 0.575$, which were run up to the final distances $z_f = 10^4$, $z_f = 1.5 \times 10^4$, and $z_f = 3 \times 10^4$, respectively. We also briefly discuss our simulations results for $m = 0.25$, $m = 0.57$, and $m = 0.573$. The accuracy of the numerical simulations is verified by monitoring the numerical value of $M$. We find that the deviation of $M$ from the expected value of 2.6 is always smaller than 0.0095. According to the analysis and the simulations in section IV A, the pulse energies should exhibit stable limit-cycle oscillations with period 1 for $m = 0.35$, stable limit-cycle oscillations with period 2 for $m = 0.55$, and transition to chaos for $m = 0.575$. Similar to the setup of the simulations with Eq. (21), the value of $\epsilon_3\tilde{b}_{12}$ is 1.0. Therefore, the weak perturbation assumption of the derivation of the LV model (24) is not satisfied.

The $z$-dependences of the pulse energies obtained by the numerical simulations with the shifted 1989 coupled-NLS model (26) for $m = 0.35$, $m = 0.55$, and $m = 0.575$ are shown in Fig. 13 along with the results of the shifted 1989 Di Cera model (24). We observe excellent agreement between the results of the coupled-NLS and LV models over the entire ultra-long propagation intervals for all three values of $m$. Thus, the coupled-NLS simulations confirm that the $U_j(z)$ exhibit stable limit-cycle oscillations with period 1 and period 2 for $m = 0.35$ and $m = 0.55$, and transition to chaos for $m = 0.575$. Similar high-quality agreement between the results of the coupled-NLS and LV models is obtained for $m = 0.25$, $m = 0.57$, and $m = 0.573$. Accordingly, the numerically obtained $U_j(z)$ exhibit decaying oscillations that approach the equilibrium values of Eq. (30) for $m = 0.25$, stable limit-cycle oscillations with period 4 for $m = 0.57$, and transition to chaos for $m = 0.573$. These results confirm that the pulse energies obtained with the shifted coupled-NLS model (26) exhibit chaotic dynamics, where chaos emerges via Hopf and PD bifurcations, as predicted by the shifted LV model (24). We point out that the high-quality agreement between the results of the shifted coupled-NLS and LV models is very surprising due to the following reasons. First, the weak perturbation assumption, which was used in the derivation of the LV model (24), is not satisfied. Second, as we show in the following paragraphs, the numerically obtained pulse patterns undergo severe distortion during the propagation, and as a result, additional assumptions of the LV model's derivation break down. We also note that a comparison of the results in Figs. 13 and 10 strongly indicates that the shifting transformation $\mathbf{U}' = \mathbf{U} + \mathbf{a}$, which was used in section III to derive the LV and coupled-NLS models (24) and (26), is a key factor in enabling the demonstration of strongly nonlinear and chaotic dynamics of pulse energies in the waveguide array systems considered in our paper.

TABLE II: The $\eta_j(0)$ values and the actual $U_j(0)$ values used in the simulations with Eqs. (26) and (24).

| $j$ | $\tilde{U}_j(0)$ | $\eta_j(0)$ | $U_j(0)$ |
|---|---|---|---|
| 1 | 0.95 | 0.9499995974247213 | 0.9500000000007992 |
| 2 | 0.55 | 0.5502116741799910 | 0.5500000000004857 |
| 3 | 0.55 | 0.5502116741799910 | 0.5500000000004857 |
| 4 | 0.55 | 0.5495950786435060 | 0.5500000000007771 |

Important insight into the dynamics is gained by following the evolution of the pulse patterns with propagation distance $z$. Fig. 14 shows the pulse patterns $|\psi_j(t, z)|$ obtained by numerical solution of Eq. (26) for $m = 0.35$ at four representative distances together with the theoretical predictions of Eq. (A1). We observe that the pulse patterns of sequences 1 and 4 are strongly distorted already at $z = 1000$, where distortions appear as fast tem-poral oscillations (fast modulations). These distortions persist throughout the propagation, and consequently, the numerically obtained results for $|\psi_1(t, z)|$ and $|\psi_4(t, z)|$ strongly disagree with the theoretical predictions of Eq. (A1) on the entire interval $1000 \leq z \leq 10^4$. In contrast, no significant pulse pattern distortions are observed for sequences 2 and 3 for $0 \leq z < 5000$ [see Figs. 14(a) and 14(b)]. As a result, there is very good agreement be-tween the numerical and theoretical results for $|\psi_2(t,z)|$ and $|\psi_3(t,z)|$ at these distances. For $z \geq 5000$, we observe significant shifts in the positions of the solitons in sequences 2 and 3 relative to their expected positions in fully periodic sequences [see Figs. 14(c) and 14(d)]. However, even at these larger distances, no oscillating radiative tails or other strong distortion characteristics are observed for these sequences, which is very remarkable, considering the violation of the weak perturbation assumption in the coupled-NLS simulations.

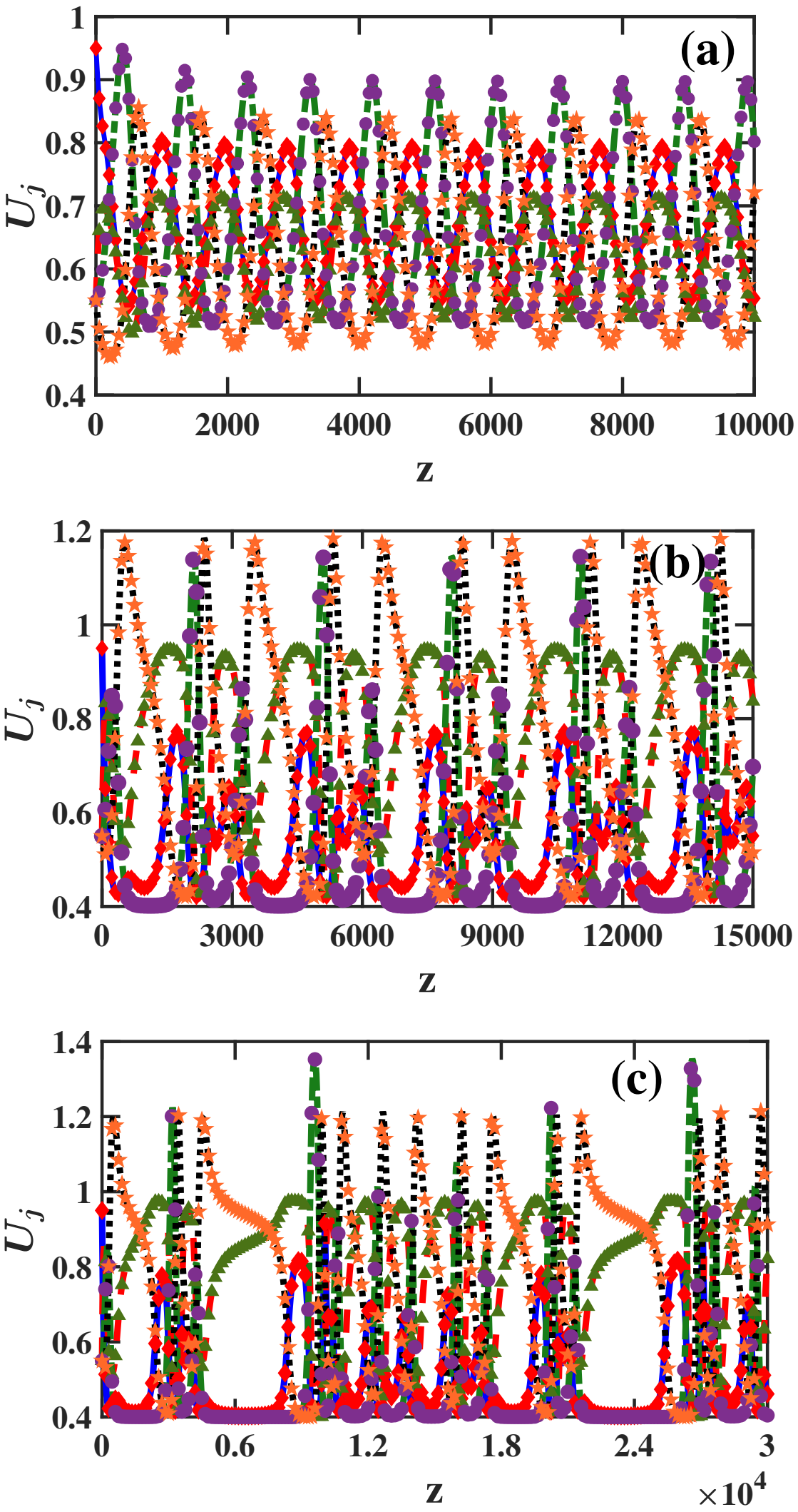

FIG. 13: The $z$-dependences of the pulse energies $U_j$ obtained with the shifted 1989 coupled-NLS and Di Cera models (26) and (24) for $m = 0.35$ (a), $m = 0.55$ (b), and $m = 0.575$ (c). The solid blue, dashed red, dashed-dotted green, and dotted black curves represent $U_j(z)$ with $j = 1, 2, 3, 4$, obtained by the simulations with Eq. (26). The red diamonds, green triangles, purple circles, and orange stars represent $U_j(z)$ with $j = 1, 2, 3, 4$, obtained with Eq. (24).

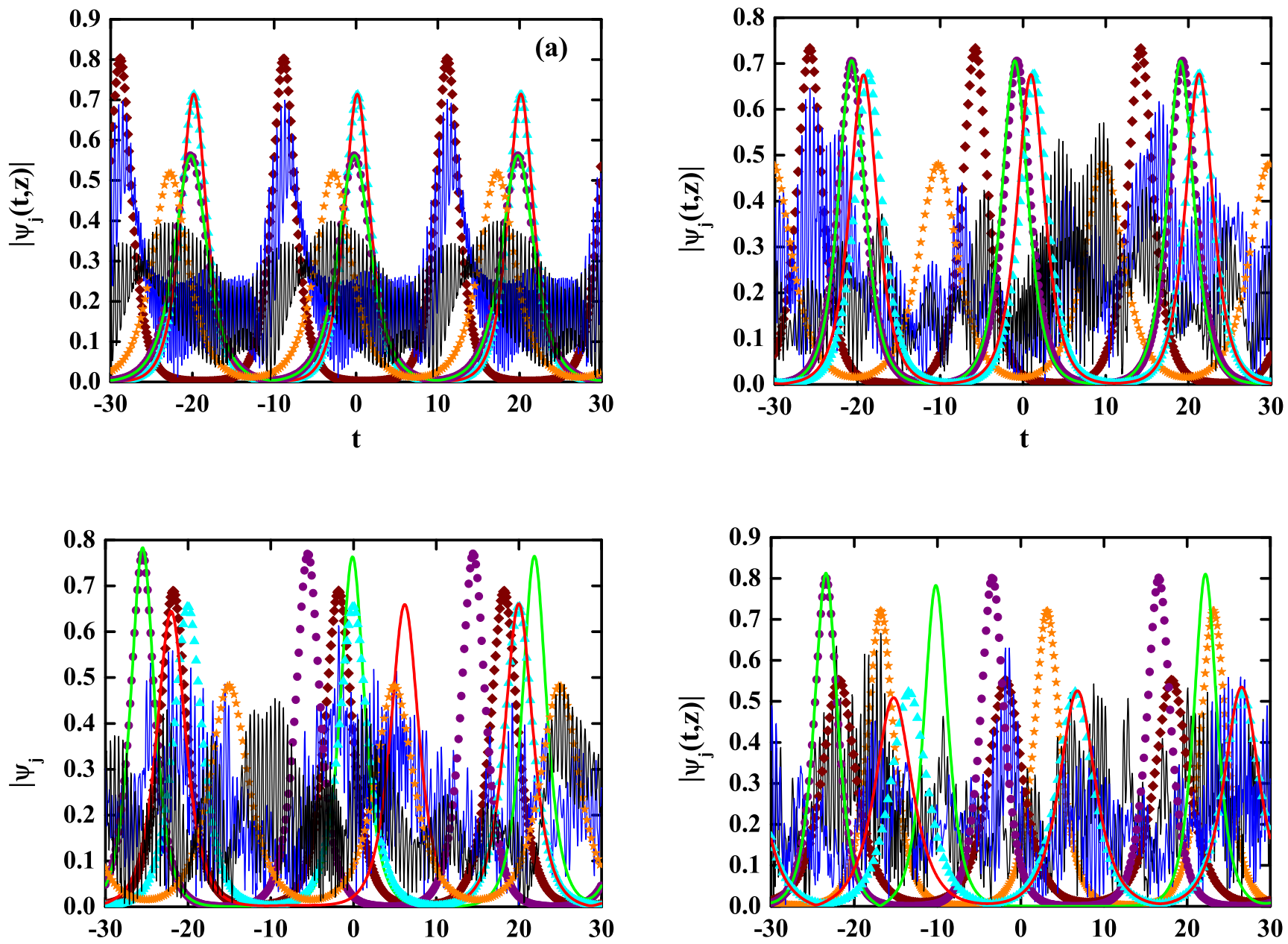

FIG. 14: The pulse patterns $|\psi_j(t,z)|$ obtained with the shifted 1989 coupled-NLS model (26) for $m = 0.35$ at $z = 1000$ (a), $z = 4000$ (b), $z = 5000$ (c), and $z = 10^4$ (d). The solid blue, solid red, solid green, and solid black curves represent $|\psi_j(t,z)|$ with $j = 1, 2, 3, 4$, obtained by numerical solution of Eq. (26). The brown diamonds, cyan triangles, purple circles, and orange stars represent the theoretical predictions for $|\psi_j(t,z)|$ with $j = 1, 2, 3, 4$, obtained with Eq. (A1).

The dynamics of the pulse patterns $|\psi_j(t,z)|$, which is obtained for $m = 0.575$, i.e., in the chaotic regime of energy dynamics, is equally striking. Fig. 15 shows the $|\psi_j(t,z)|$ obtained in the simulation with Eq. (26) along with the theoretical predictions of Eq. (A1). As seen in Fig. 15(a), at $z = 1000$, the soliton sequences retain their shape, and as a result, the agreement between the theoretical predictions and the numerical results for the $|\psi_j(t,z)|$ is very good for all four sequences. However, at larger distances, $z \geq 5000$, the pulse patterns develop strong distortions, and the numerically obtained results for the pulse patterns deviate significantly from the theoretical predictions [see Figs. 15(b), 15(c), and 15(d)]. In particular, sequences 1 and 4 are distorted by fast modulations of $|\psi_1(t, z)|$ and $|\psi_4(t, z)|$, similar to the fast modulations of these sequences in Figs. 11, 12, and 14. In contrast, no distortions due to fast modulations are observed for sequences 2 and 3. Instead, pulse pattern distortions of these sequences are mainly due to significant shifts of the soliton positions relative to their expected positions in completely periodic sequences. Additionally, in each of the sequences 2 and 3 we observe solitons with different amplitude values. The spatiotemporal dynamics of the pulse patterns, which is observed for $m = 0.25$, $m = 0.55$, $m = 0.57$, and for $m = 0.573$ (in the chaotic regime of energy dynamics) is similar to the one seen in Figs. 14 and 15.

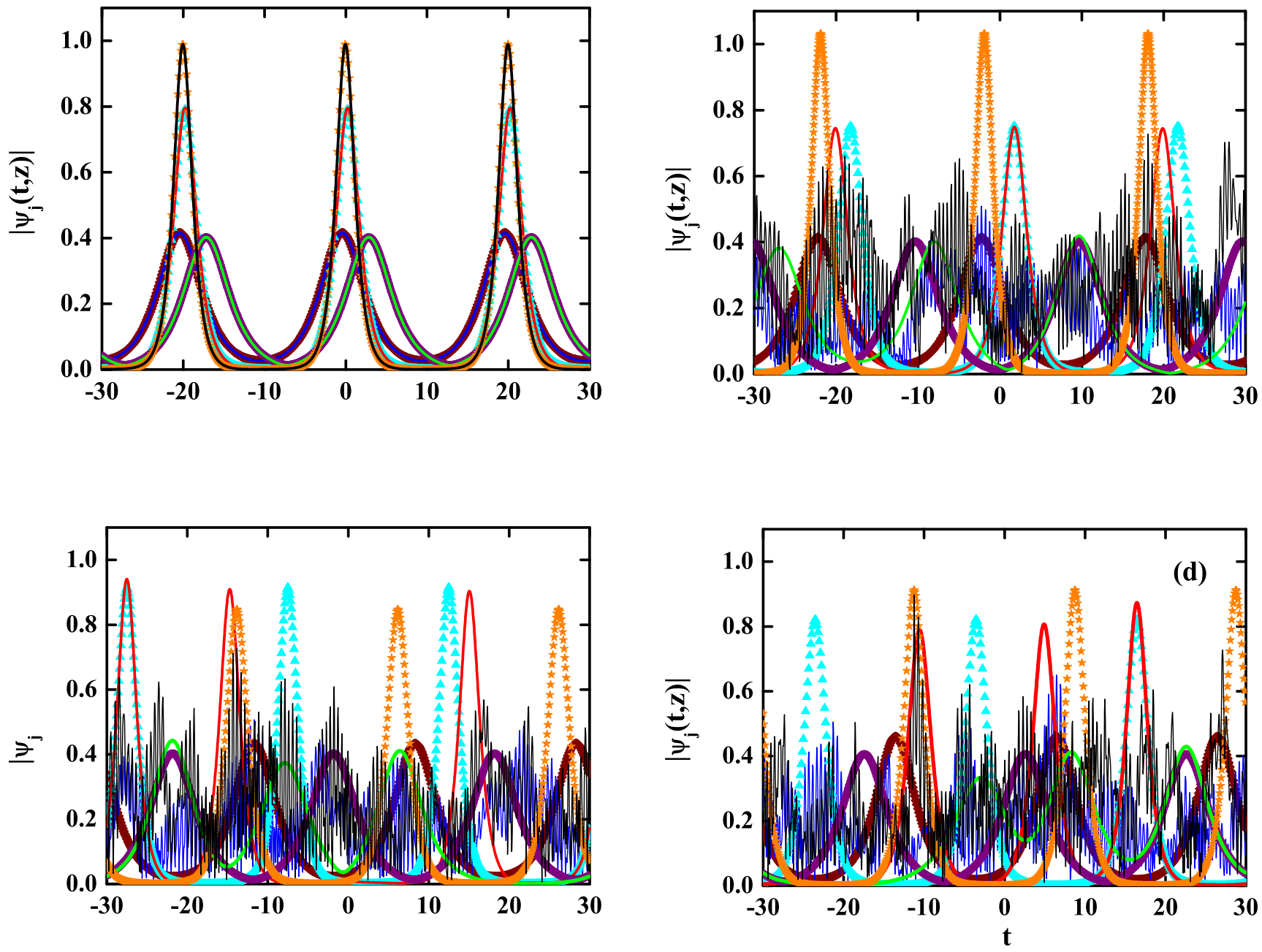

FIG. 15: The pulse patterns $|\psi_j(t, z)|$ obtained with the shifted 1989 coupled-NLS model (26) for $m = 0.575$ at $z = 1000$ (a), $z = 5000$ (b), $z = 1.5 \times 10^4$ (c), and $z = 3 \times 10^4$ (d). The solid blue, solid red, solid green, and solid black curves represent $|\psi_j(t, z)|$ with $j = 1, 2, 3, 4$, obtained by numerical solution of Eq. (26). The brown diamonds, cyan triangles, purple circles, and orange stars represent the theoretical predictions for $|\psi_j(t, z)|$ with $j = 1, 2, 3, 4$, obtained with Eq. (A1).

The results in Figs. 13 - 15 and similar results, which are obtained with other parameter setups, clearly demonstrate that the pulse sequences in the nonlinear waveguide array systems of Eq. (26) exhibit transition to dissipative spatiotemporal chaos. These results are highly remarkable since the chaotic dynamics of pulse energies is demonstrated with high precision despite the severe pulse pattern distortions experienced by the four pulse sequences. Additionally, a comparison of the results in Figs. 13 - 15 with the results in Figs. 10 - 12 strongly indicates that the shifting transformation $\mathbf{U}' = \mathbf{U} + \mathbf{a}$, which was introduced in section III, is essential for the realization of spatiotemporal chaos in the nonlinear waveguide array systems considered in our paper.

### C. Simulations with the two 1990 coupled-NLS models

#### 1. *Simulations results for the unshifted 1990 coupled-NLS model*

The parameter setup that is used in the numerical simulations with the *unshifted* 1990 coupled-NLS model (22) is similar to the one used in section IV B for the unshifted 1990 Di Cera model (17). This choice facilitates a clear comparison between the results of the two models for energy dynamics. In accordance with this choice, the values of the $\tilde{b}_{jk}$ coefficients are $\tilde{b}_{12} = \tilde{b}_{34} = 50$, and $\tilde{b}_{14} = \tilde{b}_{23} = 5$. Using Eq. (29), we see that these values correspond to the values $k_1 = k_3 = 10.0$, and $k_2 = k_4 = 1.0$, which were considered in section IV B 2. The preliminary initial energies $\tilde{U}_j(0)$ are identical to the initial energies used in section IV B 2. That is, we use: $\tilde{U}_1(0) = 0.12$, $\tilde{U}_2(0) = 0.16$, $\tilde{U}_3(0) = 0.2$, and $\tilde{U}_4(0) = 0.52$. The corresponding $\eta_j(0)$ values and the actual $U_j(0)$ values are listed in Table 3. We run the simulations with $m = 0.52$ and $m = 0.58$ up to the final distance $z_f = 10^4$, and verify the simulations accuracy by monitoring the numerically obtained value of $M$. We find that the deviation of $M$ from the expected value of 1.0 is always smaller than 0.013. According to the analysis and the simulations in section IV B, the pulse energies should exhibit decaying oscillations that approach equilibrium values for $m = 0.52$, and transition to chaos for $m = 0.58$. Since $\epsilon_3\tilde{b}_{12} = \epsilon_3\tilde{b}_{34} = 1.0$, the weak perturbation assumption, used in the derivation of the LV model (17), is not satisfied.

Fig. 16 shows the $z$-dependences of the pulse energies obtained by numerical solution of the unshifted coupled-NLS model (22) for $m = 0.52$ and $m = 0.58$. Also shown are the predictions of the unshifted LV model (17). In both cases, we observe very good agreement between the results of the coupled-NLS and LV models up to intermediate propagation distances ($z \sim 4000$ for $m = 0.52$, and $z \sim 5000$ for $m = 0.58$). However, at larger distances, the coupled-NLS model's results deviate significantly from the LV model's results. Furthermore, in the chaotic regime of the LV model, i.e., at $m = 0.58$, the deviations keep growing with increasing $z$, and therefore, we observe complete mismatch between the results of the two models for $z > 6000$. These findings along with similar findings for other sets of physical parameter values strongly suggest that chaotic dynamics of pulse energies cannot be realized in the unshifted 1990 coupled-NLS model (22). Thus, both unshifted coupled-NLS models (21) and (22) fail to exhibit the chaotic dynamics of pulse energies that is predicted by the two corresponding reduced LV models.

TABLE III: The $\eta_j(0)$ values and the actual $U_j(0)$ values used in the simulations with Eqs. (22) and (17).

| $j$ | $\tilde{U}_j(0)$ | $\eta_j(0)$ | $U_j(0)$ |
|---|---|---|---|
| 1 | 0.12 | 0.0078518607857404 | 0.1200000000072364 |
| 2 | 0.16 | 0.1938220365300367 | 0.1599999999996530 |
| 3 | 0.2 | 0.2242774011341681 | 0.2000000000003804 |
| 4 | 0.52 | 0.5193377705372768 | 0.5200000000008452 |

The evolution of the pulse patterns $|\psi_j(t, z)|$ with propagation distance $z$ at $m = 0.52$ is shown in Fig. 17. More specifically, this figure shows the coupled-NLS simulation's result for the $|\psi_j(t, z)|$ at four representative distances together with the theoretical prediction of Eq. (A1). We observe strong disagreement between the coupled-NLS simulation's result and the theoretical prediction already at $z = 1000$. Indeed, the numerically obtained pulse patterns of sequences 1 and 2 consist of only one and two pulses, respectively, instead of three. Additionally, significant position shifts and unequal amplitudes are observed for the pulses in sequences 3 and 4. At larger distances, $4000 \le z \le 10^4$, stronger pulse pattern distortions in the form of fast modulations appear, first in sequence 1 (at $z = 4000$), and then in sequence 2 (at $z = 7000$). Additionally, sequences 3 and 4 are still distorted by significant position shifts, unequal pulse amplitudes, and radiative tails. However, no significant pulse pattern modulations are observed for the latter two sequences even at $z_f = 10^4$.

The spatiotemporal dynamics of the pulse patterns, which is obtained for $m = 0.58$, i.e., in the chaotic regime of the LV model (17), is equally interesting. Fig. 18 shows the pulse patterns obtained by the simulation with the coupled-NLS model (22) together with the theoretical prediction of Eq. (A1). At $z = 1000$, we observe significant pulse pattern distortions of sequences 1 and 2 with no significant fast modulations, while sequences 3 and 4 are only slightly distorted due to position shifts and unequal pulse amplitudes. At $z = 4000$, the distortions of sequences 3 and 4 become significant, and sequence 4 develops an oscillating (modulated) radiative tail. At larger distances, $7000 \le z \le 10^4$, sequences 1, 2, and 4 are strongly distorted by pronounced radiative modulations, while sequence 3 remains significantly distorted but with no observable fast modulations. Additionally, at $z = 7000$, we observe the temporary generation of a single high-peak soliton in sequence 2.

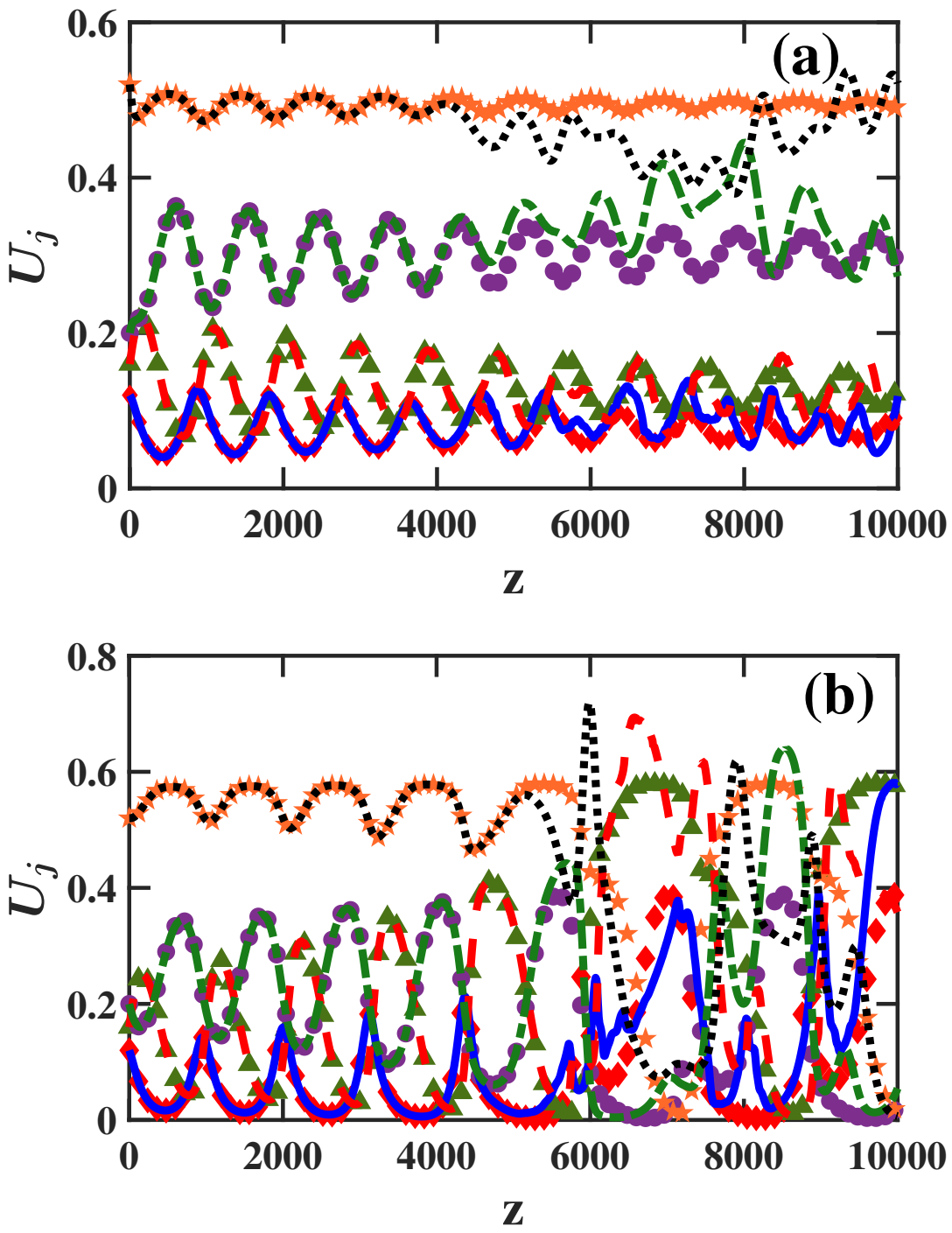

FIG. 16: The $z$-dependences of the pulse energies $U_j$ obtained with the unshifted 1990 coupled-NLS and Di Cera models (22) and (17) for $m = 0.52$ (a) and $m = 0.58$ (b). The solid blue, dashed red, dashed-dotted green, and dotted black curves represent $U_j(z)$ with $j = 1, 2, 3, 4$, obtained by numerical solution of Eq. (22). The red diamonds, green triangles, purple circles, and orange stars represent $U_j(z)$ with $j = 1, 2, 3, 4$, obtained with Eq. (17).

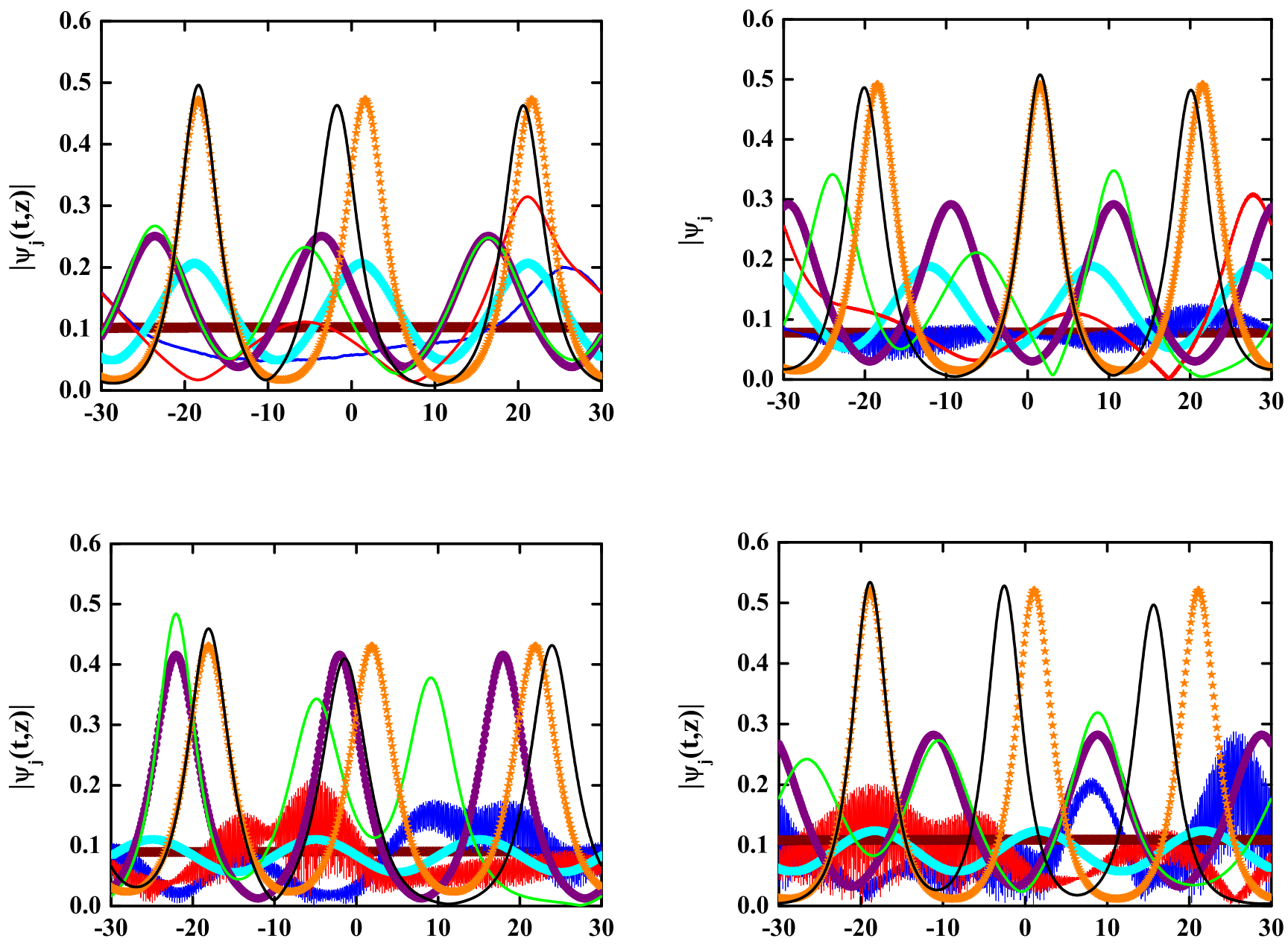

FIG. 17: The pulse patterns $|\psi_j(t, z)|$ obtained with the unshifted 1990 coupled-NLS model (22) for $m = 0.52$ at $z = 1000$ (a), $z = 4000$ (b), $z = 7000$ (c), and $z = 10^4$ (d). The solid blue, solid red, solid green, and solid black curves represent $|\psi_j(t, z)|$ with $j = 1, 2, 3, 4$, obtained by numerical solution of Eq. (22). The brown diamonds, cyan triangles, purple circles, and orange stars represent the theoretical predictions for $|\psi_j(t, z)|$ with $j = 1, 2, 3, 4$, obtained with Eq. (A1).

### 2. *Simulations results for the shifted 1990 coupled-NLS model*

Since chaotic dynamics is not realized in the unshifted coupled-NLS model (22), we must turn to investigate the energy and pulse pattern dynamics in the *shifted* coupled-NLS model (27). The values of the $\tilde{b}_{jk}$ coefficients in the numerical simulations with Eq. (27) are the same as the ones used in the simulations with Eq. (22). Additionally, the values of the shift parameters are $a_j = 0.4$ for $1 \le j \le 4$, such that the preliminary initial energy values are increased by 0.4 relative to the values used in the simulations with Eq. (22). Thus, we now use $\tilde{U}_1(0) = 0.52$, $\tilde{U}_2(0) = 0.56$, $\tilde{U}_3(0) = 0.6$, and $\tilde{U}_4(0) = 0.92$. The corresponding $\eta_j(0)$ values and the actual $U_j(0)$ values are listed in Table 4. We present here the simulations results for $m = 0.52$, $m = 0.6$, $m = 0.61$, and $m = 0.7$, which were run up to the distances $z_f = 10^4$, $z_f = 3 \times 10^4$, $z_f = 3 \times 10^4$, and $z_f = 10^4$, respectively. We also concisely discuss our simulations results for $m = 0.58$ and $m = 0.8$. We verify the the accuracy of the numerical simulations by monitoring the numerical value of $M$. We find that the deviation of $M$ from the expected value of 2.6 is always smaller than 0.008. According to the analysis and the simulations in section IV B, the pulse energies should exhibit decaying oscillations that approach stable equilibrium values for $m = 0.52$, transition to chaos for $m = 0.6$ and $m = 0.61$, and approach to the equilibrium values of one of the stable fuzzy equilibrium points of the LV model (25) for $m = 0.7$. Similar to the setup of the simulations with Eq. (22), $\epsilon_3\tilde{b}_{12} = \epsilon_3\tilde{b}_{34} = 1.0$. As a result, the weak perturbation assumption of the derivation of the LV model (25) is not satisfied in the simulations.

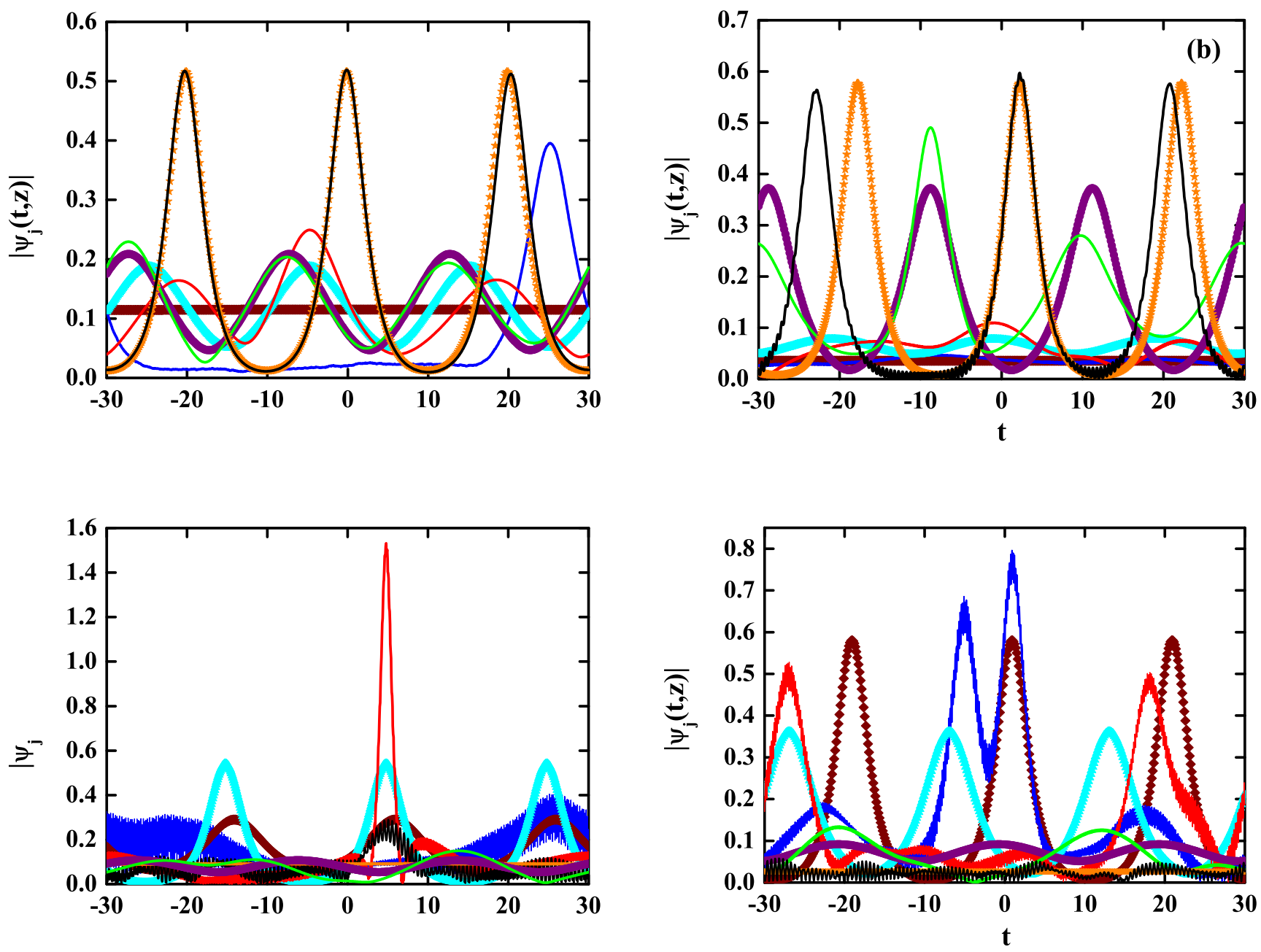

FIG. 18: The pulse patterns $|\psi_j(t,z)|$ obtained with the unshifted 1990 coupled-NLS model (22) for $m = 0.58$ at $z = 1000$ (a), $z = 4000$ (b), $z = 7000$ (c), and $z = 10^4$ (d). The solid blue, solid red, solid green, and solid black curves represent $|\psi_j(t,z)|$ with $j = 1, 2, 3, 4$, obtained by numerical solution of Eq. (22). The brown diamonds, cyan triangles, purple circles, and orange stars represent the theoretical predictions for $|\psi_j(t,z)|$ with $j = 1, 2, 3, 4$, obtained with Eq. (A1).

The dynamics of the pulse energies obtained in the simulations with the shifted coupled-NLS model (27) for $m = 0.52$, $m = 0.6$, $m = 0.61$, and $m = 0.7$ is shown in Fig. 19. Also shown are the predictions of the corresponding shifted LV model (25). The agreement between the results of the coupled-NLS and LV models is excellent over the entire propagation intervals for all four $m$ values. This means that the coupled-NLS simulations confirm that the $U_j(z)$ exhibit decaying oscillations that approach stable equilibrium values for $m = 0.52$, transition to chaos for $m = 0.6$ and $m = 0.61$, and a single oscillation that is followed by a decay to the equilibrium values of a stable fuzzy equilibrium point for $m = 0.7$. Similar excellent agreement between the results of the two models is found for $m = 0.58$ (transition to chaos) and for $m = 0.8$ (decay to a stable fuzzy equilibrium point). These findings are very surprising, considering the clear violation of the weak perturbation assumption in the coupled-NLS simulations and the strongly nonlinear character of the dynamics. Furthermore, based on the results in Figs. 16 and 19, we conclude that the shifting transformation $\mathbf{U}' = \mathbf{U} + \mathbf{a}$, which was used in section III to derive the LV and coupled-NLS models (25) and (27), is a crucial factor in enabling the demonstration of chaotic dynamics of pulse energies in the waveguide array systems that are studied in our paper.

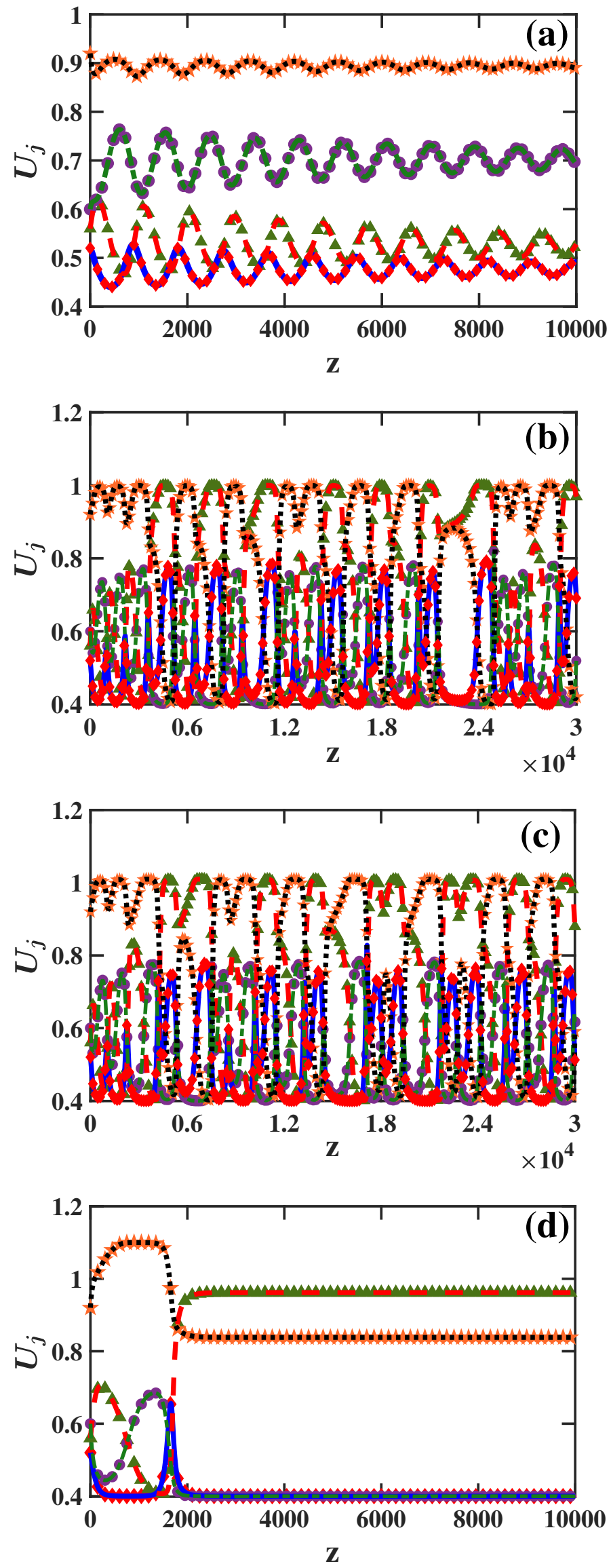

FIG. 19: The $z$-dependences of the pulse energies $U_j$ obtained with the shifted 1990 coupled-NLS and Di Cera models (27) and (25) for $m = 0.52$ (a), $m = 0.6$ (b), $m = 0.61$ (c), and $m = 0.7$ (d). The solid blue, dashed red, dashed-dotted green, and dotted black curves represent $U_j(z)$ with $j = 1, 2, 3, 4$, obtained by the simulations with Eq. (27). The red diamonds, green triangles, purple circles, and orange stars represent $U_j(z)$ with $j = 1, 2, 3, 4$, obtained with Eq. (25).

TABLE IV: The $\eta_j(0)$ values and the actual $U_j(0)$ values used in the simulations with Eqs. (27) and (25).

| $j$ | $\tilde{U}_j(0)$ | $\eta_j(0)$ | $U_j(0)$ |
|---|---|---|---|
| 1 | 0.52 | 0.5193377705372768 | 0.5200000000008452 |
| 2 | 0.56 | 0.5601797542776694 | 0.5600000000004587 |
| 3 | 0.6 | 0.6000928485309487 | 0.5999999999990997 |
| 4 | 0.92 | 0.9199993120519139 | 0.9199999999994120 |

Deeper understanding of the dynamics is achieved by following the evolution of the pulse patterns with propagation distance $z$. Fig. 20 shows the pulse patterns $|\psi_j(t, z)|$ obtained by the simulation with Eq. (27) for $m = 0.52$ at four representative distances together with the theoretical prediction of Eq. (A1). No significant pulse pattern distortions are observed at $z = 1000$, and as a result, the agreement between the theoretical prediction and the coupled-NLS simulation's result is excellent for all four sequences at this distance. At larger distances, sequences 1 and 2 develop significant pulse pattern distortions due to substantial position shifts and unequal amplitudes. In contrast, no significant pulse pattern distortions are observed for sequences 3 and 4 even at $z = 4000$ and $z = 7000$, while at $z = 10^4$, these sequences exhibit weak distortions due to small position shifts.

The dynamics of the pulse patterns, which is observed in the chaotic regime of energy dynamics, is even more striking. Fig. 21 shows the pulse patterns $|\psi_j(t, z)|$ obtained by the numerical simulation with Eq. (27) for $m = 0.61$ along with the corresponding theoretical prediction of Eq. (A1). We see that at $z = 1000$, there are no significant pulse pattern distortions, and therefore, the agreement between the theoretical prediction and the simulation's result is excellent for all four sequences. However, at larger distances, the four sequences develop strong pulse pattern distortions. The distortions in sequences 1 and 4 are dominated by high-frequency modulations of the $|\psi_j(t,z)|$. On the other hand, the distortions in sequences 2 and 3 are due to substantial position shifts and unequal pulse amplitudes, but without any significant high-frequency modulations of the $|\psi_j(t,z)|$. The spatiotemporal dynamics that is observed for the other $m$ vales in the chaotic regime of energy dynamics ($m = 0.58$ and $m = 0.6$) is similar to the one seen in Fig. 21. In contrast, the pulse pattern dynamics that is observed for the other $m$ values outside of the chaotic regime of energy dynamics ($m = 0.7$ and $m = 0.8$) is similar to the one in Fig. 20.

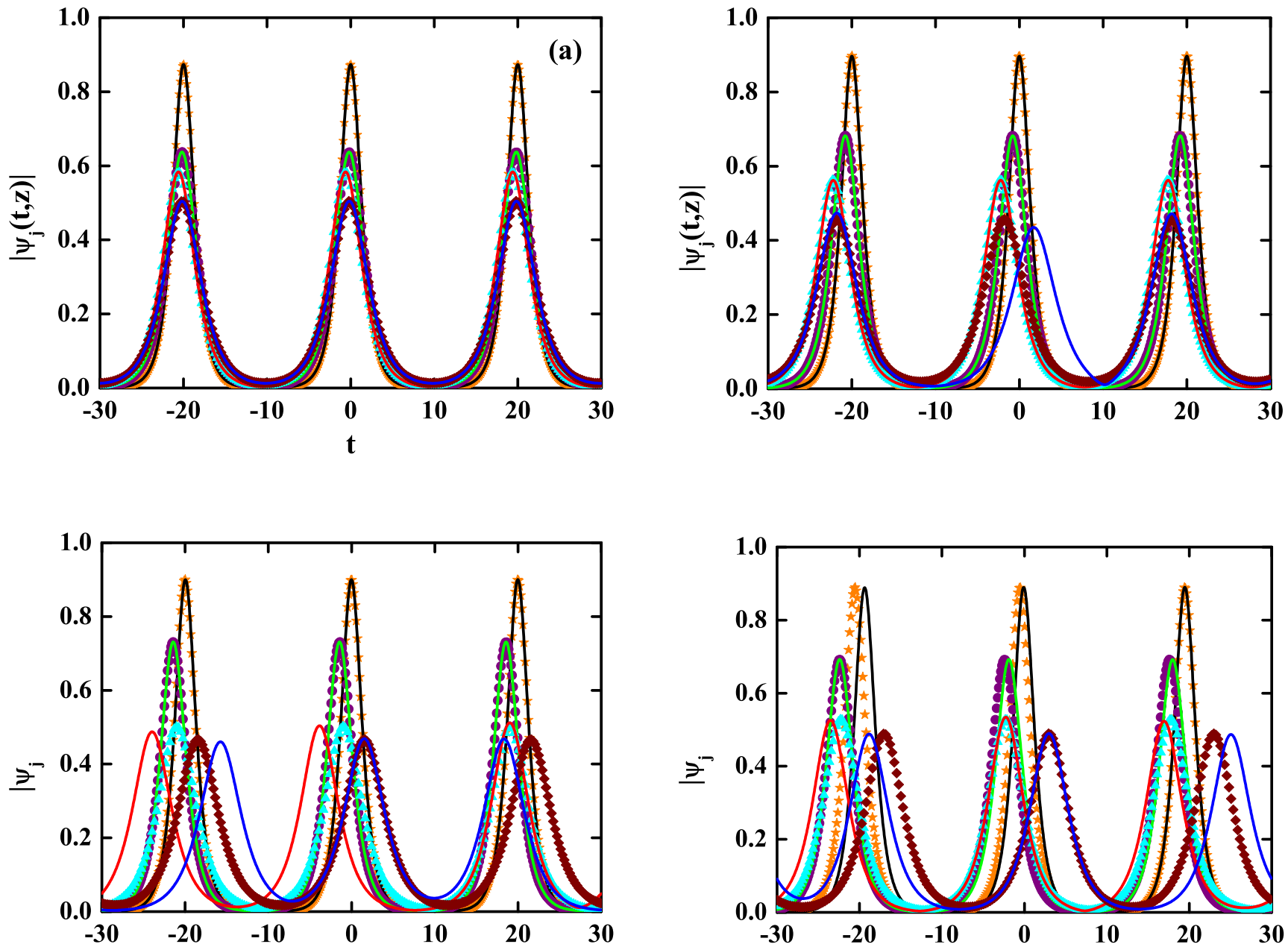

FIG. 20: The pulse patterns $|\psi_j(t,z)|$ obtained with the shifted 1990 coupled-NLS model (27) for $m = 0.52$ at $z = 1000$ (a), $z = 4000$ (b), $z = 7000$ (c), and $z = 10^4$ (d). The solid blue, solid red, solid green, and solid black curves represent $|\psi_j(t,z)|$ with $j = 1, 2, 3, 4$, obtained by numerical solution of Eq. (27). The brown diamonds, cyan triangles, purple circles, and orange stars represent the theoretical predictions for $|\psi_j(t,z)|$ with $j = 1, 2, 3, 4$, obtained with Eq. (A1).

The results in Figs. 19 - 21 and similar results, which are achieved with other parameter setups, clearly show that the pulse sequences in the nonlinear waveguide array systems of Eq. (27) exhibit transition to dissipative spatiotemporal chaos. Thus, our study demonstrate transition to spatiotemporal chaos with high accuracy in two distinct types of waveguide array systems with conservation of the total energy, which are described by Eqs. (26) and (27). These findings are very surprising, considering the strong pulse pattern distortions that are observed in the coupled-NLS simulations. Furthermore, our findings also strongly suggest that the shifting transformation $\mathbf{U}' = \mathbf{U} + \mathbf{a}$, which was introduced in section III, is crucial for the realization of spatiotemporal chaos in these nonlinear waveguide array systems.

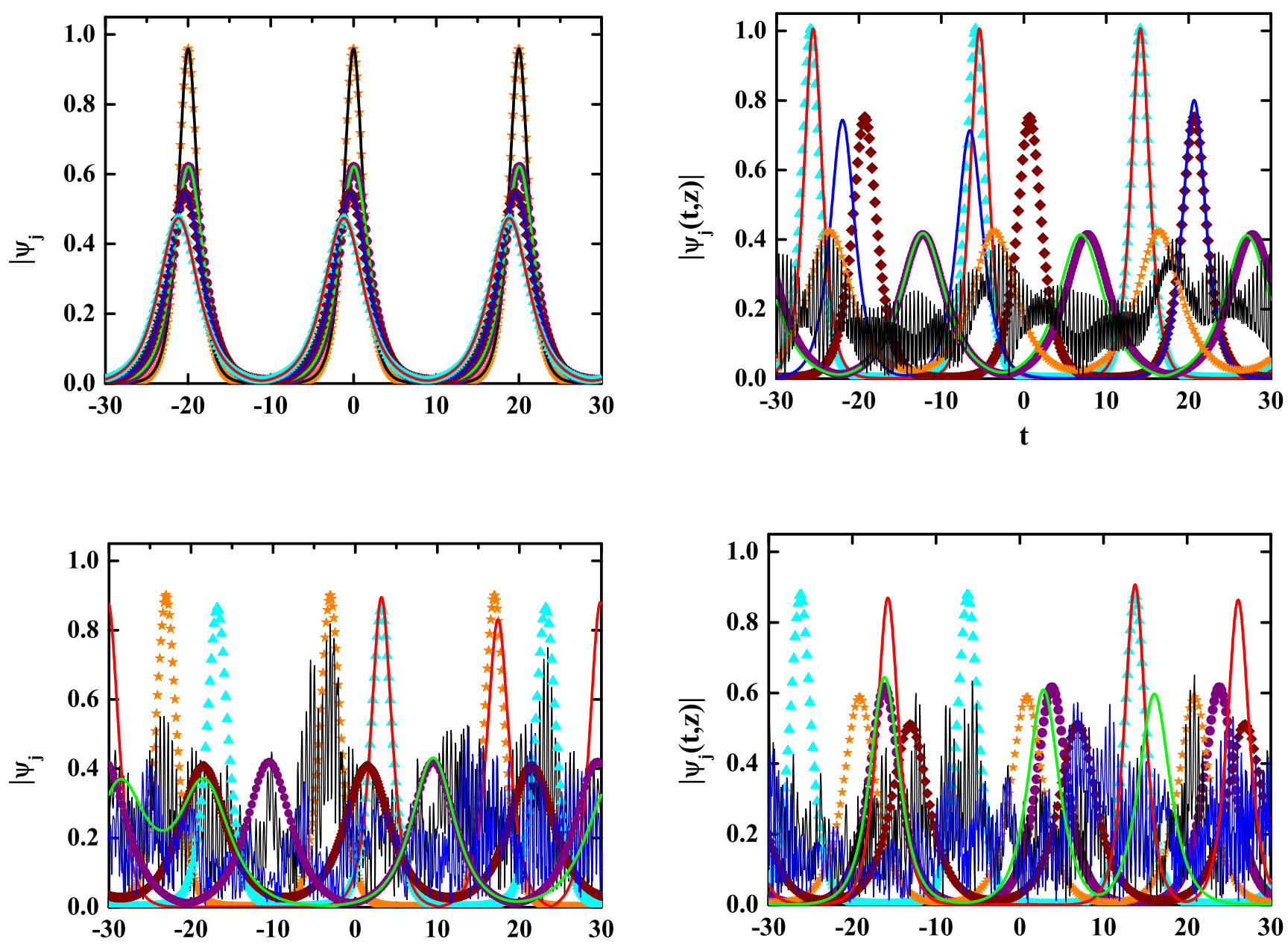

FIG. 21: The pulse patterns $|\psi_j(t,z)|$ obtained with the shifted 1990 coupled-NLS model (27) for $m = 0.61$ at $z = 1000$ (a), $z = 5000$ (b), $z = 1.5 \times 10^4$ (c), and $z = 3 \times 10^4$ (d). The solid blue, solid red, solid green, and solid black curves represent $|\psi_j(t,z)|$ with $j = 1, 2, 3, 4$, obtained by numerical solution of Eq. (27). The brown diamonds, cyan triangles, purple circles, and orange stars represent the theoretical predictions for $|\psi_j(t,z)|$ with $j = 1, 2, 3, 4$, obtained with Eq. (A1).

## VI. CONCLUSIONS

We studied the propagation of multiple sequences of optical pulses in two distinct types of nonlinear waveguide array systems with cubic gain and loss. By employing a perturbation theory for the fundamental soliton of the cubic NLS equation, we showed that the dynamics of pulse energies in these systems is described by two types of generalized LV models with conservation of the total energy, the unshifted 1989 Di Cera model, and the unshifted 1990 Di Cera model. These generalized LV models were used in the past to describe chemical concentration dynamics in networks of autocatalytic chemical reactions [67, 68]. More importantly, the models exhibit chaotic dynamics in a wide region in parameter space. We checked the predictions of the two unshifted LV models by extensive numerical simulations with two unshifted perturbed coupled cubic NLS models. We found that the predictions of the LV models and the results of the simulations with the unshifted coupled-NLS models for energy dynamics strongly disagree in the chaotic regime of the LV models. These findings strongly indicate that spatiotemporal chaos cannot be realized in systems described by the unshifted coupled-NLS models.

The deficiencies of the unshifted LV and coupled-NLS models stem mainly from the fact that in the chaotic regime, the pulse energies attain very small values at some distances. As a result, at these distances, the pulse patterns of some of the sequences are rather flat, and are subject to strong modulation instabilities. To overcome these deficiencies, we introduced a shifting transformation of the pulse energies of the form $U'_j = U_j + a_j$, where $0 < a_j < 1$ for $1 \le j \le 4$. For any initial condition $U_j(0) > 0$ for $1 \le j \le 4$, the $U'_j(z)$ satisfy $U'_j(z) \ge a_j > 0$ for all $z$, and therefore, the problem of small $U_j(z)$ values is circumvented. We refer to the new LV and coupled-NLS models obtained by this shifting transformation as the shifted 1989 and 1990 Di Cera models, and the shifted 1989 and 1990 coupled-NLS models, respectively. Energy dynamics in the two shifted Di Cera models is just a shifted version (in phase space) of energy dynamics in the two unshifted Di Cera models. As a result, the two shifted Di Cera models exhibit the same chaotic dynamics as the two unshifted Di Cera models.

Our extensive numerical simulations with the two shifted coupled-NLS models showed excellent agreement between the shifted LV and coupled-NLS models results for energy dynamics both in the chaotic regime and outside of the chaotic regime. Thus, our study provided the first demonstration of transition to spatiotemporal chaos with multiple colliding pulse sequences in systems described by perturbed coupled-NLS models. Our findings are very surprising since chaotic dynamics of pulse energies was realized with high precision despite the severe pulse pattern distortions experienced by the pulse sequences, and despite the violation of the weak perturbation assumption, which was used in the derivation of the LV models. Furthermore, due to the central roles of the cubic NLS and LV models in nonlinear science and engineering, our results are of broad interest in the context of research on spatiotemporal chaos in general.

**Acknowledgments**

The authors are thankful to Toan T. Huynh for valuable help in the initial stages of this research project. D.C. is grateful to the Department of Mathematics and Technology at Kean University for providing technological support for the numerical computations.

## APPENDIX A: THE THEORETICAL PREDICTION FOR THE $|\psi_j(t, z)|$

The theoretical prediction for the $|\psi_j(t, z)|$, which is used extensively in section V, is based on the adiabatic perturbation theory for the fundamental solitons of the cubic NLS equation, and on the assumptions that the soliton sequences remain periodic and are only weakly distorted. By the adiabatic perturbation theory, we can write the solution to the perturbed NLS equation as $\psi_j(t, z) = \psi_{sj}(t, z) + \nu_{rj}(t, z)$, where $\psi_{sj}(t, z)$ is the soliton part, and $\nu_{rj}(t, z)$ is the radiation part [23, 24, 61, 78]. Since we assume that the soliton sequences are only weakly distorted, we can neglect the radiation part, and take the theoretical prediction as the soliton part: $\psi_j^{(th)}(t, z) \equiv \psi_{sj}(t, z)$ [36, 39]. Additionally, since we assume that the soliton sequences are periodic and are weakly distorted, we can write $\psi_{sj}$ as the sum of $2N + 1$ fundamental solitons of the unperturbed cubic NLS equation with slowly varying parameters, whose peaks are separated by a constant integer multiple of $T$ [24, 36, 39]. It follows that the theoretical prediction for $\psi_j(t, z)$ is given by [36, 39]:

$$\psi_j^{(th)}(t, z) = \eta_j(z) e^{i\theta_j(z)} \sum_{n=-N-N_s}^{N+N_s} \frac{\exp\{-i\beta_j(z)\,[t - y_j(z) - nT]\}}{\cosh\{\eta_j(z)\,[t - y_j(z) - nT]\}}, \tag{A1}$$

where $t_{min} \le t \le t_{max}$, $\eta_j(z)$ is the common amplitude of the solitons in the theoretical $j$th sequence, $\beta_j(z)$ is the common frequency, and $\theta_j(z)$ is the common overall phase. Additionally, $y_j(z)$ is the position of the "central" soliton in the theoretical $j$th sequence, i.e., the soliton that is located in the interval $-T/2 \le t \le T/2$. Note that we enforce the periodicity of the theoretical pulse sequences with good accuracy by adding $N_s$ solitons on both sides of the computational domain. This addition is also in full accordance with the periodic boundary conditions that are used in the coupled-NLS simulations.

The theoretical prediction for the $j$th pulse pattern at distance $z$, $|\psi_j^{(th)}(t,z)|$, is calculated by using Eq. (A1) with values of $\eta_j(z)$, $\beta_j(z)$, and $y_j(z)$, which are determined in the following manner.

1. $\beta_j(z) = \beta_j(0)$.
2. To find $y_j(z)$, we first locate the global maximum of $|\psi_j^{(num)}(t,z)|$ in the computational domain $[t_{min}, t_{max}]$ and denote it by $y_{jm}(z)$. The quantity $y_j(z)$ is then given by:

$$y_j(z) = \begin{cases} y_{jm}(z) + NT & \text{if } -\left(N+\frac{1}{2}\right)T \le y_{jm}(z) < -\left(N-\frac{1}{2}\right)T, \\ y_{jm}(z) + (N-1)T & \text{if } -\left(N-\frac{1}{2}\right)T \le y_{jm}(z) < -\left(N-\frac{3}{2}\right)T, \\ \vdots & \vdots \\ y_{jm}(z) & \text{if } -\frac{1}{2}T \le y_{jm}(z) \le \frac{1}{2}T, \\ \vdots & \vdots \\ y_{jm}(z) - (N-1)T & \text{if } \left(N-\frac{3}{2}\right)T < y_{jm}(z) \le \left(N-\frac{1}{2}\right)T, \\ y_{jm}(z) - NT & \text{if } \left(N-\frac{1}{2}\right)T < y_{jm}(z) \le \left(N+\frac{1}{2}\right)T. \end{cases} \tag{A2}$$

3. The $\eta_j(z)$ values are calculated from the numerical $U_j(z)$ values, $U_j^{(num)}(z)$, by the following two-step procedure, which is justified by arguments similar to the ones presented in the third and fourth paragraphs in section V A.

   (a) In the first step, we determine the $U_j^{(num)}(z)$ values by using Eq. (58).

   (b) In the second step, we determine $\eta_j(z)$ by numerical solution of the equation

$$(4N+2)^{-1} \int_{t_{min}}^{t_{max}} dt\, |\psi_j^{(th)}(t,z)|^2 = U_j^{(num)}(z), \tag{A3}$$

   where $\psi_j^{(th)}(t,z)$ is given by Eq. (A1).

---

[1] G.P. Agrawal, Nonlinear Fiber Optics, Academic, San Diego, CA, 2019.

[2] A. Hasegawa, Y. Kodama, Solitons in Optical Communications, Clarendon, Oxford, 1995.

[3] E. Iannone, F. Matera, A. Mecozzi, M. Settembre, Nonlinear Optical Communication Networks, Wiley, New York, 1998.

[4] L.F. Mollenauer, J.P. Gordon, Solitons in Optical Fibers: Fundamentals and Applications, Academic, San Diego, CA, 2006.

[5] F. Dalfovo, S. Giorgini, L.P. Pitaevskii, S. Stringari, Rev. Mod. Phys. 71 (1999) 463.

[6] R. Carretero-González, D.J. Frantzeskakis, P.G. Kevrekidis, Nonlinearity 21 (2008) R139.

[7] Y.S. Kivshar, B.A. Malomed, Rev. Mod. Phys. 61 (1989) 763.

[8] N. Asano, T. Taniuti, N. Yajima, J. Math. Phys. 10 (1969) 2020.

[9] W. Horton, Y.H. Ichikawa, Chaos and Structure in Nonlinear Plasmas, World Scientific, Singapore, 1996.

[10] A.C. Newell, Solitons in Mathematics and Physics, SIAM, Philadelphia, 1985.

[11] M.J. Ablowitz, P.A. Clarkson, Solitons, Nonlinear Evolution Equations and Inverse Scattering, Cambridge University Press, Cambridge, 1991.

[12] A.R. Osborne, Nonlinear Ocean Waves and the Inverse Scattering Transform, Elsevier, Amsterdam, 2010.

[13] G.P. Agrawal, Applications of Nonlinear Fiber Optics, Academic, San Diego, CA, 2020.

[14] Y. Kodama, A. Hasegawa, IEEE J. Quantum Electron. 23 (1987) 510.

[15] Q. Lin, O.J. Painter, G.P. Agrawal, Opt. Express 15 (2007) 16604.

[16] R. Dekker, N. Usechak, M. Först, A. Driessen, J. Phys. D (2007) 40 R249.

[17] M. Borghi, C. Castellan, S. Signorini, A. Trenti, L. Pavesi, J. Opt. 19 (2017) 093002.

[18] A. Hayat, A. Nevet, M. Orenstein, Phys. Rev. Lett. 102 (2009) 183002.

[19] A. Hayat, A. Nevet, P. Ginzburg, M. Orenstein, Semicond. Sci. Technol. 26 (2011) 083001.

[20] M. Reichert, A.L. Smirl, G. Salamo, D.J. Hagan, E.W. Van Stryland, Phys. Rev. Lett. 117 (2016) 073602.

[21] Y. Kodama, J. Stat. Phys. 39 (1985) 597.

[22] J.N. Elgin, T. Brabec, S.M.J. Kelly, Opt. Commun. 114 (1995) 321.

[23] D.J. Kaup, Phys. Rev. A 42 (1990) 5689.

[24] M. Chertkov, Y. Chung, A. Dyachenko, I. Gabitov, I. Kolokolov, V. Lebedev, Phys. Rev. E 67 (2003) 036615.

[25] A. Peleg, D. Chakraborty, Phys. Rev. A 98 (2018) 013853.

[26] E.A. Kuznetsov, A.V. Mikhailov, and I.A. Shimokhin, Physica D 87 (1995) 201.

[27] K. Smith, L.F. Mollenauer, Opt. Lett. 14 (1989) 1284.

[28] F. Forghieri, R.W. Tkach, A.R. Chraplyvy, in I.P. Kaminow and T.L. Koch (Eds.), Optical Fiber Telecommunications, Vol. III, Academic, San Diego, CA, 1997 (Chapter 8).

[29] R.-J. Essiambre, G. Kramer, P.J. Winzer, G.J. Foschini, B. Goebel, J. Lightwave Technol. 28

(2010) 662.

[30] Q.M. Nguyen, A. Peleg, Opt. Commun. 283 (2010) 3500.

[31] A. Peleg, Q.M. Nguyen, Y. Chung, Phys. Rev. A 82 (2010) 053830.

[32] A. Peleg, Y. Chung, Phys. Rev. A 85 (2012) 063828.

[33] D. Chakraborty, A. Peleg, J.-H. Jung, Phys. Rev. A 88 (2013) 023845.

[34] Q.M. Nguyen, A. Peleg, T.P. Tran, Phys. Rev. A 91 (2015) 013839.

[35] D. Chakraborty, A. Peleg, Q.M. Nguyen, Opt. Commun. 371 (2016) 252.

[36] A. Peleg, Q.M. Nguyen, T.P. Tran, Opt. Commun. 380 (2016) 41.

[37] A. Peleg, Q.M. Nguyen, T.T. Huynh, Eur. Phys. J. D 71 (2017) 30.

[38] A. Peleg, D. Chakraborty, Commun. Nonlinear Sci. Numer. Simulat. 63 (2018) 145.

[39] A. Peleg, T.T. Huynh, Physica D 466 (2024) 134222.

[40] D.K. Campbell, J.F. Schonfeld, C.A. Wingate, Physica D 9 (1983) 1.

[41] D.K. Campbell, M. Peyrard, P. Sodano, Physica D 19 (1986) 165.

[42] R.H. Goodman, R. Haberman, Phys. Rev. Lett. 98 (2007) 104103.

[43] Y. Zhu, R. Haberman, J. Yang, Phys. Rev. Lett. 100 14 (2008) 143901.

[44] I.S. Aranson, L. Kramer, Rev. Mod. Phys. 74 (2002) 99.

[45] B.I. Shraiman, A. Pumir, W. van Saarloos, P.C. Hohenberg, H. Chaté, M. Holen, Physica D 57 (1992) 241.

[46] R.J. Deissler, H.R. Brand, Phys. Rev. Lett. 72 (1994) 478.

[47] M. Howard, M. van Hecke, Phys. Rev. E 68 (2003) 026213.

[48] J.M. Soto-Crespo, N. Akhmediev, Phys. Rev. Lett. 95 (2005) 024101.

[49] A. Peleg, Phys. Lett. A 360 (2007) 533.

[50] Y. Chung, A. Peleg, Phys. Rev. A 77 (2008) 063835.

[51] A. Peleg, Phys. Lett. A 373 (2009) 2734.

[52] The dimensionless distance $z$ in Eq. (1) is $z = X/(2L_D)$, where $X$ is the dimensional distance, $L_D = \tau_0^2/|\tilde{\beta}_2|$ is the dispersion length, $\tau_0$ is the soliton width, and $\tilde{\beta}_2$ is the second-order dispersion coefficient. The dimensionless time is $t = \tau/\tau_0$, where $\tau$ is time. $\psi_j = (\gamma_3\tau_0^2/|\tilde{\beta}_2|)^{1/2}E_j$, where $E_j$ is the electric field of the $j$th pulse and $\gamma_3$ is the cubic nonlinearity coefficient.

[53] R.A. Negres, J.M. Hales, A. Kobyakov, D.J. Hagan, E.W. Van Stryland, IEEE J. Quantum Electron. 38 (2002) 1205.

[54] C.M. Cirloganu, L.A. Padilha, D.A. Fishman, S.Webster, D.J. Hagan, E.W. Van Stryland,

Opt. Express 19 (2011) 22951.

[55] D.A. Fishman, C.M. Cirloganu, S. Webster, L.A. Padilha, M. Monroe, D.J. Hagan, E.W. Van Stryland, Nat. Photonics 5 (2011) 561.

[56] D.C. Hutchings, M. Sheik-Bahae, D.J. Hagan, E.W. Van Stryland, Opt. Quantum Electron. 24 (1992) 1.

[57] M. Sheik-Bahae, Nonlinear Optics Basics: Kramers-Krönig Relations in Nonlinear Optics, in B.D. Guenther, D.G. Steel (Eds.), Encyclopedia of Modern Optics, Academic, London, UK, 2004.

[58] The parameter $\epsilon_3$ in Eq. (3) is related to the dimensional cubic loss or gain parameter $\rho_3$ by $\epsilon_3 = 2\rho_3/\gamma_3$.

[59] A. Peleg, M. Chertkov, I. Gabitov, Phys. Rev. E 68 (2003) 026605.

[60] J. Soneson, A. Peleg, Physica D 195 (2004) 123.

[61] Y. Chung, A. Peleg, Nonlinearity 18 (2005) 1555.

[62] In the actual calculation it is convenient to take $\beta_1 = 0$, such that $\Delta\beta = \beta_2$, as was done in Refs. [31, 59–61].

[63] L.F. Mollenauer, P.V. Mamyshev, IEEE J. Quantum Electron. 34 (1998) 2089.

[64] M. Nakazawa, IEEE J. Sel. Top. Quant. Electron. 6 (2000) 1332.

[65] A. Peleg, Q.M. Nguyen, T.T. Huynh, Eur. Phys. J. D 71 (2017) 315.

[66] A.H. Gnauck, P.J. Winzer, J. Lightwave Technol. 23 (2005) 115.

[67] E. Di Cera, P.E. Phillipson, J. Wyman, Proceedings of the National Academy of Sciences 86 (1989) 142.

[68] S. Mori, E. Di Cera, Phys. Lett. A 143 (1990) 369.

[69] Notice that the notations in Eqs. (18)-(20) are the same as the ones used in Eqs. (1) and (3).

[70] Here we also provide some corrections to the analysis presented in p. 144 of Ref. [67].

[71] M. Lakshmanan, S. Rajasekar, Nonlinear Dynamics, Springer, Berlin, 2002.

[72] S. Wiggins, Introduction to Applied Nonlinear Dynamical Systems and Chaos, Springer, New York, 2003.

[73] S.H. Strogatz, Nonlinear Dynamics and Chaos: with Applications to Physics, Biology, Chemistry, and Engineering, Westview, Cambridge, MA, 1994.

[74] E.N. Lorenz, J. Atm. Sci. 20 (1963) 130.

[75] B. Hasselblatt, A. Katok, A First Course in Dynamics: with a Panorama of Recent Develop-

ments, Cambridge University Press, Cambridge, UK, 2003.

[76] A. Wolf, J.B. Swift, H.L. Swinney, J.A. Vastano, Physica D 16 (1985) 285.

[77] J. Yang, Nonlinear Waves in Integrable and Nonintegrable Systems, SIAM, Philadelphia, 2010.

[78] A. Peleg, D. Chakraborty, Physica D 406 (2020) 132397.